\documentclass[aps,prd,amsmath,amssymb,amsfonts,showpacs,superscriptaddress,floatfix,twocolumn]{revtex4}
\usepackage{amsmath}
\usepackage{amssymb}
\usepackage{graphicx}
\usepackage{bbm}
\usepackage{caption}
\usepackage{subcaption}
\usepackage{epsfig}
\usepackage{braket}
\usepackage{slashed}
\usepackage{bm}
\usepackage[usenames,dvipsnames]{color}
\usepackage{color}
\usepackage{float}
\usepackage{hyperref}

\def\be{\begin{equation}}
\def\ee{\end{equation}}

\def\bea{\begin{eqnarray}}
\def\eea{\end{eqnarray}}
\begin{document}

\title{Finite-representation approximation of lattice gauge theories at the continuum limit with tensor networks} 

\author{Boye Buyens}
\affiliation{Department of Physics and Astronomy,
Ghent University,
  Krijgslaan 281, S9, 9000 Gent, Belgium}

\author{Simone Montangero}
  \affiliation{Institute for Complex Quantum Systems \& Center for Integrated Quantum Science and Technology (IQST), Ulm University,
  Albert-Einstein-Allee 11, D-89069 Ulm, Germany}
  \affiliation{Theoretische Physik, Universit\"at des Saarlandes, D-66123 Saarbr\"ucken, Germany}

\author{Jutho Haegeman}
 \affiliation{Department of Physics and Astronomy,
Ghent
 University,
  Krijgslaan 281, S9, 9000 Gent, Belgium}
 
\author{Frank Verstraete}
  \affiliation{Department of Physics and Astronomy,
Ghent
 University,
  Krijgslaan 281, S9, 9000 Gent, Belgium}
  \affiliation{Vienna Center for Quantum Science and Technology, Faculty of Physics, University of Vienna, Boltzmanngasse 5, 1090 Vienna, Austria}

 \author{Karel Van Acoleyen}
\affiliation{Department of Physics and Astronomy,
Ghent
 University,
  Krijgslaan 281, S9, 9000 Gent, Belgium}

\begin{abstract}
\noindent  
It has been established that Matrix Product States can be used to compute the ground state and single-particle excitations and their properties of lattice gauge theories at the continuum limit. However, by construction, in this formalism the Hilbert space of the gauge fields is truncated to a finite number of irreducible representations of the gauge group. We investigate quantitatively the influence of the truncation of the infinite number of representations in the Schwinger model, one-flavour QED$_2$, with a uniform electric background field. We compute the two-site reduced density matrix of the ground state and the weight of each of the representations. We find that this weight decays exponentially with the quadratic Casimir invariant of the representation which justifies the approach of truncating the Hilbert space of the gauge fields. Finally, we compute the single-particle spectrum of the model as a function of the electric background field.
\end{abstract}

\maketitle
\section{Introduction}
Wilsons' famous paper `Confinement of quarks' \cite{Wilson1974} has led to a big breakthrough for quantum chromodynamics (QCD), the theory describing strong interactions. Not only did Wilson offer an explanation why no free quarks appear in Nature, he also introduced his so-called Wilsonian path integral which enables to numerically compute expectation values using the Monte-Carlo method \cite{Kogut1983}. With the increasing computing power, this method has since its first results at the end of the Seventies \cite{Creutz1979} produced by far the most impressive results for QCD \cite{Ukawa2015,Bali1999}. Examples include the determination of the light hadron masses \cite{Fodor2012}, the determination of the quark masses \cite{McNeile2013} and obtaining the phase diagram at finite temperature \cite{Philipsen2013}. Despite its success this method is troubled by the sign-problem for finite fermion densities and, as defined on an Euclidean lattice, does not enable to perform real-time evolution. 

One year later, Kogut and Susskind presented their so-called Kogut-Susskind Hamiltonian \cite{Kogut1975} which corresponds to the Wilsonian path integral in the transfer matrix formalism \cite{Creutz1977,Baaquie1986}. As a Hamiltonian method, this approach overcomes in principle the sign problem and enables out-of-equilibrium simulations. A new problem that arises is the many-body problem: the dimension of the Hilbert space increases exponentially with the number of sites. This problem is not specific to QCD only, but holds for any strongly correlated many-body system: the Hilbert space describing the space of states is too large to simulate on a classical computer. 

Fortunately, often one is only interested in the low-energy states of a system and it turns out that the area law for entanglement entropy \cite{Hastings2007,Eisert2008,Masanes2009} gives a universal identification of the physically relevant tiny corner of Hilbert space for these states. This is where Tensor Network States (TNS) \cite{Orus2004,Cirac2009} come into play. They constitute a variational class of states that efficiently represent general low-energy states, by encoding the wave function into a set of tensors whose interconnections capture the proper entanglement behavior. The most famous example of TNS are the Matrix Product States (MPS) \cite{Schollwoeck2011} in one spatial dimension, which underlie White's Density Matrix Renormalization Group (DMRG) \cite{White1992}. Since the formulation of DMRG in terms of MPS, the number of MPS algorithms for many-body systems has increased rapidly. In particular for lattice gauge theories they have been applied successfully in many different contexts \cite{Banuls2013a,Banuls2016,Byrnes2002a,Byrnes2002b,Byrnes2003,Sugihara2005,Rico2014,Silvi2014,Kuehn2014,Kuehn2015,Pichler2015b,Silvi2016,Milsted2015,Dalmonte2016,Banuls2016b}.

In the Kogut-Susskind formalism the Hilbert space is defined by all the irreducible representations of the Lie-algebra underlying the gauge group. If the gauge group has an infinite number of irreducible representations, a natural question one could ask is whether we can safely truncate the infinite number of irreducible representations, defining the Hilbert space of the gauge fields, to a manageable number of representations. In particular, when approaching the continuum limit or a phase transition it is not obvious at all whether this is possible. In this paper we answer this question for (1+1) dimensional  quantum electrodynamics (QED) also known as the massive Schwinger model \cite{Schwinger1962}. Despite its simplicity as an abelian gauge theory in one spatial dimension, it has many interesting physical features like for instance confinement and chiral symmetry breaking. This made this model very attractive to test analytical and numerical methods \cite{Kogut1974,Coleman1975,Coleman1976,Banks1976,Fischler1979,Rothe1979,Hamer1982,Bender1985,Hetrick1988,Iso1990,Abdalla1991,Sachs1991,Kluger1992,Steele1995,Grignani1996,Hosotani1996,Rodriguez1996,Adam1997,Adam1997b,Adam1998,Schmidt1998,Armoni1999,Hosotani1998,Byrnes2002a,Byrnes2002b,Byrnes2003,Korcyl2012,Cichy2012,Banuls2013a,Hebenstreit2013,Hebenstreit2014,Kuehn2014,Banuls2016,Buyens2013,Buyens2014,Buyens2015,Buyens2016,Buyens2016b}. This model also gained interest from the experimentalists in the context of quantum simulators, see  \cite{Hauke2013,Wiese2013,Martinez2016,Kasper2015} and references therein. As a $U(1)$-gauge theory, all the irreducible representations are one-dimensional and can be labeled by an integer $p \in \mathbb{Z}$. As we will show in Sec.~\ref{sec:FCS2ev}, we will only need to retain a few of these representations to obtain reliable results in the continuum limit. 

Besides the fermion mass $m$ and the charge $g$, the Schwinger model also depends on the electric background field $\alpha \in [0,1[$. It has many interesting equivalent interpretations ranging from labeling the different vacua in the massless Schwinger model \cite{Adam1997} to the charge between an external quark-antiquark pair introduced in the empty vacuum \cite{Coleman1975}. Here we determine the single-particle excitations for different values of $\alpha$. Surprisingly, earlier numerical studies on the spectrum of the Schwinger model in the non-perturbative regime exclusively focussed on the cases $\alpha = 0$ \cite{Banuls2013a,Byrnes2003,Buyens2013} and $\alpha = 1/2$ \cite{Byrnes2002a,Byrnes2002b,Byrnes2003}. An overview of the low-energy spectrum is for instance useful to have a better understanding of the dynamics induced by a quench in the form of an electric field. Indeed, in \cite{Buyens2016b} we found that the behavior for small quenches can be understood by looking at the single-particle excitations of the Hamiltonian, even beyond linear response theory.

The paper is organized as follows. For the sake of completeness, in Sec.~\ref{sec:setup} we discuss the setup for the simulations: the Kogut-Susskind formulation of the Schwinger model, gauge invariant MPS and optimization methods for MPS. The reader familiar with these subjects can skip it and start directly from Sec.~\ref{sec:FCS2ev} where we introduce the systematics on how to obtain field expectation values from our simulations at finite lattice spacing. We properly address the issue on the needed variational freedom to faithfully approximate the low-energy states when approaching the continuum limit and the phase transition. We quantify the contribution of each of the irreducible $U(1)$-representations to the ground-state expectation values by investigating the two-site reduced density matrix. We also explain there how to extrapolate the expectation values at finite lattice spacing to the continuum limit. Finally, in Sec.~\ref{sec:SPspectrum} we report the results on the single-particle spectrum as a function of the electric background field. 

\section{Setup}\label{sec:setup}
\subsection{Kogut-Susskind Hamiltonian}
The massive Schwinger model is $(1+1)$-dimensional QED with one fermion flavor and, hence, is described by the Lagrangian density
\be \mathcal{L} = \bar{\psi}\left(\gamma^\mu(i\partial_\mu+g A_\mu) - m\right) \psi - \frac{1}{4}F_{\mu\nu}F^{\mu\nu}\,.\label{Lagrangian} \ee
Here, $\psi$ is a two-component fermion field, $A_\mu$ $(\mu = 0,1)$ denotes the $U(1)$ gauge field and $F_{\mu\nu}=\partial_\mu A_\nu-\partial_\nu A_\mu$ is the corresponding field strength tensor.

In the following, we employ a lattice regularization \`{a} la Kogut-Susskind \cite{Kogut1975}. Therefore the two-component fermions are decomposed into their particle and antiparticle components which reside on a staggered lattice. These staggered fermions are converted to quantum spins $1/2$ by a Jordan-Wigner transformation with the local Hilbert space basis $\{\ket{s_n}_n: s_n \in \{-1,1\} \}$ of $\sigma_z(n)$ at site $n$. The charge $-g$ `electrons' reside on the odd lattice sites, where spin down ($s=-1$) denotes an occupied site whereas spin up ($s=+1$) corresponds to an unoccupied site. Conversely, the even sites are related to charge $+g$ `positrons' for which spin down/up corresponds to an unoccupied/occupied sites, respectively.

Moreover, we introduce the compact gauge field $\theta(n) = a g A_1(n)$, which lives on the link that connects neighboring lattice sites, and its conjugate momentum $E(n)$, which correspond to the electric field. The commutation relation $[\theta(n),E(n')]=ig\delta_{n,n'}$ determines the spectrum of $E(n)$ up to a constant: $E(n)/g = L(n) + \alpha$. Here, $L(n)$ denotes the angular operator with integer spectrum and $\alpha \in \mathbb{R}$ corresponds to the background electric field. Any of the integer eigenvalues $p \in \mathbb{Z}$ of the angular operator $L(n)$ corresponds to an irreducible one-dimensional representations of the $U(1)$ gauge group. One of the main goals of this paper is to investigate how one can deal with this infinite number of representations in numerical simulations, this is treated in more detail in subsections \ref{subsec:errD} and \ref{subsec:varNeededMPS}. 

In this formulation the gauged spin Hamiltonian derived from the Lagrangian density Eq.~(\ref{Lagrangian}) reads (see \cite{Banks1976,Kogut1975} for more details):
\bea\label{equationH0} H&=& \frac{g}{2\sqrt{x}}\Biggl(\sum_{n \in \mathbb{Z}} \frac{1}{g^2} E(n)^2 + \frac{\sqrt{x}}{g} m \sum_{n \in \mathbb{Z}}(-1)^n\sigma_z(n) \nonumber
\\ &&+ x \sum_{n \in \mathbb{Z}}(\sigma^+ (n)e^{i\theta(n)}\sigma^-(n + 1) + h.c.)\biggl) \eea
where $\sigma^{\pm} = (1/2)(\sigma_x \pm i \sigma_y)$ are the ladder operators. Here we have introduced the parameter $x$ as the inverse squared lattice spacing in units of $g$: $x \equiv 1/(g^2a^2)$. The continuum limit then corresponds to $x\rightarrow \infty$.  
Notice the different second (mass) term in the Hamiltonian for even and odd sites which originates from the staggered formulation of the fermions. 

In the time-like axial gauge the Hamiltonian is still invariant under the residual time-independent local gauge transformations generated by:
\begin{align} g\mathcal{G}(n) = & E(n)-E(n-1)-\frac{g}{2}( \sigma_z(n) + (-1)^n )\,.\label{gauss} \end{align}
As a consequence, if we restrict ourselves to physical gauge invariant operators $O$, with $[O,\mathcal{G}(n)]=0$, the Hilbert space decomposes into dynamically disconnected superselection sectors, corresponding to the different eigenvalues of $G(n)$. In the absence of any background charge the physical sector then corresponds to the $\mathcal{G}(n)=0$ sector. Imposing this condition (for every $n$) on the physical states is also referred to as the Gauss law constraint, as this is indeed the discretized version of $\partial_z E - \rho=0$, where $\rho$ is the charge density of the dynamical fermions. 

The other superselection sectors correspond to states with background charges. Specifically, if we want to consider two probe charges, one with charge $-gQ$ at site $m_L$ and one with opposite charge $+gQ$ at site $m_R$, we have to restrict ourselves to the sector:
\begin{align} g\mathcal{G}(n) =  gQ(\delta_{n,m_L} - \delta_{n,m_R})\label{spingauss0}\,.\end{align} Notice that we consider both integer and non-integer (fractional) charges $Q$.  

As in the continuum case \cite{Coleman1975}, we can absorb the probe charges into a background electric field string that connects the two sites. This amounts to the substitution $E(n) = g[L(n) + \alpha(n)]$ where $\alpha(n)$ is only nonzero in between the sites: $\alpha(n) = -Q\Theta(0 \leq n < k)$; and $L(n)$ has an integer spectrum: $L(n)=p\in \mathbb{Z}$. In terms of $L(n)$ the Gauss constraint now reads:
 \be\label{eq:spingauss} G(n) = L(n) - L(n-1) -\frac{ \sigma_z(n) + (-1)^n }{2} = 0\,,  \ee and we finally find the Hamiltonian:
 \bea\label{eq:equationH0} H&=& \frac{g}{2\sqrt{x}}\Biggl(\sum_{n \in \mathbb{Z}} [L(n) + \alpha(n)]^2 + \frac{\sqrt{x}}{g} m \sum_{n \in \mathbb{Z}}(-1)^n\sigma_z(n) \nonumber
\\ &&+ x \sum_{n \in \mathbb{Z}}(\sigma^+ (n)e^{i\theta(n)}\sigma^-(n + 1) + h.c.)\biggl),\eea
in accordance with the continuum result of \cite{Coleman1976}. For our purpose we consider the Schwinger model in the thermodynamic limit in a uniform electric background field ($\alpha(n) = \alpha,\forall n$), hence the Hamiltonian reads
 \bea\label{eq:equationH} H_\alpha&=& \frac{g}{2\sqrt{x}}\Biggl(\sum_{n \in \mathbb{Z}} [L(n) + \alpha]^2 + \frac{\sqrt{x}}{g} m \sum_{n \in \mathbb{Z}}(-1)^n\sigma_z(n) \nonumber
\\ &&+ x \sum_{n \in \mathbb{Z}}(\sigma^+ (n)e^{i\theta(n)}\sigma^-(n + 1) + h.c.)\biggl).\eea
Note that we explicitly denoted the $\alpha-$dependence in $H_\alpha$.

\subsection{Phase diagram and single-particle spectrum for the Schwinger model}\label{subsec:phaseDiagSchwinger}
Before turning our attention to the numerics, we briefly discuss the phase diagram and the the single-particle spectrum that we can expect for the Schwinger model. This is based on analytical studies in the weak-coupling limit ($m/g \gg 1$) and the strong-coupling limit ($m/g \ll 1$), numerical studies in the non-perturbative regime in earlier studies and also the new results that are discussed in detail in Sec.~\ref{sec:SPspectrum}. In units $g=1$, there are two free parameters: $m/g$ and $\alpha$. Moreover, the model is periodic in $\alpha$ with period 1 and physics for $\alpha \in [0,1/2[$ can be mapped to physics for $\alpha \in [1/2,1]$ by the following transformation:
\begin{subequations}\label{eq:CTsymm}
\be \label{eq:CTelF} L(n) \rightarrow -1 - L(n+1), \theta(n) \rightarrow - \theta(n+1), \ee
\be \sigma^{\pm}(n) \rightarrow \sigma^{\mp}(n+1), \sigma_z(n) \rightarrow - \sigma_z(n+1). \ee
\end{subequations}
Indeed, under this transformation we find that $H_\alpha$ is mapped to $H_{1-\alpha}$. For $\alpha = 1/2$, it follows that this is actually a symmetry of the Hamiltonian: the so-called $CT$ symmetry (`C' charge conjugation, `T' translation over one site). As we discuss below, this symmetry plays a special role as there is a critical value $(m/g)_c$ of $(m/g)$ above which this symmetry is spontaneously broken. Also, for $\alpha = 0$ the Hamiltonian has also a $CT$ symmetry but now with $L(n) \rightarrow - L(n+1)$ instead of eq. (\ref{eq:CTelF}). In this case, this symmetry is not spontaneously broken for all values of $m/g$.  \\
\\
For $m/g = 0$, the model is exactly solvable and can be mapped to a Klein-Gordon field describing the so-called Schwinger boson with mass $g/\sqrt{\pi}$. Historically, this was the main motivation why Schwinger considered this model \cite{Schwinger1962}: the model is an example where a massless gauge field, the photon, acquires mass \cite{Schwinger1961} and as such it was in fact a pioneer for the Higgs mechanism. 
\\
\\When $m/g = 0$, physics is independent from $\alpha$. In contrast, when $m/g \neq 0$ it does depend on $\alpha$. The cases $\alpha = 0$ and $\alpha = 1/2$ are somehow special as the model exhibits in that case the $CT$ symmetry. Therefore, we first discuss the more generic case $0 < \alpha < 1/2$, afterwards we treat the cases $\alpha = 0$ and $\alpha = 1/2$. 
\\
\\ \textit{1. The case $0 < \alpha < 1/2$. }In mass perturbation theory \cite{Coleman1976,Adam1997}, $m/g \ll 1$,  there are two single-particle excitations for $\alpha \leq 0.25$ \footnote{With a single-particle or elementary excitation, we mean an energy eigenstate of the Hamiltonian that is separated by a gap from the rest of the spectrum (for a fixed momentum) \cite{Haegeman2013b}. In general it is believed that the spectrum of a quantum field theory consists of a number of single-particle excitations with energies $\mathcal{E}_1,\mathcal{E}_2,\ldots$ and a continuum spectrum starting from $2\mathcal{E}_1$. The physical picture is that eigenstates in the continuum spectrum are multi-particle scattering states of these elementary excitations which can decay into two or more of these elementary excitations. Therefore, they are refered to as `not stable'. It is possible that some eigenstates with energy larger than $2\mathcal{E}_1$ cannot decay into two or more elementary excitations due to additional symmetries in the Hamiltonian. For the Schwinger model it is believed that this only occurs for $\alpha = 0$ where the $CT$ symmetry prevents the decay of a particle with energy $\mathcal{E}_3$ with $2\mathcal{E}_1 \leq \mathcal{E}_3 < \mathcal{E}_1 +\mathcal{E}_2$. Such eigenstates are also refered to as elementary or single-particle excitations.}. The first single-particle excitations with energy $\mathcal{E}_1$ corresponds to the Schwinger boson in the limit $m/g \rightarrow 0$ while the second single-particle excitation with energy $\mathcal{E}_1$ is easiest interpreted as a bound state of two Schwinger bosons. When $\alpha \geq 0.25$, the energy for the second eigenvalue becomes larger than or equal to 2$\mathcal{E}_1$ and, therefore, it is not stable anymore. On the other hand, in the weak-coupling limit ($m/g \gg 1$) the number of single-particle excitations grows approximately with $(m/g)^2/(1/2 - \alpha)$ for $\alpha < 1/2$ \cite{Coleman1976}. 

\begin{figure}[t]
\includegraphics[width= 200 pt]{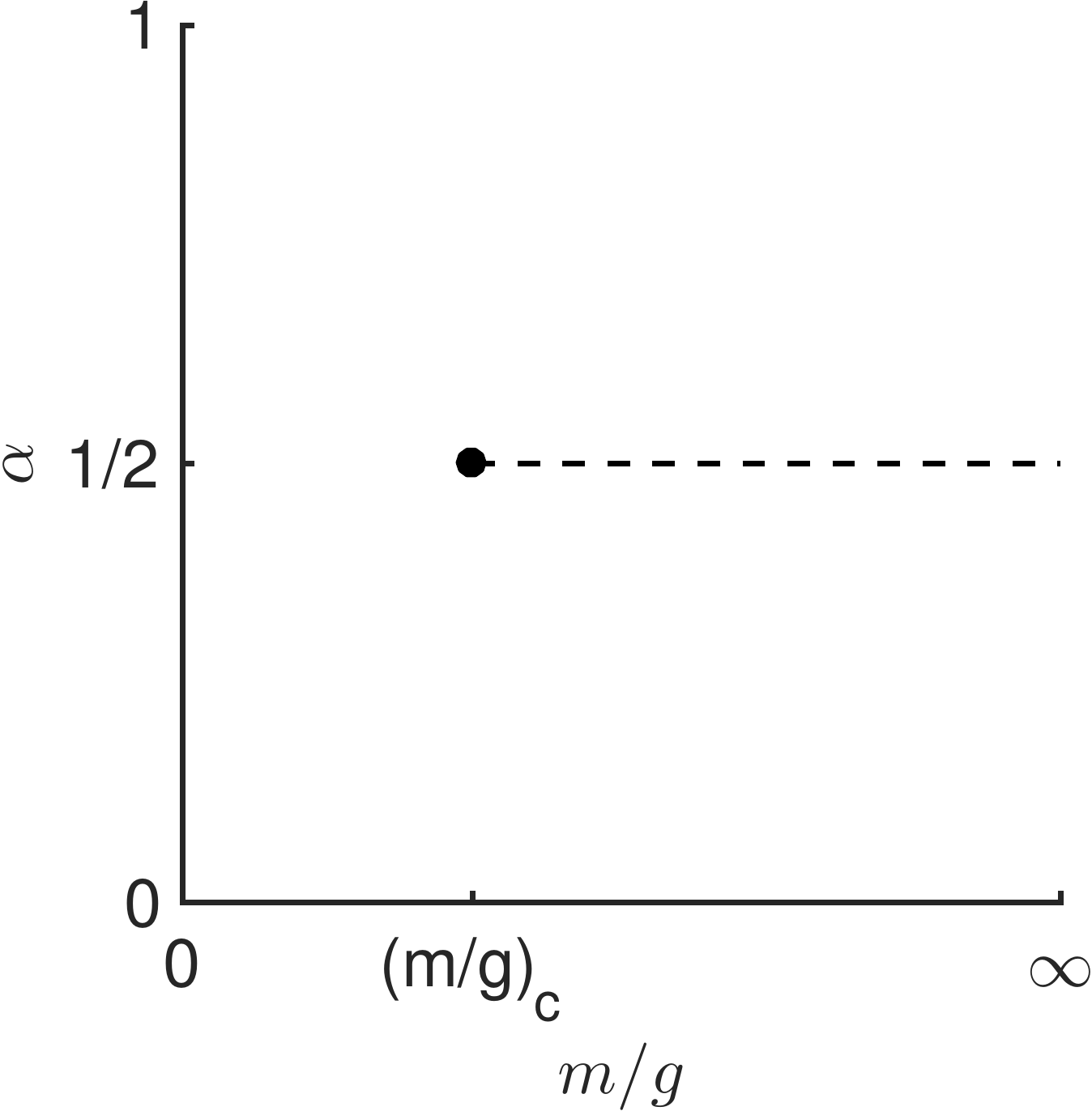}
\captionsetup{justification=raggedright}
\caption{\label{fig:PhaseDiag} The phase diagram of the Schwinger model. For $\alpha = 1/2$ there is a phase transition at $m/g = (m/g)_c$ related to the $CT$ symmetry. When $m/g < (m/g)_c$, the symmetry is not spontaneously broken while for $m/g > (m/g)_c$ the symmetry is spontaneously broken.}
\end{figure}

As we see in Sec.~\ref{sec:SPspectrum}, the behavior in the non-perturbative regime ($m/g \sim \mathcal{O}(1)$) interpolates between the strong- and the weak-coupling limit. For $m/g \lesssim 0.3$ we find the existence of a value $\alpha_c$ below which there are two single-particle excitations and above which there is only one single-particle excitation. This value of $\alpha_c$ comes closer to $1/2$ when $m/g$ increases. When $m/g \gtrsim 0.5$ we find that there are at least three single-particle excitations for $\alpha < 0.5$. Furthermore, our simulations suggest that the number of stable excitations increases when $\alpha$ tends to $1/2$, although this should be confirmed by other studies. This would then agree qualitatively with the behavior in the weak-coupling limit. 
\\
\\ \textit{2. The case $\alpha = 0$. }For $\alpha = 0$ the Hamiltonian has the $CT$ symmetry Eq.~(\ref{eq:CTsymm}), but where now $L(n) \rightarrow - L(n+1)$. Numerical simulations \cite{Byrnes2003,Banuls2013a,Buyens2013} pointed out that, for all values of $m/g$, this symmetry is not spontaneously broken. As a consequence, the energy eigenstates are divided into vector excitations, which flip sign under a $CT$ transformation, and scalar excitations, which are invariant under $CT$. The ground state and the second single-particle excitation with energy $\mathcal{E}_2$ behaves as a scalar under $CT$, while the first single-particle excitation with energy $\mathcal{E}_1$ transforms as a vector under $CT$. Furthermore, there is another single-particle excitation with energy $\mathcal{E}_3$. This excitation is best interpreted as a bound state of the excitations with energy $\mathcal{E}_1$ and $\mathcal{E}_2$. For $m/g \lesssim 0.3$, we found that  this vector excitation is only stable due to symmetry considerations (a vector excitation cannot decay into two vector excitations) and, hence, disappears form the single-particle spectrum for $\alpha \neq 0$. Similar to the case $\alpha > 0$, the number of scalar and vector single-particle excitations grows with $(m/g)^2$ when $m/g$ is large. 
\\
\\ \textit{3. The case $\alpha =  1/2$.}
As already mentioned before, for $\alpha = 0$ the $CT$ transformation Eq.~(\ref{eq:CTsymm}) is a symmetry of the Hamiltonian. Already in 1975, Coleman predicted the existence of a critical mass $(m/g)_c$ below which the ground state has the $CT$ symmetry and above which the $CT$ symmetry is spontaneously broken \cite{Coleman1976}, see Fig.~\ref{fig:PhaseDiag}. The most precise value for this critical mass has been found with MPS simulations by Byrnes \cite{Byrnes2002a,Byrnes2002b,Byrnes2003} and he found that $(m/g)_c = 0.3335(2)$. Byrnes also conjectured that the corresponding phase transition falls in the university class of the Ising model. When approaching the phase transition from below, $m/g \leq (m/g)_c$, the mass gap decreases and becomes zero at the phase transition. When $m/g \geq (m/g)_c$, the vacuum is two-fold degenerate and the elementary excitations are kinks connecting these two vacua. They were also predicted by Coleman \cite{Coleman1976} and the most precise estimates for their masses were found by Byrnes \cite{Byrnes2002a,Byrnes2002b,Byrnes2003}.

\subsection{Gauge invariant MPS}
Consider now the lattice spin-gauge system Eq.~(\ref{eq:equationH}) on $2N$ sites. On site $n$ the matter fields are represented by the spin operators with basis $\{ \ket{s_n}_n: s_n \in \{-1,1\}\}$. The gauge fields live on the links and on link $n$ their Hilbert space is spanned by the eigenkets $\{\ket{p_n}_{n}: p_n \in \mathbb{Z}\}$ of the angular operator $L(n)$. But notice that for our numerical scheme we only retain a finite range: $p_{min}(n+1) \leq p_n \leq p_{max}(n+1)$. We address the issue of which values to take for $p_{min}(n+1)$ and $p_{max}(n+1)$ in subsection \ref{subsec:varNeededMPS}. Furthermore, it is convenient to block site $n$ and link $n$ into one effective site with local Hilbert space spanned by $\{\ket{s_n,p_n}_n \}$. 
Writing $ \kappa_n = (s_n,p_n)$ we introduce the multi-index 
$$\bm{\kappa} = \bigl((s_1,p_1),(s_2,p_2),\ldots,(s_{2N},p_{2N})\bigl) = (\kappa_1,\ldots,\kappa_{2N}).$$ 
With these notations we have that the effective site $n$ is spanned by $\{\ket{\kappa_n}_n \}$. Therefore the Hilbert space of the full system of $2N$ sites and $2N$ links, which is the tensor product of the local Hilbert spaces, has basis $\{\ket{\bm{\kappa}}= \ket{\kappa_1}_1\ldots \ket{\kappa_{2N}}_{2N} \}$ and a general state $\ket{\Psi}$ is thus a linear combination of these $\ket{\bm{\kappa}}$: 
$$\ket{\Psi} = \sum_{\bm{\kappa}} C_{\kappa_1,\ldots,\kappa_{2N}}\ket{\bm{\kappa}}$$
 with basis coefficients $C_{\kappa_1,\ldots,\kappa_{2N}} \in \mathbb{C}$.\\
\\ 
A general MPS $\ket{\Psi[A]}$ now assumes a specific form for the basis coefficients \cite{Fannes1992}:  
\be\ket{\Psi[A]} = \sum_{\bm{\kappa}}v_L^\dagger A_{\kappa_1}(1) A_{\kappa_2}(2)\ldots A_{\kappa_{2N}}(2N)v_R \ket{\bm{\kappa}},\label{eq:MPS}\ee
where $A_{\kappa_n}(n)$ is a complex $D(n)\times D(n+1)$ matrix with components $[A_{\kappa_n}(n)]_{\alpha \beta}$ and where $v_L \in \mathbb{C}^{D(1)\times 1}, v_R \in \mathbb{C}^{D(2N+1)\times 1}$ are boundary vectors. The MPS ansatz thus associates with each site $n$ and every local basis state $\ket{\kappa_n}_n =\ket{s_n,p_n}_n$ a matrix $A_{\kappa_n}(n)= A_{s_n,p_n}(n)$. The indices $\alpha$ and $\beta$ are referred to as virtual indices, and $D = \max_n [D(n)]$ is called the bond dimension. 

To better understand the role of the bond dimension in MPS simulations it is useful to consider the Schmidt decomposition with respect to the bipartition of the lattice consisting of the two regions $\mathcal{A}_1(n) = \mathbb{Z}[1, \ldots, n]$ and $\mathcal{A}_2(n) = \mathbb{Z}[n+1,\ldots, 2N]$ \cite{Schollwoeck2011}:
\be \label{eq:MPSschmidt} \ket{\Psi[A]} = \sum_{\alpha = 1}^{D(n+1)} \sqrt{\sigma_{\alpha}(n)} \Ket{\psi_\alpha^{\mathcal{A}_1(n)}}\Ket{\psi_\alpha^{\mathcal{A}_2(n)}}. \ee
Here $\Ket{\Psi_\alpha^{\mathcal{A}_1(n)}}$ (resp. $\Ket{\Psi_\alpha^{\mathcal{A}_2(n)}}$) are orthonormal unit vectors living in the tensor product of the local Hilbert spaces belonging to the region $\mathcal{A}_1(n)$ (resp. $\mathcal{A}_2(n)$) and $\sigma_{\alpha}(n)$, called the Schmidt values, are non-negative numbers that sum to one. One can easily deduce that for a general MPS of the form Eq.~(\ref{eq:MPS}) at most $D(n+1)$ Schmidt values are nonzero (for the cut at site $n$ Eq.~(\ref{eq:MPSschmidt})). Hence, we see that taking a finite bond dimension for the MPS corresponds to a truncation in the Schmidt spectrum of a state. The success of MPS is then explained by the fact that ground states of local gapped Hamiltonians can indeed be approximated very efficiently in $D$ \cite{Hastings2007} and that the computation time for expectation values of local observables scales only with $D^3$, allowing for reliable simulations on an ordinary desktop.

To parameterize gauge invariant MPS, i.e. states that obey $G(n)\ket{\Psi[A]}=0$ for every $n$, it is convenient to give the virtual indices a multiple index structure $\alpha\rightarrow (q,\alpha_q); \beta \rightarrow (r,\beta_r)$, where $q$ resp. $r$ labels the eigenvalues of $L(n-1)$ resp. $L(n)$. In \cite{Buyens2013} it is proven that the condition $G(n)=0$, Eq.~(\ref{eq:spingauss}), then imposes the following form on the matrices:
\be
{[A_{s,p}(n)]}_{(q,\alpha_q),(r,\beta_r)} =  {[a_{q,s}(n)]}_{\alpha_q,\beta_r}\delta_{q+(s+(-1)^n)/2,r}\delta_{r,p}
\label{eq:gaugeMPS},\ee
where $\alpha_q = 1\ldots D_q(n)$, $\beta_r = 1 \ldots D_r(n+1)$. The first Kronecker delta is Gauss' law, $G(n) = 0$, on the virtual level while the second Kronecker delta connects the virtual index $r$ with the physical eigenvalue $p$ of $L(n)$. Because the indices $q$ (resp. $r$) label the eigenvalues of $L(n-1)$ (resp. $L(n)$) and we only retain the eigenvalues of $L(n-1)$ in the interval $\mathbb{Z}[p_{min}(n),p_{max}(n)]$ (resp. of $L(n)$ in the interval $\mathbb{Z}[p_{min}(n+1),p_{max}(n+1)]$), we have that $D_q(n) = 0$ for $q > p_{max}(n)$ and $q < p_{min}(n)$. The formal total bond dimension of this MPS is $D(n) = \sum_{q = p_{min}(n)}^{p_{max}(n)} D_q(n)$, but notice that, as Eq.~(\ref{eq:gaugeMPS}) takes a very specific form, the true variational freedom lies within the matrices $a_{q,s}(n) \in \mathbb{C}^{D_q(n) \times D_r(n+1)}$. 

Gauge invariance Eq.~(\ref{eq:spingauss}) is of course also reflected in the Schmidt decomposition Eq.~(\ref{eq:MPSschmidt}): for states of the form Eq.~(\ref{eq:gaugeMPS}) the Schmidt values can be labeled with the same double index $\alpha \rightarrow (q,\alpha_q)$. More specifically, the Schmidt decomposition Eq.~(\ref{eq:MPSschmidt}) now reads:
\be \label{eq:MPSschmidtGauge} \ket{\Psi[A]} = \hspace{-0.2cm}\sum_{q = p_{min}(n+1)}^{p_{max}(n+1)} \sum_{\alpha_q=1}^{D_q(n+1)} \sqrt{\sigma_{q,\alpha_q}(n)} \Ket{\psi_{q,\alpha_q}^{\mathcal{A}_1(n) }}\Ket{\psi_{q,\alpha_q}^{\mathcal{A}_2(n)}}. \ee

Another advantage of MPS simulations is that one can work directly in the thermodynamic limit $N \rightarrow \infty$, see \cite{Haegeman2011,Haegeman2013,Zauner2017}, bypassing any possible finite size artifacts. In the following we work in this limit. As in this limit the Hamiltonian is invariant under translations over two sites, $a_{q,s}(n)$ only depends on the parity of $n$. In particular it follows that the MPS ansatz eqs. (\ref{eq:MPS}) and (\ref{eq:gaugeMPS}), depends on a finite number of parameters. Similar as in \cite{Buyens2015} we block site $2n-1$ and $2n$ into one effective site $n$. Hence, the MPS ansatz for the ground state reads:
\begin{subequations}\label{eq:gaugeMPSFinalForm}
\be\ket{\Psi[a]} = \sum_{\bm{\kappa}}v_L^\dagger \left(\prod_{n = 1}^N A_{\kappa_{2n-1},\kappa_{2n}}\right)v_R \ket{\bm{\kappa}},\ee
($N \rightarrow + \infty$) with
\bea [A_{s_1,p_1,s_2,p_2}]_{(q,\alpha_q);(r,\beta_r)} &= \delta_{p_1,q + (s_1 -1)/2} \delta_{p_2,q + (s_1+s_2)/2} \nonumber
\\ & \delta_{p_2,r} [a_{q,s_1,s_2}]_{\alpha_q,\beta_r}\eea
\end{subequations}
where $[a_{q,s_1,s_2}]_{\alpha_q,\beta_r} \in \mathbb{C}^{D_q \times D_r}$ ($D_q = D_q(1)$); $s_1,s_2 = \pm 1$ and $q,p_2 \in \mathbb{Z}[p_{min},p_{max}]$ ($p_{min/max} = p_{min/max}(1)$). 

Finally, we note that, in the thermodynamic limit, the expectation values of local observables are independent of the boundary vectors $v_L$ and $v_R$. 

\subsection{TDVP for ground state}\label{subsec:TDVP}
The Time-Dependent Variational Principle (TDVP), introduced in \cite{Dirac1930}, provides a tool to evolve the Schr\"{o}dinger equation (SE) within a variational manifold in a global optimal way. Starting from the action principle for the SE, applying the Euler-Lagrange equations with respect to the variational parameters gives the TDVP equations. They have also a nice geometric interpretation \cite{Kramer1980}. Note that recently it has been shown that the TDVP unifies a lot of optimization methods for MPS such as the Density Renormalization Group algorithm and the infinite Time Evolving Block Decimation algorithm \cite{Haegeman2014a,Haegeman2016}.
\\ Here we use the framework of \cite{Haegeman2011,Haegeman2012} to apply the TDVP to the manifold of MPS of the form Eq.~(\ref{eq:gaugeMPSFinalForm}) with a fixed bond dimension. The TDVP replaces the SE, $i\partial_t\ket{\Psi[A]} = H_\alpha\ket{\Psi[A]}$, by
$$i \dot{a}_{q,s_1,s_2} = b_{q,s_1,s_2}[a], q \in \mathbb{Z}[p_{min},p_{max}]; s_1,s_2 \in \{-1,1\}, $$
where $b_{q,s_1,s_2}[a] \in \mathbb{C}^{D_q \times D_{q + (s_1+s_2)/2}}$ is a (quite complicated) expression, depending on all $a_{q,s_1,s_2} \in \mathbb{C}^{D_q \times D_{q + (s_1+s_2)/2}}$ and $H_\alpha$, which can be computed efficiently \cite{Haegeman2012}. To obtain an MPS approximation $\ket{\Psi[a]}$ for the ground state and the ground-state energy $\mathcal{E}_{0,\alpha}$, the evolution is performed in imaginary time $\tau$ ($d\tau = i dt$). A first-order Euler algorithm yields the following update-scheme
\be\label{eq:TDVPFlow} a_{q,s_1,s_2} (\tau + d\tau) = a_{q,s_1,s_2}(\tau) - b_{q,s_1,s_2}[a(\tau)] d\tau. \ee

Starting from an initial guess $a_{q,s_1,s_2}(0)$ and after sufficient iterations with $\vert d\tau \vert \ll 1$, this scheme provides the $a_{q,s_1,s_2}$ that yields the optimal MPS approximation $\ket{\Psi[a]}$ of the ground state of $H_\alpha$ within the class of states Eq.~(\ref{eq:gaugeMPSFinalForm}). Note that although the TDVP equation does not yield a steepest descent in parameter space, it produces the best approximation to a gradient descent in the full Hilbert space. In particular, we can also compute $\eta = \sqrt{\Braket{\Phi[\overline{b},\overline{a}]\vert\Phi[b,a]}}$, with
$$\Ket{\Phi[b,a]} = \frac{d}{d\tau}\Ket{\Psi[a + b \tau]}\Biggl\vert_{\tau = 0},$$ 
which yields a notion of the norm of the gradient in full Hilbert space. In our computations we halt the algorithm when $\eta = 10^{-9}$. Due to the infinite size of the lattice, $2N \rightarrow + \infty$, the ground-state energy is infrared divergent: 
\be\label{eq:gsIRdiv} \mathcal{E}_{0,\alpha} = 2N\tilde{\mathcal{E}}_{0,\alpha}, \ee
with $\tilde{\mathcal{E}}_{0,\alpha}$ the finite energy per site which can be obtained from the TDVP algorithm. Finally, we note that this steepest descent can also be extended to a naive variational conjugate gradient method, see \cite{Milsted2013} for an example.

\subsection{Rayleigh-Ritz for single-particle excitations}\label{subsec:RRforSPE}
In the previous section we discussed how one can use the TDVP to find an optimal MPS approximation $\ket{\Psi[a]}$, Eq.~(\ref{eq:gaugeMPSFinalForm}), for the ground state of $H_\alpha$. For the single-particle excitations with momentum $k \in [-\pi\sqrt{x},\pi\sqrt{x}]$ we now use the ansatz \cite{Haegeman2012}:
\begin{subequations}\label{eq:excAnsatz}
\begin{multline} \ket{\Phi_k[b,a]} = \sum_{n  = 1}^{N} e^{2ikn/\sqrt{x}} 
\\ \sum_{\{\zeta_n\}} v_L^\dagger \left(\prod_{m< n} A_{\zeta_m}\right)B_{\zeta_{n}} \left(\prod_{m > n} A_{\zeta_m}\right)v_R \ket{\bm{\zeta}},\end{multline} 
where $\zeta_m = (\kappa_{2m-1},\kappa_{2m}) = (s_{2m-1},p_{2m-1},s_{2m},p_{2m})$, $s_k \in \{-1,1\}$, $p_k \in \mathbb{Z}[p_{min},p_{max}]$; $\ket{\bm{\zeta}} = \ket{\zeta_1,\ldots,\zeta_N}$ and $A_\zeta$ corresponds to the ground state Eq.~(\ref{eq:gaugeMPSFinalForm}) of $H_\alpha$. Gauge invariance is imposed by
\bea [B_{s_1,p_1,s_2,p_2}]_{(q,\alpha_q);(r,\beta_r)} &= \delta_{p_1,q + (s_1 -1)/2} \delta_{p_2,q + (s_1+s_2)/2} \nonumber
\\ & \delta_{p_2,r} [b_{q,s_1,s_2}]_{\alpha_q,\beta_r}\eea
with $b_{q,s_1,s_2} \in \mathbb{C}^{D_q \times D_r}$.
\end{subequations}

The ansatz is an extension of the Feynman-Bijl ansatz \cite{Bijl1941,Feynman1956}, the single mode approximation \cite{Arovas1988} and the Rommer-\"Ostlund ansatz \cite{Ostlund1995} for single-particle excitations to the thermodynamic limit. Motivated by \cite{Zimmermann1958,Haegeman2013b}, where it is proven that the momentum-$k$ eigenstates with energy separated from the rest of the spectrum in that momentum sector can be created by acting with local operators on the vacuum, we expect that the states Eq.~(\ref{eq:excAnsatz}) provide a good ansatz for bound states as long as their energies are separated sufficiently far from the other eigenstates in their momentum sector.  

As the matrices $a_{q,s_1,s_2}$ in $\ket{\Phi_k[b,a]}$ are already fixed by the requirement that they correspond to the optimal approximation Eq.~(\ref{eq:gaugeMPSFinalForm}) for the ground state of $H_\alpha$, we only need to optimize the matrices $b_{q,s_1,s_2}$ such that
$$\frac{\Braket{\Phi_k[\overline{b},\overline{a}] \vert H_\alpha \vert \Phi_k[b,a]}}{\Braket{\Phi_k[\overline{b},\overline{a}] \vert \Phi_k[b,a]}} $$
is minimal with the requirement that $\ket{\Phi_k[b,a]}$ is orthogonal to $\ket{\Psi[a]}$. As the ground-state energy is infrared divergent, see Eq.~(\ref{eq:gsIRdiv}), we subtract its contribution from $H_\alpha$, i.e. we consider $H_\alpha \leftarrow H_\alpha -\mathcal{E}_{0,\alpha}$. As discussed in \cite{Haegeman2012}, this boils down to a generalized eigenvalue equation of the form
\be \label{eq:genEigExc} H_{eff}(k)\cdot {\bm{b}} = \mathcal{E}(k)\; N_{eff}(k) \cdot{\bm{b}} \ee 
with ${\bm{b}}$ the vector containing all the elements $b_{q,s_1,s_2}$ for $q \in \mathbb{Z}[p_{min},p_{max}]$, $s_k \in \{-1,1\}$. Here $H_{eff}$ and $N_{eff}$ are expressions depending on $H_\alpha$ and $a_{q,s_1,s_2}$ for which the action on ${\bm{b}}$ can be computed efficiently. Hence, using an iterative eigenvalue solver we obtain approximations $\ket{\Phi_k[b,a]}$ for the low-energy eigenstates with momentum $k$ and their energies $\mathcal{E}(k)$.

\section{From MPS to field expectation values}\label{sec:FCS2ev}
To obtain ground state expectation values and excitation energies for the Schwinger model, we have two tasks: 
\begin{itemize}
\item [T1.] Computing reliable MPS approximations for the ground state and single-particle excitations for several values of the lattice spacing $1/g\sqrt{x}$. 
\item[T2.] Extrapolating the results at non-zero lattice spacing to the continuum limit $x \rightarrow + \infty$. 
\end{itemize}

For T1, we compute MPS approximations of the form eqs. (\ref{eq:gaugeMPSFinalForm}) and (\ref{eq:excAnsatz}) to the ground state and the single-particle excitations for $x = \{9,16,25,36,50,60,75,90,100\}$. These are then used to compute the expectation values. However, as already noted, these MPS approximations are an effective truncation in the Schmidt spectrum associated to a half chain cut of the lattice and we only recover the exact ground state in the limit $D_q \rightarrow + \infty$ and $p_{max} \rightarrow + \infty$, see Eq.~(\ref{eq:MPSschmidtGauge}). In subsection \ref{subsec:errD}, we develop a systematic way to choose $D_q$ and $p_{max}$ according to the distribution of the Schmidt values among the eigenvalue sectors $q$ of $L(n)$. Then we assign an error on our results, originating from taking finite values for $D_q$ and $p_{max}$.  We show that our results are reliable up to $10^{-6}$ for the ground state expectation values and up to order $10^{-3}$ for the energies of the single-particle excitations. In subsection \ref{subsec:varNeededMPS}, we perform a detailed analysis on how the needed number of variational parameters changes (i.e., $D_q$ and $p_{min/max}$) as a function of $\alpha$ and $m/g$. In particular, we find that this number grows when approaching the continuum limit and the phase transition. However, even close to these limits we are still able to obtain accurate results with a manageable number of parameters. Moreover, we argue that we only need to retain a small number of irreducible representations of the $U(1)$ group which represent the Hilbert space of the gauge fields.

T2 is performed in subsection \ref{subsec:continuumLimit}. We explain there how to extrapolate the results for $x = 9,16,25,36,50,60,75,90,100$ to the continuum limit by fitting the data against polynomials in $1/\sqrt{x}$ and assign a proper error to our results originating from the choice of fitting interval and fitting function. As a check, we perform for  $m/g = 0.125$ an independent continuum extrapolation by using the results for $x = 90,100,150,200,250,300,350,400$ and show that the continuum estimates are in agreement with the ones obtained from $x = 9,16,25,36,50,60,75,90,100$.\\
\\
The results for $\alpha = 0$ have already been obtained in \cite{Buyens2013}. Here, we perform computations for  
$\alpha = 0.05,0.10,0.15,0.20,\dots, 
0.40,0.45,0.47,0.48,0.50$
and use interpolating fits to obtain the results for $\alpha \in [0,1/2]$. The results for all values of $\alpha$ follow from the $CT$ transformation Eq.~(\ref{eq:CTsymm}) and periodicity in $\alpha$ with period 1, see subsection \ref{subsec:phaseDiagSchwinger}.
\\
\\The observables that are considered here are the ground-state energy per unit of length $\sqrt{x}\mathcal{E}_{0,\alpha}/2N= \sqrt{x}\braket{H_\alpha}_0/2N$, the electric field $E_\alpha  = \Braket{ E }_0$, the chiral condensate $\Sigma_\alpha =  \Braket{ \bar{\psi}\psi }_0$ and the axial fermion current density $\Gamma_\alpha^5 = i \Braket{\bar{\psi}\gamma^5 \psi}_0$. Here $\Braket{\ldots}_0$ denotes the expectation value with respect to the ground state of $H_\alpha$. We refer to Eq.~(\ref{eq:quantDisc}) in appendix \ref{sec:appQuantLatVer} for the discretized versions of these quantities. 

Both the electric field and the axial fermion current density transform as vectors under a $CT$ transformation. Hence, they serve as an order parameter for the spontaneous symmetry breaking of the $CT$ symmetry at $\alpha = 1/2$. Also, as for $\alpha = 0$ the $CT$ symmetry is not spontaneously broken, they are then always zero: $E_{\alpha = 0} = \Gamma_{\alpha = 0}^5 = 0$. Finally, we note that these quantities are UV-finite. 

The chiral condensate is a scalar under the $CT$ transformation. Note however that, for $m/g \neq 0$, the chiral condensate is a UV-divergent quantity. In \cite{Banuls2016,Buyens2014} it is shown that for $\alpha = 0$ this divergence originates from the free chiral condensate (i.e. the chiral condensate for $g=0$). Here we remove the divergence by subtracting the chiral condensate for $\alpha = 0$, i.e. we consider
$$\Delta \Sigma_\alpha = \Sigma_\alpha - \Sigma_{\alpha = 0},$$
which is also UV finite. The energy per unit of length is UV divergent as well, and similar as for the chiral condensate, we obtain a UV-finite quantity by considering the so-called string tension $\sigma_\alpha$:
$$\sigma_\alpha =  \sqrt{x}\left(\frac{\mathcal{E}_{0,\alpha}-\mathcal{E}_{0,\alpha = 0}}{2N}\right).$$
This nomenclature stems from the investigation of confinement where $\sigma_\alpha$ indeed corresponds to the string tension (asymptotic force per unit of length) between an external quark-antiquark pair with charge $\alpha$ \cite{Coleman1975,Buyens2015}. 

Finally, we will consider the energy of the excited states (with respect to the ground-state energy), obtained via the method in subsection \ref{subsec:RRforSPE}. The energies are denoted by $\mathcal{E}_{1,\alpha}, \mathcal{E}_{2,\alpha}, \ldots$ with $\mathcal{E}_{1,\alpha} \leq  \mathcal{E}_{2,\alpha}\leq  \ldots$

\subsection{The limits $D_q \rightarrow + \infty$ and $p_{max} \rightarrow + \infty$}\label{subsec:errD}
\subsubsection{Ground state}
Here we discuss how to fix $D_q$ and $p_{max}$ in the numerical simulations and estimate the errors that this introduces. Taking a finite bond dimension $D_q$ corresponds to a truncation in the Schmidt decomposition Eq.~(\ref{eq:MPSschmidtGauge}):
\be \label{eq:MPSschmidtGaugeTsqInv} \ket{\Psi[a]} = \sum_{q = p_{min}}^{p_{max}} \sum_{\alpha_q=1}^{D_q} \sqrt{\sigma_{q,\alpha_q}} \Ket{\psi_{q,\alpha_q}^{\mathcal{A}_1(2n) }}\Ket{\psi_{q,\alpha_q}^{\mathcal{A}_2(2n)}} \ee
where we take into account translation invariance over two sites and where the half-chain cut is taken between an even site and an odd site.

\begin{figure}[t]
\begin{subfigure}[b]{.24\textwidth}
\includegraphics[width=\textwidth]{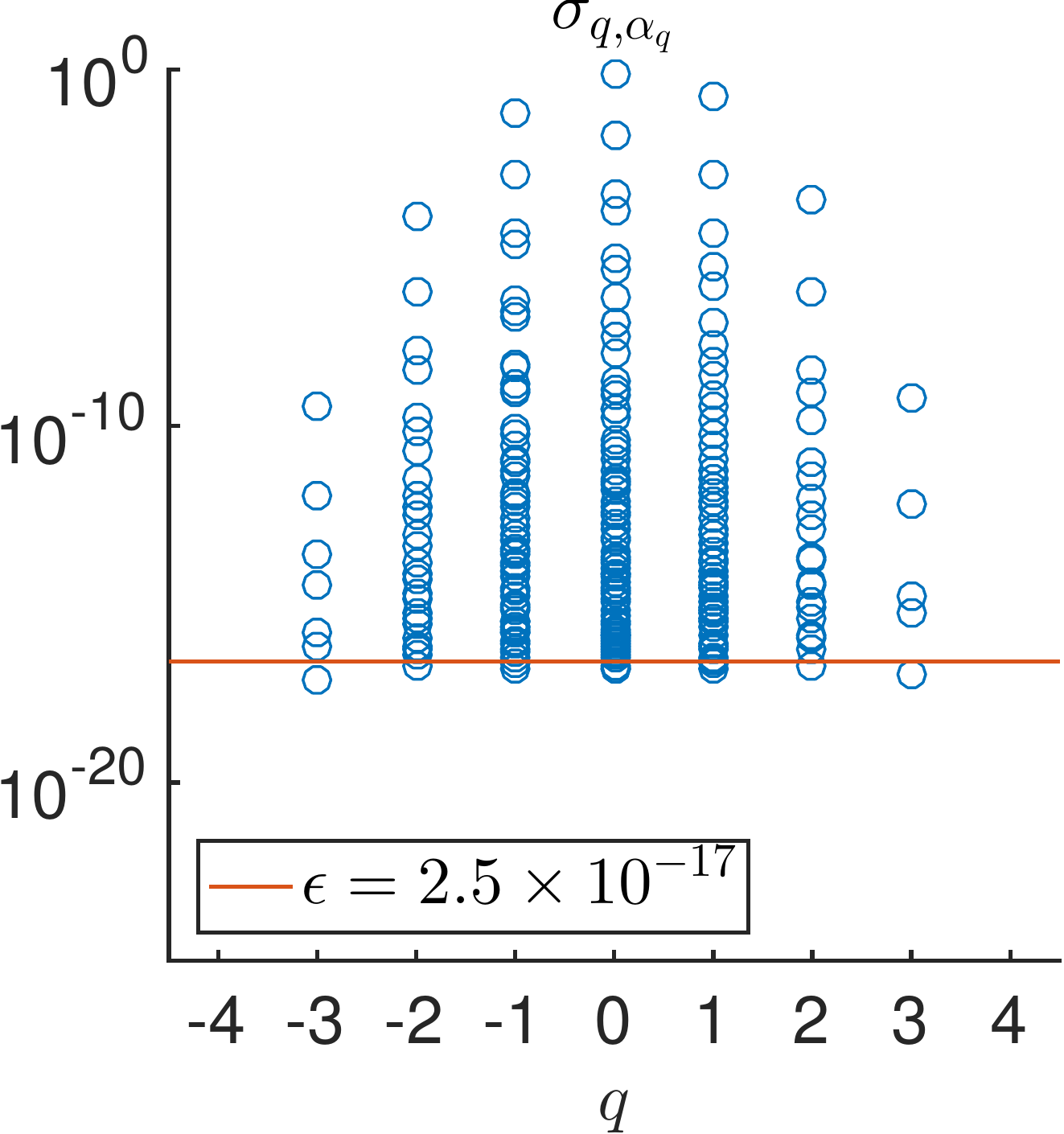}
\caption{\label{fig:SchmidtDistra}}
\end{subfigure}\hfill
\begin{subfigure}[b]{.24\textwidth}
\includegraphics[width=\textwidth]{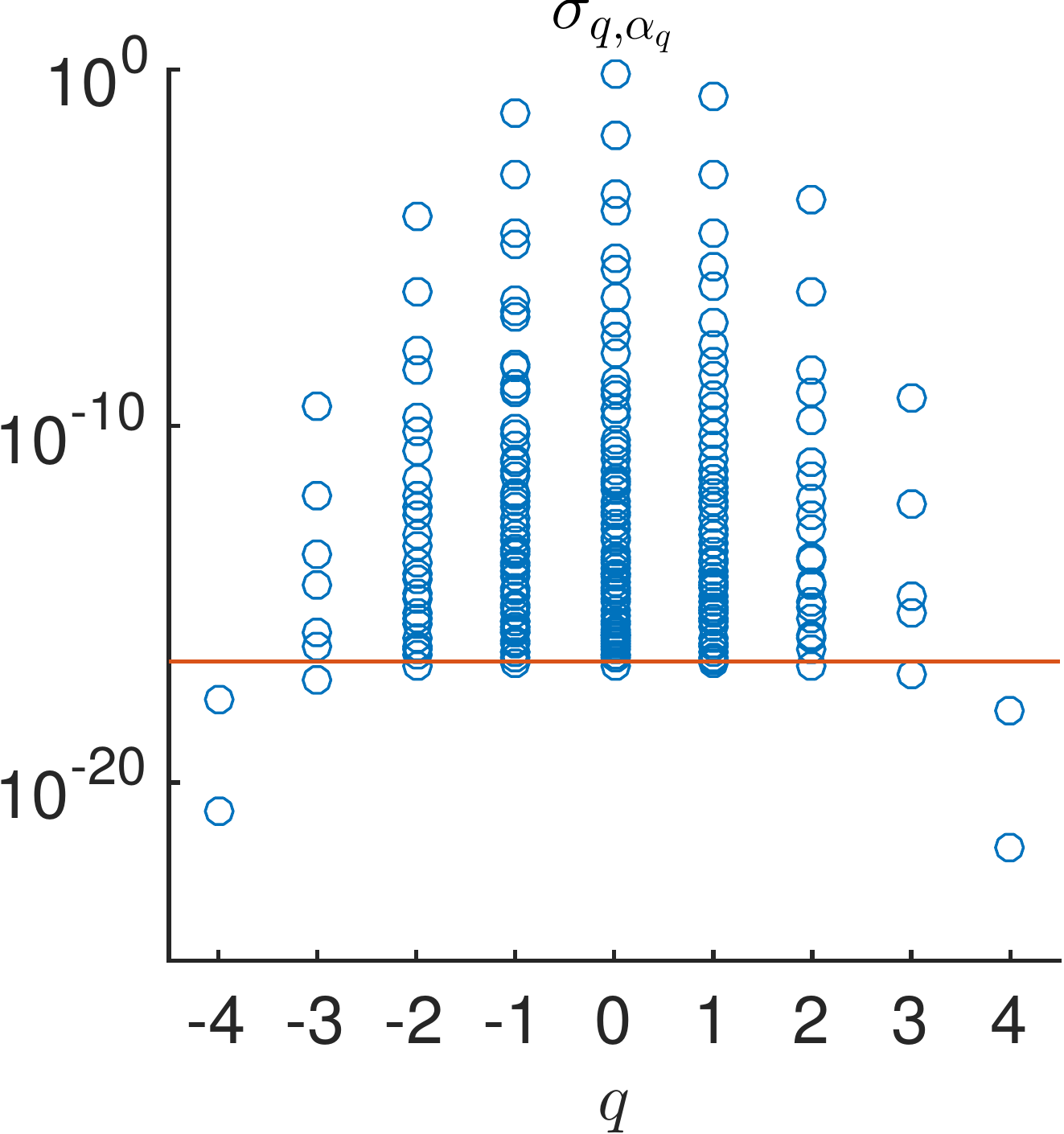}
\caption{\label{fig:SchmidtDistrb}}
\end{subfigure}\vskip\baselineskip
\captionsetup{justification=raggedright}
\caption{\label{fig:SchmidtDistr} $m/g= 0.3, x =100, \alpha = 0.4$. $D_q$ is chosen such that the smallest Schmidt values in each eigenvalue sector of $L(n)$ equals approximately $\epsilon = 2.5 \times 10^{-17}$.  We have set everywhere $p_{min} = - p_{max}$. (a) $p_{max} = 3$.  (b) $p_{max} = 4$.}
\end{figure}

The distribution of $D_q$ is chosen by looking at the Schmidt coefficients $\sigma_{q,\alpha_q}$ and demanding that the smallest coefficients of each sector approximately equal a preset tolerance $\epsilon$. In Fig.~\ref{fig:SchmidtDistr} we show this for an example with $\epsilon = 2.5 \times 10^{-17}$. Furthermore, a particular choice of $p_{min}$ and $p_{max}$ implies taking $D_q = 0$ for $q \notin \mathbb{Z}[p_{min},p_{max}]$ and, hence, also corresponds to a truncation in the Schmidt spectrum. Similarly as in Fig.~\ref{fig:SchmidtDistr}, we find in general the relevant eigenvalues sectors of $L(n)$ to be centered around $p_0 = 0$ for $\vert \alpha \vert \lesssim 0.5$. Physically this is explained by the first term in the Hamiltonian Eq.~(\ref{eq:equationH}) which punishes large expectation values for the electric field. The largest Schmidt value in each $q$-sector decreases as we move farther away from $q = p_0$. For instance, from Fig.~\ref{fig:SchmidtDistrb} we clearly observe that the eigenvalue sectors $q = \pm 4$ are redundant for $\epsilon = 2.5 \times 10^{-17}$, i.e. $\forall \alpha_q = 1\ldots D_q: \sigma_{q,\alpha_q} \leq \epsilon$ for $\vert q \vert \geq 4$. In general, we found for $\vert q \vert \gtrsim 5$ that all the Schmidt values $\sigma_{q,\alpha_q}$ were sufficiently small, even when approaching the continuum limit, and we could safely take $D_q = 0$ for these values of $q$.\\
\\From Eq.~(\ref{eq:MPSschmidtGaugeTsqInv}) it is clear that by taking smaller and smaller values for $\epsilon$, the threshold below which we discard the Schmidt values in Eq.~(\ref{eq:MPSschmidtGaugeTsqInv}), our MPS approximation $\ket{\Psi[a]}$ becomes closer to the real ground state. As our reference state we take the MPS approximation $\ket{\Psi[a_0]}$ with 
\begin{subequations}\label{eq:epsandpmax}
\be\label{eq:epsandpmaxa} \epsilon = 2.5 \times 10^{-17} \mbox{ and }p_{max} = -p_{min} = 4.\ee
To check whether this value for $\epsilon$ is sufficiently small, we perform additional simulations with resp.
\be\epsilon = 2.5 \times 10^{-17} \mbox{ and } p_{max}  = -p_{min} = 3,\ee
\be\epsilon = 10^{-16} \mbox{ and } p_{max}  = -p_{min} = 4,\ee
\be\epsilon = 10^{-16}\mbox{ and } p_{max}  = -p_{min} = 3\ee
\end{subequations}
leading to the MPS approximations resp. $\ket{\Psi[a_1]}$, $\ket{\Psi[a_2]}$, and $\ket{\Psi[a_3]}$ and check how the results differ among the simulations. 

\begin{figure}[t]
\begin{subfigure}[b]{.24\textwidth}
\includegraphics[width=\textwidth]{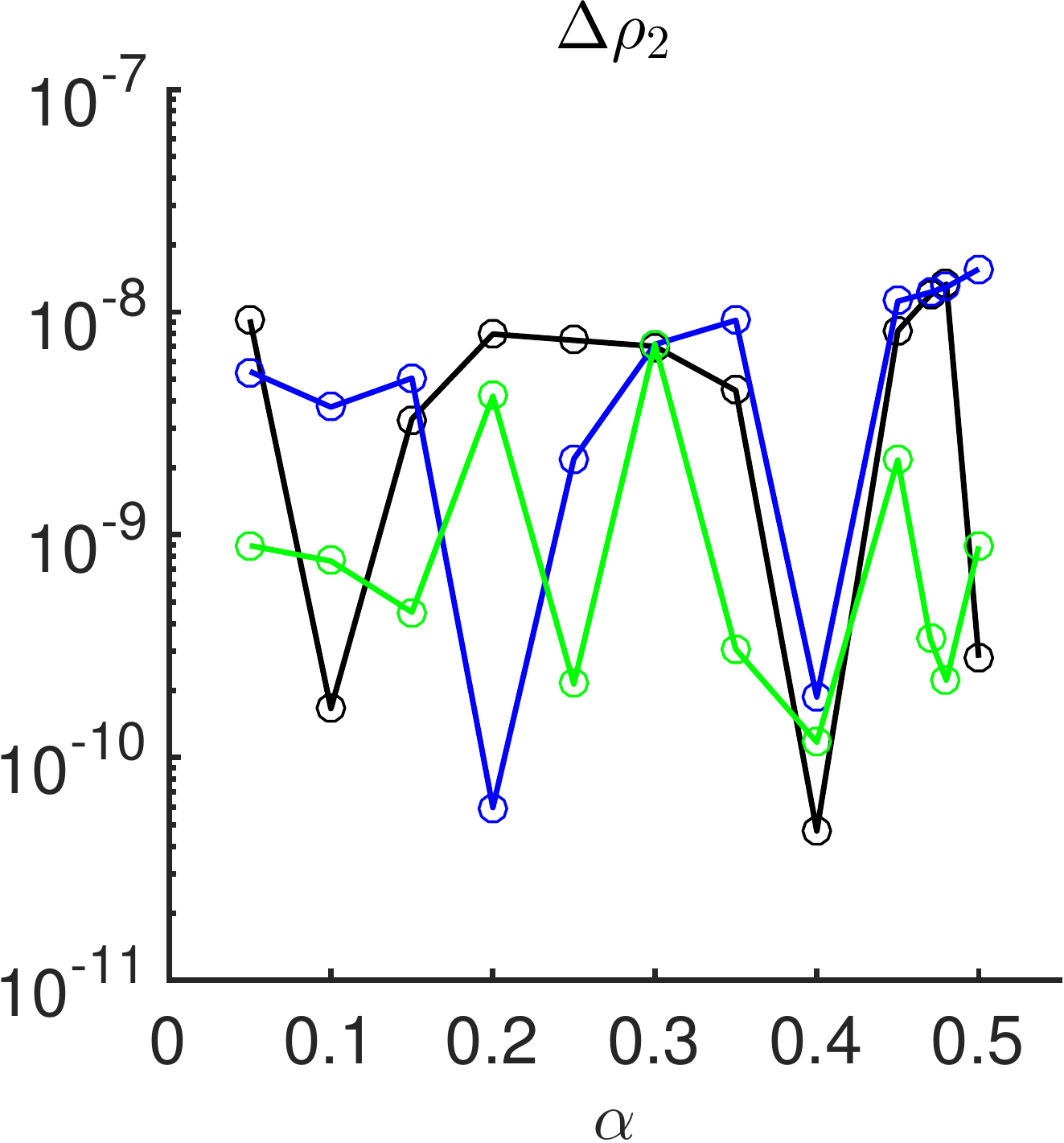}
\caption{\label{fig:deltaRho2}}
\end{subfigure}\hfill
\begin{subfigure}[b]{.24\textwidth}
\includegraphics[width=\textwidth]{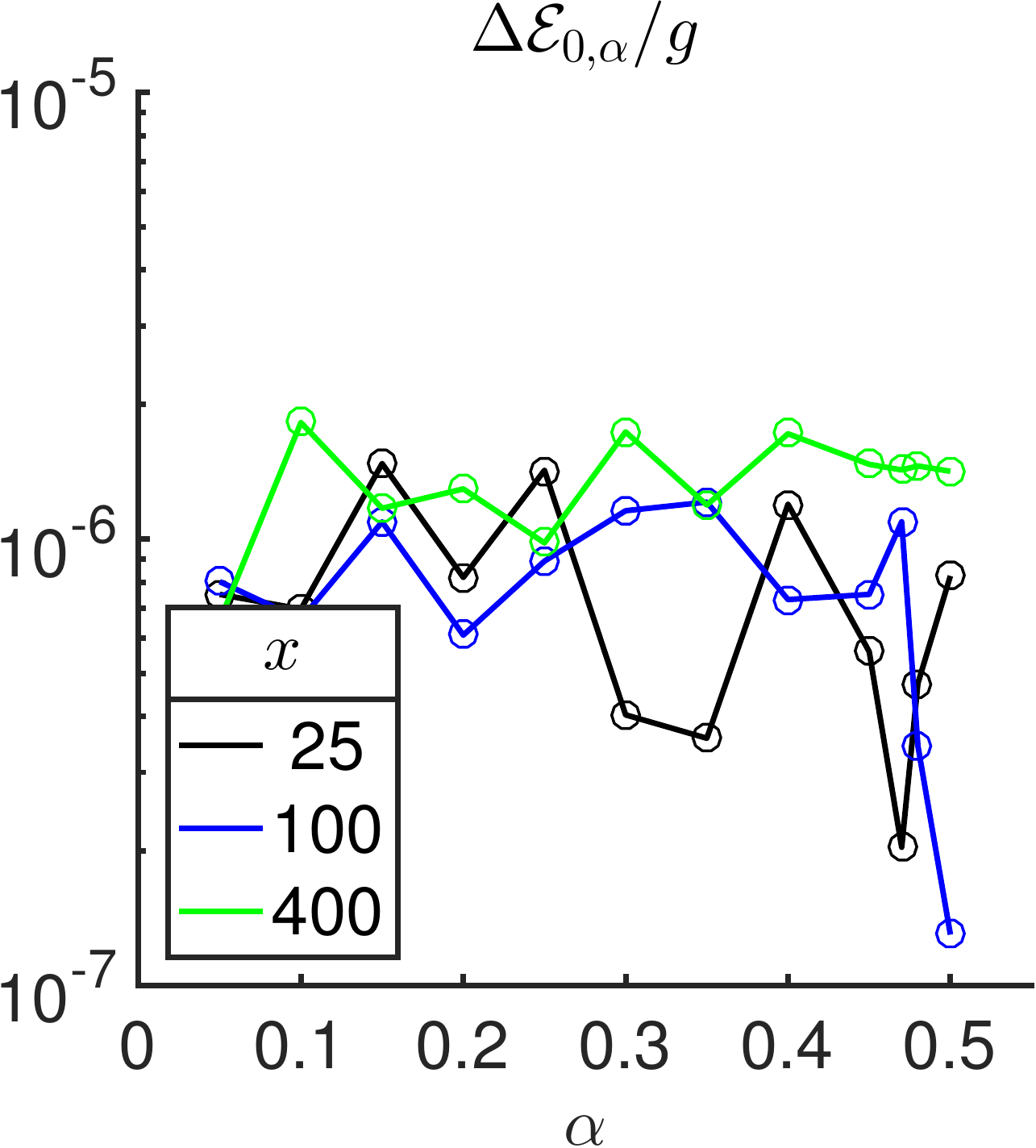}
\caption{\label{fig:deltaE0sq}}
\end{subfigure}\vskip\baselineskip
\captionsetup{justification=raggedright}
\caption{\label{fig:errGroundState} $m/g= 0.125$, $x=25,100,400$. (a) $\Delta\rho_2$ (defined in Eq.~(\ref{eq:deltarho2}) as the differences in the two-site reduced density matrices between the MPS result $\ket{\Psi[a_0]}$ and the other MPS approximations $\ket{\Psi[a_n]}$ ($n \geq 1$) with less precision) as a function of $\alpha$. (b) Variance $\Delta\mathcal{E}_{0,\alpha}$ of $H_\alpha$, Eq.~(\ref{eq:defHsqE0}), with respect to the MPS approximation $\ket{\Psi[a_0]}$ of the ground state. }
\end{figure}

The observables of interest take the form
$$O = \sum_{n=1}^{N-1} T^{2n-2}oT^{-2n+2},$$
where $o$ is an operator with support on the effective sites $1$ and $2$ (consisting of the physical sites and links $1,2,3,4$) and $T$ is the translation over one site. For the expectation value per site $O[a]$ with respect to $\ket{\Psi[a]}$ we have that
\be\ O[a] = \frac{1}{2N}\Braket{\Psi[a] \vert O \vert \Psi[\bar{a}]}  = \mbox{tr}\left(\rho_2[a] \cdot o\right)  \ee
where $\rho_2$ is the two-site reduced density matrix of $\ket{\Psi[a]}$ (see Appendix \ref{app:RedDensMPSSchwingerModel} for the details). As is shown in Appendix \ref{app:RedDensMPSSchwingerModel}, gauge invariance of $O$, $[O,G(n)] = 0$, implies that 
\be\label{eq:expTwoSiteO}  O[a]  = \mbox{tr}\left(\rho_{2}[a] \cdot o\right)  =  \displaystyle{\sum_{q = p_{min}}^{p_{max}} \mbox{tr}\left(\rho_{2,q}[a]\cdot o_q \right) }\ee
where $\rho_{2,q}[a]$ and $o_q$ can be found in Eq.~(\ref{eq:redDensMatr2app}) in Appendix \ref{app:RedDensMPSSchwingerModel}.

When comparing the expectation values of two different MPS approximations $\ket{\Psi[a]}$ and $\ket{\Psi[a']}$ for the ground state ($a_{q,s_1,s_2} \in \mathbb{C}^{D_q \times D_{q+(s_1+s_2)/2}},a'_{q,s_1,s_2} \in \mathbb{C}^{D'_q \times D'_{q+(s_1+s_2)/2}}$), we note that H\"older's inequality implies that
\begin{multline} \vert O[a] - O[a'] \vert  \\ \leq  \left(\sum_{q = p_{min}}^{p_{max}}\Biggl\vert \Biggl\vert \rho_{2,q}[a] - \rho_{2,q}[a'] \Biggl\vert \Biggl\vert _1\right) \cdot  \left(\max_{p_{min}\leq q \leq p_{max}} \vert \vert o_q \vert \vert_{\infty}\right)   \end{multline}
where $\vert\vert \cdot \vert\vert_m$ denotes the $m-$Schatten norm of the operator (i.e., the $m$-norm of the vector containing the singular values). For the local variables of interest (e.g., electric field, energy,$\ldots$) $\vert \vert o_q \vert \vert_{\infty}$ is bounded by a polynomial in $q$, see Appendix \ref{app:RedDensMPSSchwingerModel}. Hence, 
$$\Delta \rho_2[a,a'] =  \sum_{q = p_{min}}^{p_{max}}\Biggl\vert \Biggl\vert \rho_{2,q}[a] - \rho_{2,q}[a'] \Biggl\vert \Biggl\vert_1$$
is a good measure to compare two different MPS approximations $\ket{\Psi[a]}$ and $\ket{\Psi[a']}$ for the same ground state. 

Now, from the MPS approximations $\ket{\Psi[a_0]}$, $\ket{\Psi[a_1]}$, $\ket{\Psi[a_2]}$, and $\ket{\Psi[a_3]}$, see Eq.~(\ref{eq:epsandpmax}), we compute the reduced density matrices $\rho_2[a_0], \rho_2[a_1]$, $\rho_2[a_2]$ and $\rho_2[a_3]$. This enables us to compute the quantity
\be \label{eq:deltarho2} \Delta \rho_2 = \max_{n = 1,2,3}\left(\sum_{q = p_{min}}^{p_{max}}\Biggl\vert \Biggl\vert \rho_{2}[a_n] - \rho_{2}[a_0] \Biggl\vert \Biggl\vert_1\right),\ee
which is shown in Fig.~\ref{fig:deltaRho2} for $m/g = 0.125, x= 25,100,400$ and $0.05 \leq \alpha \leq 0.5$. In all cases $\Delta \rho_2$ is of order $10^{-8}$ or smaller. This is in fact what we would expect because taking $\epsilon \lesssim 10^{-16}$ corresponds to discarding in the Schmidt decomposition Eq.~(\ref{eq:MPSschmidtGauge}) terms with norm smaller than $\sqrt{\epsilon} \lesssim 1\times 10^{-8}$. 

We can also compute the variance of $H_\alpha$ with respect to $\ket{\Psi[a]}$, 
\begin{subequations} \label{eq:defHsqE0} 
\bea \Delta \mathcal{E}_{0,\alpha}[a]  = & \displaystyle{\frac{1}{\sqrt{2N}}\vert \vert H_\alpha \ket{\Psi[a]} - \mathcal{E}_{0,\alpha} \ket{\Psi[a]}} \vert \vert \nonumber \\
 = & \displaystyle{\sqrt{\frac{1}{2N}\Braket{\Psi[\bar{a}] \left\vert \left(H_\alpha - \mathcal{E}_{0,\alpha}\right)^2\right\vert \Psi[a]}}}\eea
with 
\be \vert \vert \ket{\Psi} \vert \vert = \sqrt{\braket{\Psi\vert\Psi}}, \ee
which is also a good measure to quantify how good our MPS approximates the real ground state. The computation of $\Delta \mathcal{E}_{0,\alpha}[a]$ can be done efficiently using conventional MPS techniques \cite{Vanderstraeten2015}. In Fig.~\ref{fig:deltaE0sq} we show $\Delta \mathcal{E}_{0,\alpha}$ for $m/g = 0.125$ which equals 
\be \Delta \mathcal{E}_{0,\alpha} = \vert \Delta \mathcal{E}_{0,\alpha}[a_0] \vert \ee
\end{subequations}
with $a_0$ corresponding to the MPS ground-state approximation $\ket{\Psi[a_0]}$, see Eq.~(\ref{eq:epsandpmaxa}). Although this quantity is of order $10^{-6}$ or smaller, it is two order of magnitudes larger than $\Delta\rho_2$, see Fig.~\ref{fig:deltaRho2}. This is no contradiction because $\Delta \mathcal{E}_{0,\alpha}[a]$ involves the computation of the expectation value of $H_\alpha^2$ which is not a local operator and, hence, cannot be computed as in Eq.~(\ref{eq:expTwoSiteO}). 
\\
\\We conclude that the TDVP simulations with $p_{max} = -p_{min} = 4$ and $\epsilon = 2.5 \times 10^{-17}$ provides us faitfhul MPS approximations for the real ground state of $H_\alpha$. 

\begin{table}[t]
\begin{tabular}{| c| c||   c | c | c |}
\hline 
$n$& $(\epsilon,p_{max})$ & $\mathcal{E}_{1,\alpha}^{(n)}$ & $\mathcal{E}_{2,\alpha}^{(n)}$& $\mathcal{E}_{3,\alpha}^{(n)}$\\
\hline
$0$ & $(2.5 \times 10^{-17},4)$ & 0.75333   & 1.40854  & 1.631 \\
$1$ & $(2.5 \times 10^{-17},3)$ & 0.75332  &  1.40849  & 1.629\\
$2$ &$(10^{-16},4)$ & 0.75323  & 1.40849  & 1.641\\
$3$ & $(10^{-16},3)$ & 0.75323&   1.40850 &   1.639\\
\hline
\hline
& $\mathcal{E}_{m,\alpha}$ & 0.75333   & 1.40854  & 1.631\\
&$\delta \mathcal{E}_{m,\alpha}$ & $1.0 \times 10^{-4}$ & $5.3 \times 10^{-5}$ &$1.0 \times 10^{-2}$ \\
&$\Delta \mathcal{E}_{m,\alpha}$ & $6.2 \times 10^{-3}$ & $3.0 \times 10^{-2}$ &  $0.38$ \\
\hline
\end{tabular}
\captionsetup{justification=raggedright}
\caption{\label{table:Excmdivg125e3} $m/g = 0.125, \alpha = 0.15,x = 400$. We compare the three lowest eigenvalues $\mathcal{E}_{1,\alpha}$, $\mathcal{E}_{2,\alpha}$ and $\mathcal{E}_{3,\alpha}$ obtained from the generalized eigenvalue equation Eq.~(\ref{eq:genEigExc}) for different tolerances in our simulations. $\mathcal{E}_{1,\alpha}$ and $\mathcal{E}_{2,\alpha}$ correspond to single-particle excitations, while $\mathcal{E}_{3,\alpha}$ originates from a multi-particle state. Indeed, in all cases we have $\mathcal{E}_{3,\alpha} > 2\mathcal{E}_{1,\alpha}$, hence it can decay in two particles with smaller energy. From the differences between the $\mathcal{E}_{k,\alpha}^{(0)}$ with $\mathcal{E}_{k,\alpha}^{(n)}$ we can compute $\delta \mathcal{E}_{m,\alpha}$, Eq.~(\ref{eq:deltaEmsmall}), which is displayed as well. We also compute the variance $\Delta \mathcal{E}_{m,\alpha}$, Eq.~(\ref{eq:defHsqExcm}). We find that the errors on $\mathcal{E}_{3,\alpha}$ are at least one order in magnitude larger than the errors on $\mathcal{E}_{1,\alpha}$ and $\mathcal{E}_{2,\alpha}$.}
\end{table}

\subsubsection{Single-particle excitations}
As explained in subsection \ref{subsec:RRforSPE}, once we have an MPS approximation $\ket{\Psi[a]}$ for the ground state of $H_\alpha$, we can use the ansatz $\ket{\Phi_k[b,a]}$, see Eq.~(\ref{eq:excAnsatz}), to approximate the momentum-$k$ excitations. As the Schwinger model is Lorentz invariant (in the continuum limit) the excitation energies $\mathcal{E}(k)$ of the states with momentum $k$ can be obtained from the ones with zero momentum by the Einstein dispersion relation $\mathcal{E}(k) = \sqrt{\mathcal{E}(0)^2 + k^2}$. Therefore we restrict ourselves to the zero-momentum states ($k = 0$). 

Not all the solutions of the generalized eigenvalue equation Eq.~(\ref{eq:genEigExc}) correspond to single-particle excitations. For instance, the generalized eigenvalue equation Eq.~(\ref{eq:genEigExc}) also gives solutions that correspond to multi-particle excitations. Note however, that it is clear that an ansatz of the from Eq.~(\ref{eq:excAnsatz}) is not suited for these type of excitations, and, hence, that the solution of Eq.~(\ref{eq:genEigExc}) gives in fact the overlap of a state of the from Eq.~(\ref{eq:excAnsatz}) with a multi-particle eigenstate. Therefore, these solutions are not reliable. For two-particle scattering states an MPS-ansatz is introduced and discussed in \cite{Vanderstraeten2014}. 

Let us consider a specific example example from our simulations to explain how we separate the solutions corresponding to single-particle excitations from solutions corresponding to multi-particle excitations. In table \ref{table:Excmdivg125e3} we show the three lowest eigenvalues $\mathcal{E}_{1,\alpha}^{(n)}$, $\mathcal{E}_{2,\alpha}^{(n)}$ and $\mathcal{E}_{3,\alpha}^{(n)}$ of the generalized eigenvalue equation Eq.~(\ref{eq:genEigExc}) where we started from the MPS approximation $\ket{\Psi[a_n]}$ for the ground state with $\epsilon$ and $p_{max} = - p_{min}$ as in Eq.~(\ref{eq:epsandpmax}). As our final result we take the result corresponding to our simulation for $(\epsilon,p_{max}) = (2.5 \times 10^{-17},4)$, i.e. $\mathcal{E}_{m,\alpha} = \mathcal{E}_{m,\alpha}^{(0)}$ and an error $\delta \mathcal{E}_{m,\alpha}$ on this result is estimated by comparing it with the energies of the other simulations:
\be\label{eq:deltaEmsmall} \delta \mathcal{E}_{m,\alpha} = \max_{n = 1,2,3} \vert \mathcal{E}_{m,\alpha}^{(n)} - \mathcal{E}_{m,\alpha}^{(0)} \vert. \ee
From table \ref{table:Excmdivg125e3} one observes that the energies $\mathcal{E}_{1,\alpha}$ and $\mathcal{E}_{2,\alpha}$ are stable under the limit $\epsilon \rightarrow 0$ within an error of $10^{-4}$ whereas the error on $\mathcal{E}_{3,\alpha}$ is two orders in magnitude larger. 

\begin{figure}[t]
\begin{subfigure}[b]{.24\textwidth}
\includegraphics[width=\textwidth]{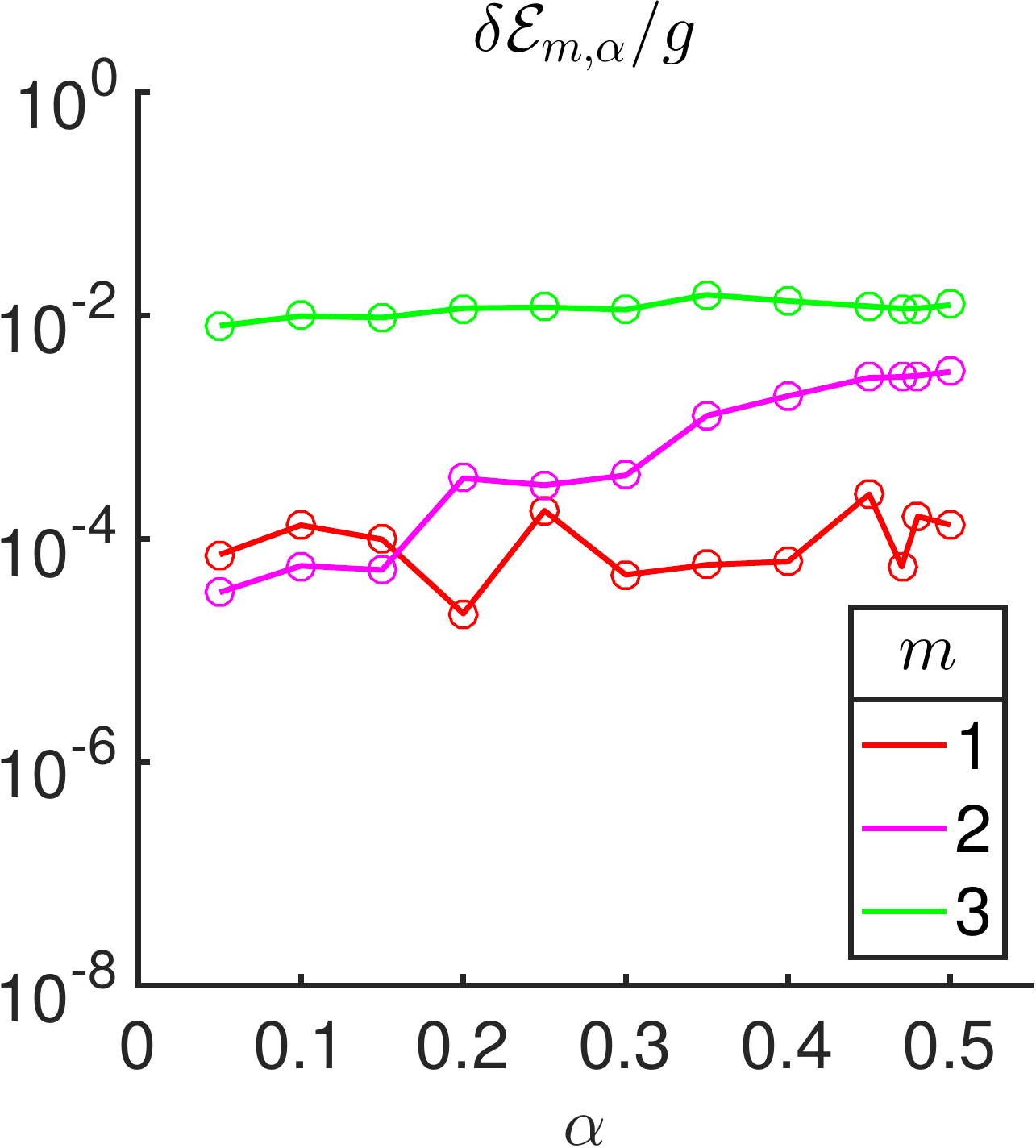}
\caption{\label{fig:deltaEn}}
\end{subfigure}\hfill
\begin{subfigure}[b]{.24\textwidth}
\includegraphics[width=\textwidth]{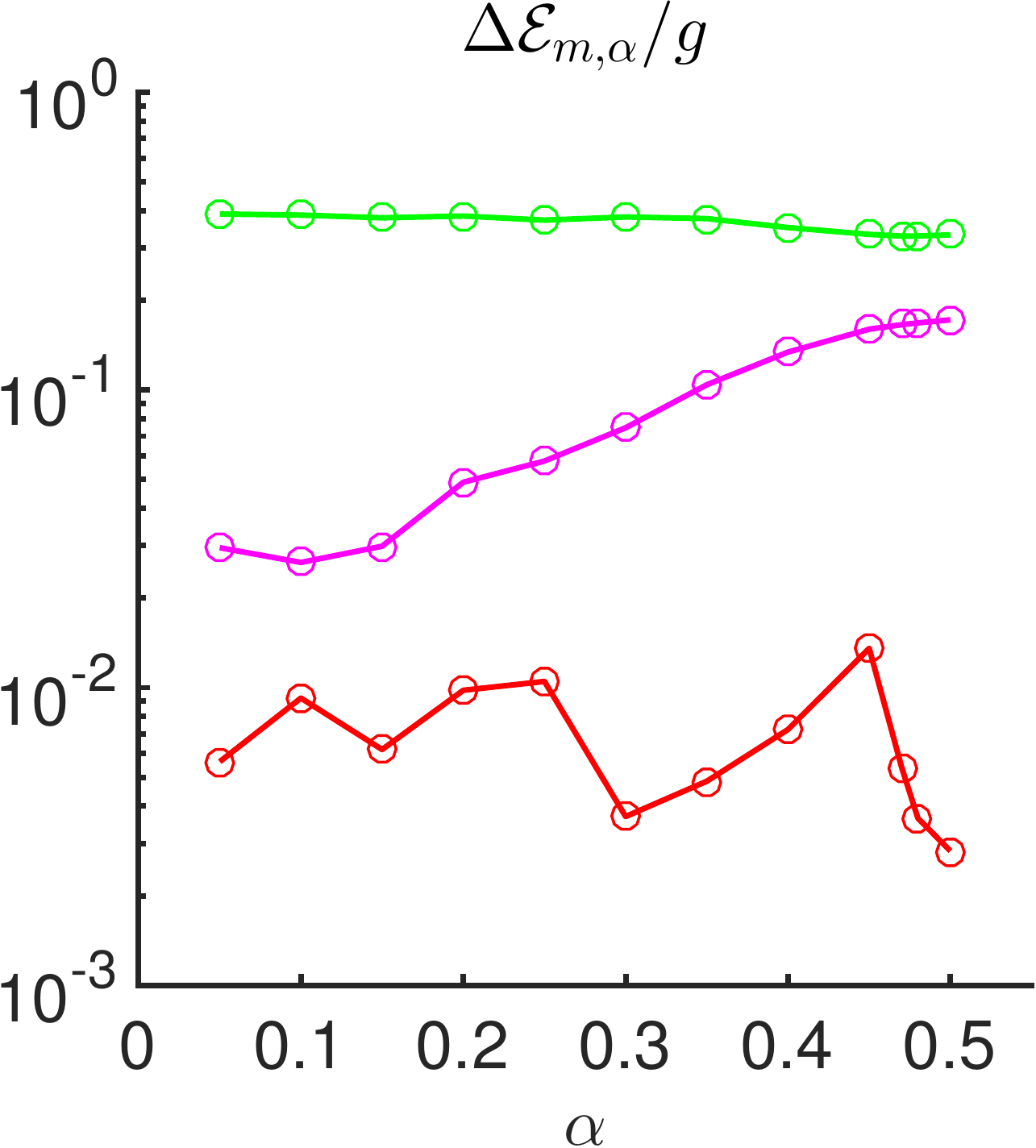}
\caption{\label{fig:deltaEnsq}}
\end{subfigure}\vskip\baselineskip
\captionsetup{justification=raggedright}
\caption{\label{fig:errExcState} $m/g= 0.125, x = 400$. Measures for the error in the excitation energies $\mathcal{E}_{1,\alpha}$ (red), $\mathcal{E}_{2,\alpha}$  (magenta) and $\mathcal{E}_{3,\alpha}$ (green) as a function of $\alpha$. (a) $\delta \mathcal{E}_{m,\alpha}$, Eq.~(\ref{eq:deltaEmsmall}), which is obtained by comparing the estimates with other estimates obtained from simulations with less precision. (b) The variance $\Delta \mathcal{E}_{m,\alpha}$, Eq.~(\ref{eq:defHsqExcm}), of $H_\alpha$ with respect to our MPS approximation for the excited state.}
\end{figure}

Note that $\mathcal{E}_{3,\alpha} \geq 2\mathcal{E}_{1,\alpha}$, hence we expect that this energy corresponds to a state Eq.~(\ref{eq:genEigExc}) which has overlap with a two-particle eigenstate of $H_\alpha$. On the other hand, we have that $\mathcal{E}_{2,\alpha} \leq 2 \mathcal{E}_{1,\alpha}$ and there is no reason why $\mathcal{E}_{2,\alpha}$ should not correspond to a single-particle excitation. In Fig.~\ref{fig:deltaEn} we show how $\delta \mathcal{E}_{m,\alpha}$ varies for different values of $\alpha$. We indeed find that the error on the lowest eigenvalue $\mathcal{E}_{1,\alpha}$ does not significantly change as a function of $\alpha$. In contrast, the error on the second eigenvalue increases. As we discuss in subsection \ref{sec:SPspectrum}, the second particle with energy $\mathcal{E}_{2,\alpha}$ indeed disappears in the multi-particle spectrum for $\alpha \gtrsim 0.35$, i.e. for $\alpha \gtrsim 0.35$ we find that $\mathcal{E}_{2,\alpha} \geq 2\mathcal{E}_{1,\alpha}$. In general we thus only consider the solutions $(\mathcal{E}_{m,\alpha}, \ket{\Phi_0[b_m,a]})$ for which $\mathcal{E}_{m,\alpha} \leq 2\mathcal{E}_{1,\alpha}$. 

Similar to Eq.~(\ref{eq:defHsqE0}), we can also compute the variance as a measure for the error,
\begin{subequations} \label{eq:defHsqExcm}
\be \Delta \mathcal{E}_{m,\alpha}[a] = \frac{1}{\sqrt{2N}}\vert \vert H_\alpha \ket{\Phi_0[b,a]} - \mathcal{E}_{m,\alpha} \ket{\Phi_0[b,a]} \vert \vert, \ee
which can be done efficiently using MPS techniques \cite{Vanderstraeten2015}. Note however, that as this quantity is a sum of negative and positive terms with comparable magnitude, there can be relatively large errors in $\Delta \mathcal{E}_{m,\alpha}[a]$ and this quantity is very likely to overestimate the error. However it can at least give a good indication whether $\ket{\Phi_0[b,a]}$ corresponds to an eigenstate of $H_\alpha$. In Fig.~\ref{fig:deltaEnsq} we show 
\be \Delta \mathcal{E}_{m,\alpha} \equiv \Delta \mathcal{E}_{m,\alpha} [a_0]\ee 
\end{subequations}
for $m/g = 0.125$ and $x = 400$ for different values of $\alpha$. We indeed find that $\Delta \mathcal{E}_{m,\alpha}$ correlates with the behavior of $\delta \mathcal{E}_{m,\alpha}$, but that it is two orders of magnitude larger than $\delta \mathcal{E}_{m,\alpha}$. 

In general we found that the errors on the excitation energies were significantly larger than the ones on the ground state expectation values but they were still under control: in general smaller than $10^{-2}$ and in most cases only of order $10^{-4}$.

\subsection{Charge sector occupation}\label{subsec:varNeededMPS}
In \cite{Buyens2015} we found that the half-chain Von Neumann entropy,
$$S = -\sum_{q}\sum_{\alpha_q}\sigma_{q,\alpha_q}\log(\sigma_{q,\alpha}), $$
scales as
$$S \sim \log(\xi\sqrt{x}) $$
with $\xi$ the correlation length and $x$ the inverse lattice spacing squared, as was predicted by Cardy and Calabrese \cite{Calabrese2004}. Given the fact that for a MPS 
$$\vert S \vert \lesssim \log(D), D = \sum_q D_q,$$
we can anticipate that the bond dimension should scale as
\be\label{eq:scaleD} D \sim \left(\xi\sqrt{x}\right)^\beta,\ee
for some power $\beta$. In particular, when approaching the continuum limit ($x \rightarrow + \infty$) or the phase transition for $m/g \rightarrow (m/g)_c$ and $\alpha \rightarrow 1/2$ ($\xi \rightarrow + \infty$), we expect to need large $D_q$. As truncating the eigenvalues of $L(n)$ between $p_{min}$ and $p_{max}$ corresponds to taking $D_q = 0$ for $q \notin \mathbb{Z}[p_{min},p_{max}]$, one expects that we would also need larger values for $\vert p_{min} \vert$ and $p_{max}$. However, as we already mentioned, we found for all our simulations that $\sigma_{q,\alpha_q} \leq 2.5\times 10^{-17}$ for $\vert q \vert \geq 5$, implying that $p_{max} = 4$ sufficies.

To quantify the weight of each of the eigenvalue sectors of $L(n)$, we consider again the MPS approximation $\ket{\Psi[a_0]}$ Eq.~(\ref{eq:gaugeMPSFinalForm}) for the ground state of $H_\alpha$ obtained by using the TDVP where
$$p_{max} = - p_{min} = 4 \mbox{ and } \epsilon = 2.5 \times 10^{-17},$$
i.e. we have chosen $D_q$ such that the smallest eigenvalue in each of the sectors equals approximately $\epsilon = 2.5 \times 10^{-17}$, see Fig.~\ref{fig:SchmidtDistr}. Then we compute the quantity $\tilde{D}_q$,
$$\tilde{D}_q = \# \{\sigma_{q,\alpha_q} \geq 10^{-16} : \alpha_q = 1\ldots D_q\},$$
which counts the number of Schmidt values larger than or equal to $10^{-16}$. It is obvious that $\tilde{D}_q$ gives a good measure for the weight of each of the eigenvalue sectors of $L(n)$ in the ground state. 
\\
\\In general, we are interested in expectation values of local gauge-invariant quantities. To identify the contribution of each of the eigenvalue sectors of $L(n)$ to these expectation values, we note that it follows from Eq.~(\ref{eq:expTwoSiteO}) that the contribution of each of the eigenvalue sectors $q$ of $L(n)$ to the expectation value with respect to $\ket{\Psi[a_0]}$ is
$$\mbox{tr}[\rho_{2,q}[a_0] \cdot o_q], $$
which is bounded by (H\"older's inequality)
$$\vert \mbox{tr}[\rho_{2,q}[a_0] \cdot o_q]  \vert \leq \vert\vert \rho_{2,q}[a_0] \vert \vert_1 \cdot \vert\vert o_q \vert\vert_{\infty},$$
where $\vert\vert \cdot \vert\vert_m$ denotes the $m-$Schatten norm of the operator. We refer to Eq.~(\ref{eq:redDensMatr2app}) in Appendix \ref{app:RedDensMPSSchwingerModel} for the explicit expressions of $\rho_{2,q}[a_0]$ and $o_q$. There we also show that for the quantities we are interested in (electric field, energy,$\ldots$), $\vert\vert o_q \vert\vert_{\infty}$ scales at most polynomially with $q$. Provided that $\vert \vert \rho_{2,q} \vert \vert_1 \equiv \vert \vert \rho_{2,q}[a_0] \vert \vert_1$ decreases fast enough (e.g. exponentially) with $q$, it follows that the contribution of the eigenvalue sectors $q$ of $L(n)$ for large $\vert q \vert$ to the ground state expectation values is negligible. Therefore we also investigate the quantity $\vert \vert \rho_{2,q} \vert \vert_1$, which is the sum of the singular values of $\rho_{2,q}[a_0]$. 

\begin{figure}[t]
\begin{subfigure}[b]{.24\textwidth}
\includegraphics[width=\textwidth]{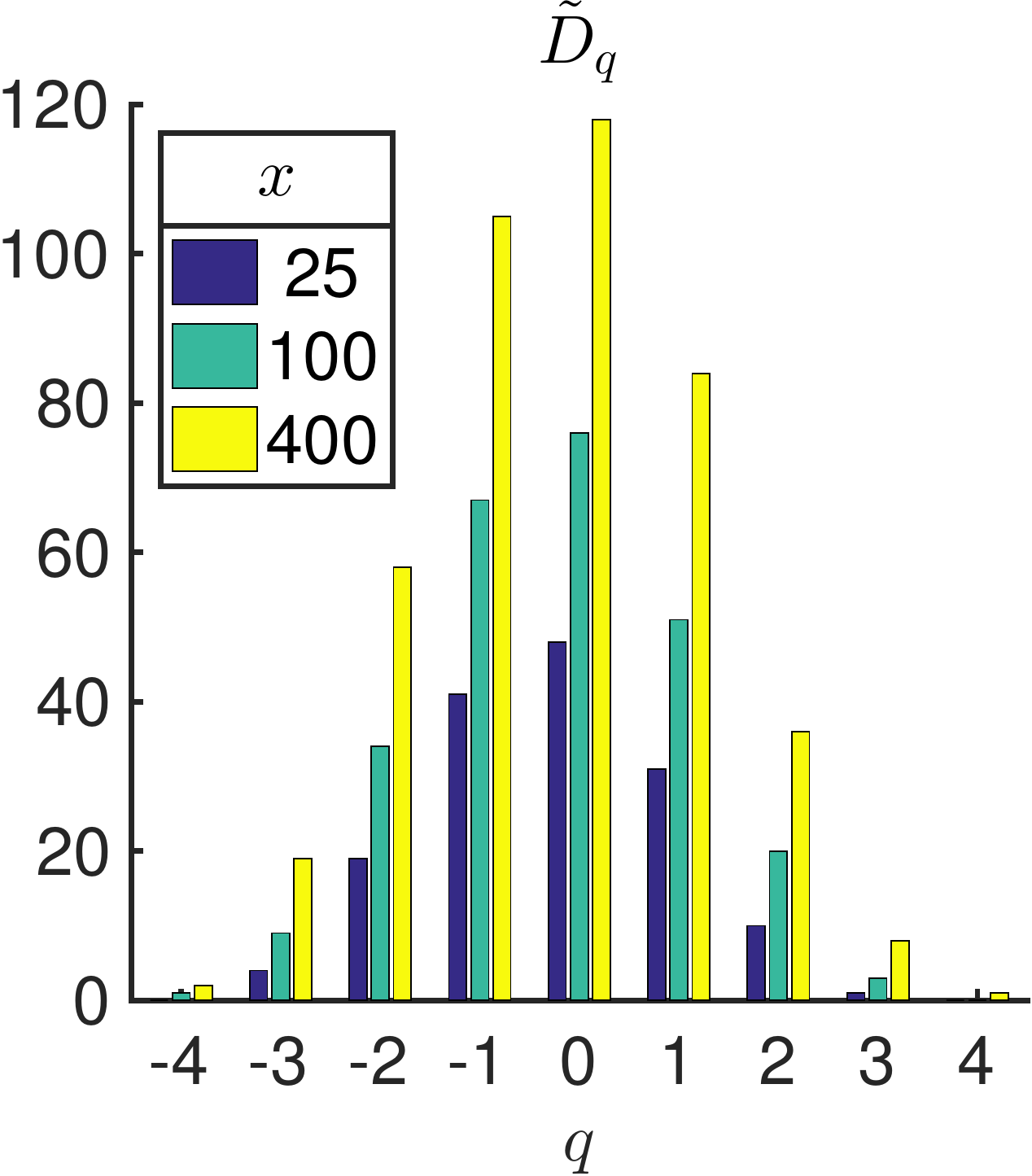}
\caption{\label{fig:DistrDqContinuum}}
\end{subfigure}\hfill
\begin{subfigure}[b]{.24\textwidth}
\includegraphics[width=\textwidth]{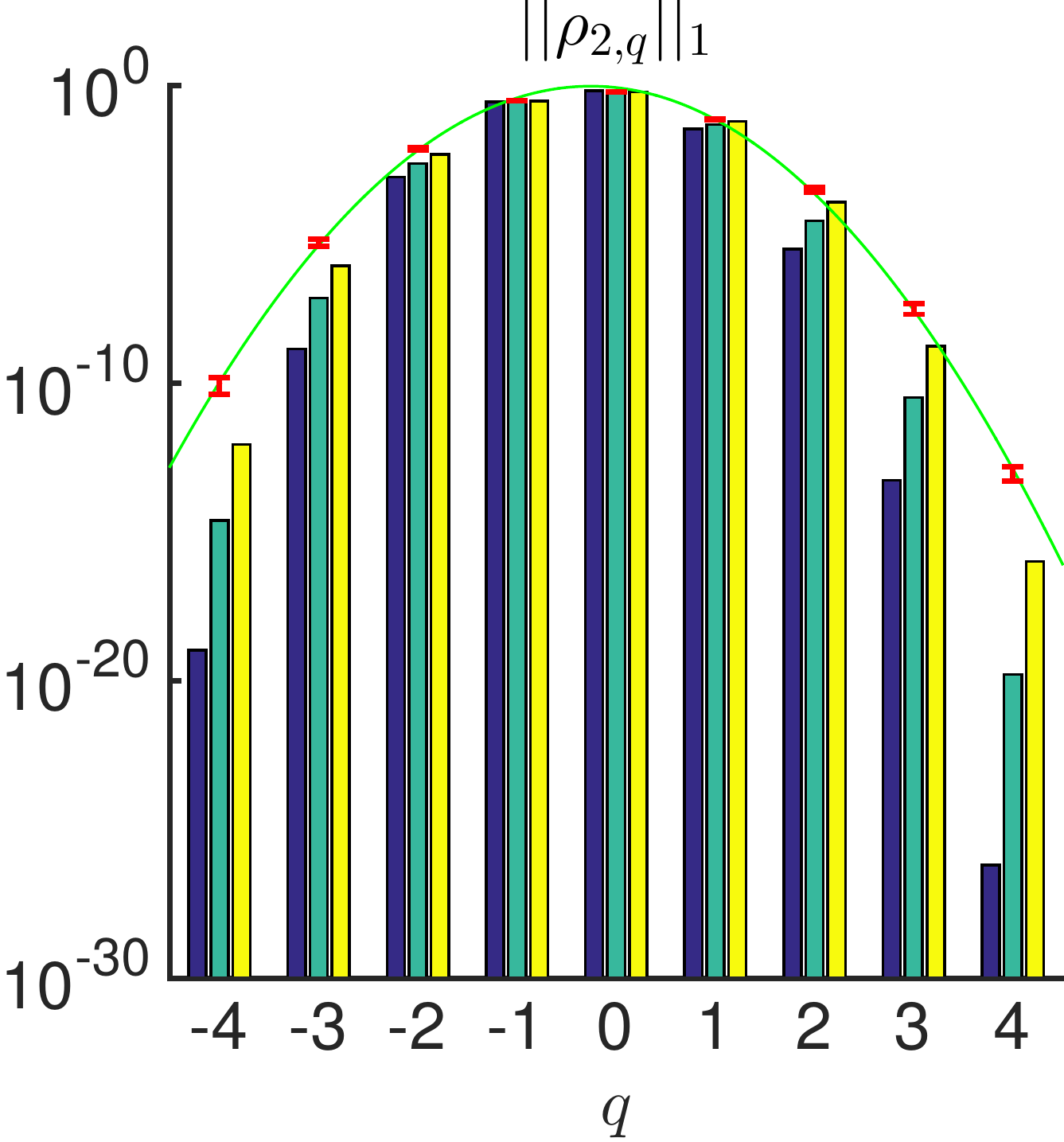}
\caption{\label{fig:DistrRhoqContinuum}}
\end{subfigure}\vskip\baselineskip
\captionsetup{justification=raggedright}
\caption{\label{fig:DistrContinuum} $m/g = 0.125, \alpha = 0.5$. Scaling of $\tilde{D}_q$ and $\vert\vert\rho_{2,q}\vert\vert_1$ to the continuum limit $x \rightarrow + \infty$. (a) $\tilde{D}_q$ increases with $\sqrt{x}$ in each of the eigenvalue sectors of $L(n)$, but falls of very fast with $\vert q \vert$. (b) By performing a polynomial extrapolation of $\log_{10}(\vert\vert\rho_{2,q}\vert\vert_1)$ in $1/\sqrt{x}$, we obtain estimates for the continuum value of $\vert\vert\rho_{2,q}\vert\vert_1$ (red error bars). The green line represents the parabolic fit through these estimates, Eq.~(\ref{eq:rho2expdec}), and shows that $\vert\vert\rho_{2,q}\vert\vert_1$ falls of exponentially with $q^2$ in the continuum limit.
}
\end{figure}

\subsubsection{From coarse to fine lattices}
Here we investigate the weight of the eigenvalue sectors of $L(n)$ when approaching the continuum limit $1/\sqrt{x} \rightarrow + \infty$. In Fig.~\ref{fig:DistrDqContinuum} we show the needed variational freedom in each of the sectors for $x = 25,100,400$ corresponding to the lattice spacings $1/\sqrt{x} = 0.2,0.1,0.05$ in units $g = 1$. The figure shows that for each of the eigenvalue sectors $q$ of $L(n)$, $D_q$ increases with $\sqrt{x}$ which is in agreement with Eq.~(\ref{eq:scaleD}).  

\begin{figure}[t]
\begin{subfigure}[b]{.24\textwidth}
\includegraphics[width=\textwidth]{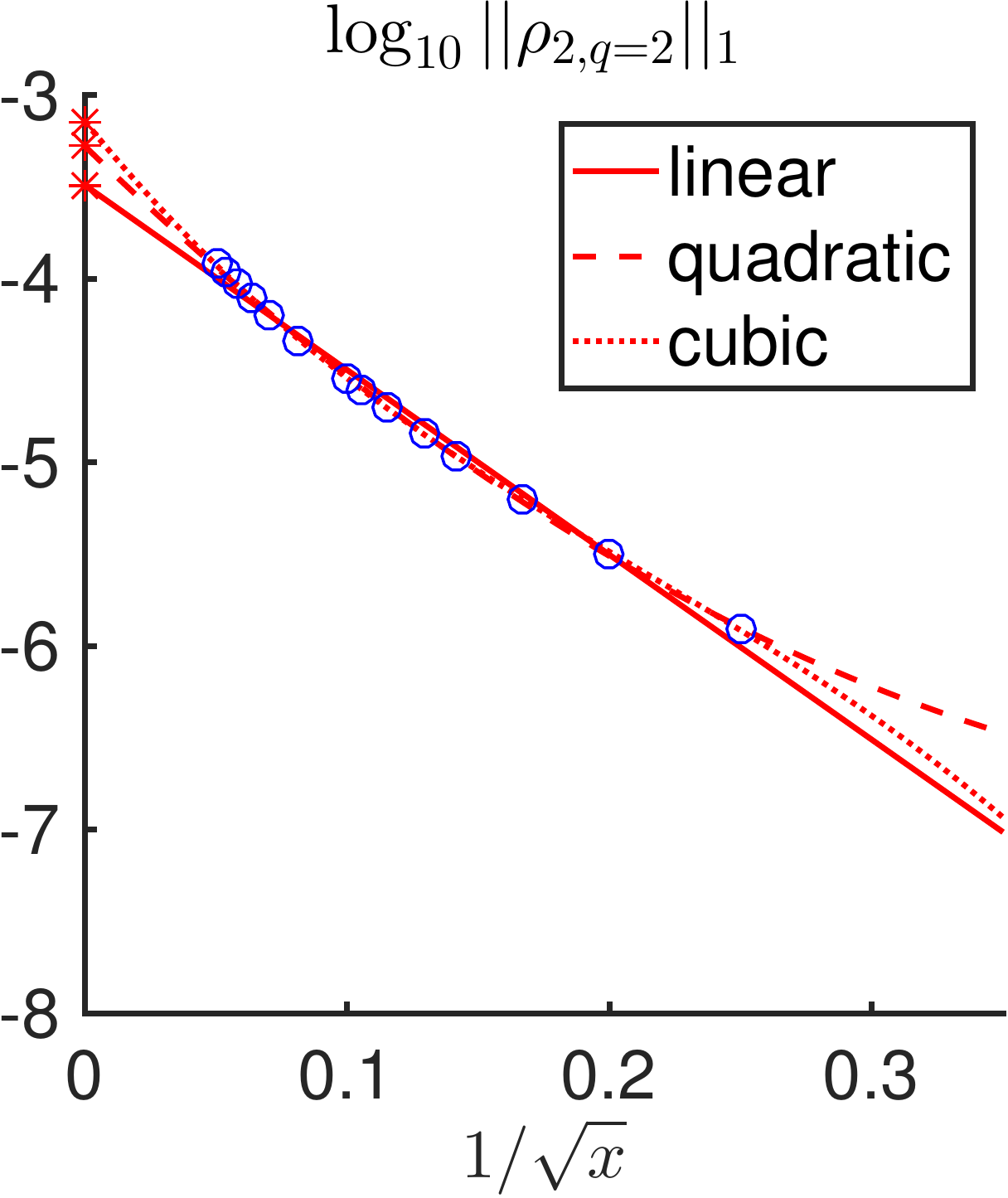}
\caption{\label{fig:log2qtoxinf}$q = 2, m/g = 0.125$.}
\end{subfigure}\hfill
\begin{subfigure}[b]{.24\textwidth}
\includegraphics[width=\textwidth]{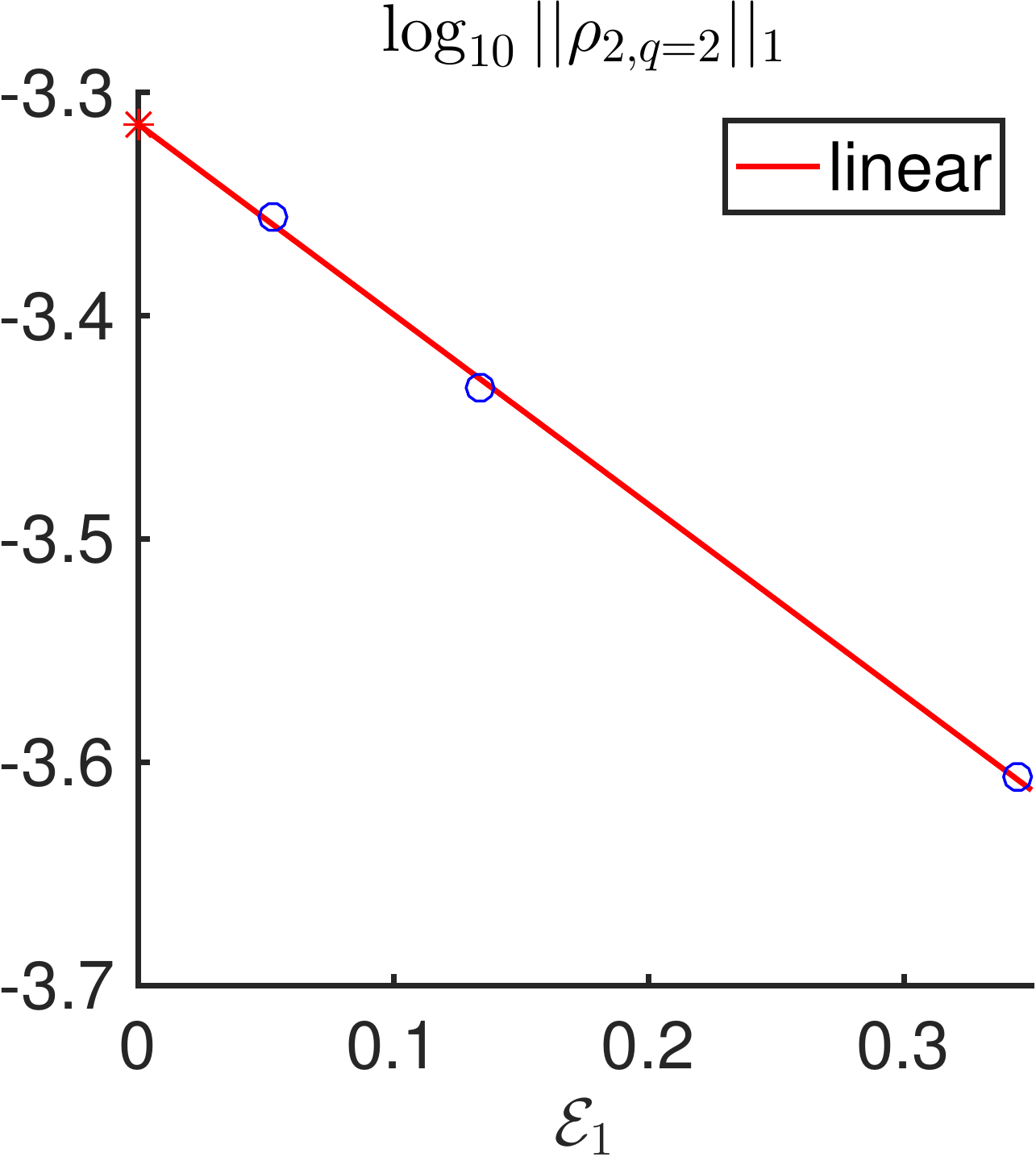}
\caption{\label{fig:log2rhoqMassGap} $q = 2, x = + \infty$. }
\end{subfigure}\vskip\baselineskip
\captionsetup{justification=raggedright}
\caption{\label{fig:ExtrapolationRedDensNorm} $\alpha  = 0.5$. (a) Linear (full line), quadratic (dashed line) and cubic fit (dotted line) of $\log\vert\vert \rho_{2,q} \vert \vert_1$ against $1/\sqrt{x}$ for $x \in [9,400]$ (blue circles). These fits allows us to obtain an estimate for $\log\vert\vert \rho_{2,q} \vert \vert_1$ in the continuum limit (stars). (b) We show here the (continuum estimates of) $\log\vert\vert \rho_{2,q} \vert \vert_1$ as a function of the mass gap $\mathcal{E}_1$. The mass gaps correspond, in increasing order, to the fermion masses $m/g = 0.3, 0.25$ and $m/g = 0.125$. The value $\mathcal{E}_1=0$ corresponds to the phase transition at $m/g = (m/g)_c \approx 0.33$. We observe an almost linear behavior (red line) which allows us to estimate $\log\vert\vert \rho_{2,q} \vert \vert_1$ at the phase transition (star). }
\end{figure}

From Fig.~\ref{fig:DistrRhoqContinuum} we find that the contribution of the eigenvalue sectors $q$ of $L(n)$ to local expectation values decreases very fast with $q$. The figure suggests that $\log(\vert \vert \rho_{2,q} \vert\vert_1)$ fits a parabola (note that the scale of the $Y-$axis is logarithmic) for all values of $x$. Moreover, we can even do a polynomial extrapolation of $\log_{10}\vert \vert \rho_{2,q} \vert \vert_1$ in $1/\sqrt{x}$ using our computations for $x = 16,25,36,50,60,75,90,100,150,200,250,300,350,400$, see Fig.~\ref{fig:log2qtoxinf} \footnote{We did linear and quadratic fitting in $1/\sqrt{x}$, to all our data, to all our data except the first point and to all our data except the first two points. Of all these estimates for the continuum value we took the linear fit to all our data as the estimate and obtained an error by computing the variance of all our estimates which are reflected in the errorbars.}. These continuum estimates are shown by the red error bars in Fig.~\ref{fig:DistrRhoqContinuum}. It is clear that they can be fitted against a quadratic function in $q$ which yields
\be\label{eq:rho2expdec} \vert \vert \rho_{2,q} \vert \vert _1\approx \exp(-1.63(5) q^2 - 0.84 (2)q - 0.1 (1)), \ee
where the errors on the coefficients are obtained by comparing with the same fit through $q = -3,-2,\ldots, 2,3$. The parabola Eq.~(\ref{eq:rho2expdec}) is shown in Fig.~\ref{fig:DistrRhoqContinuum} with the green line. For other values of $m/g$ and $\alpha$, a similar result can be obtained from our simulations for $x = 9,16,25,50,60,75,90,100$. Apparently, the dynamical gauge term $\sim \sum_nE(n)^2$ in the Hamiltonian $H_\alpha$ weights the eigenvalue sectors of $L(n)$ with a Gaussian in the ground state. 

\subsubsection{Towards the phase transition}
\begin{figure}
\begin{subfigure}[b]{.24\textwidth}
\includegraphics[width=\textwidth]{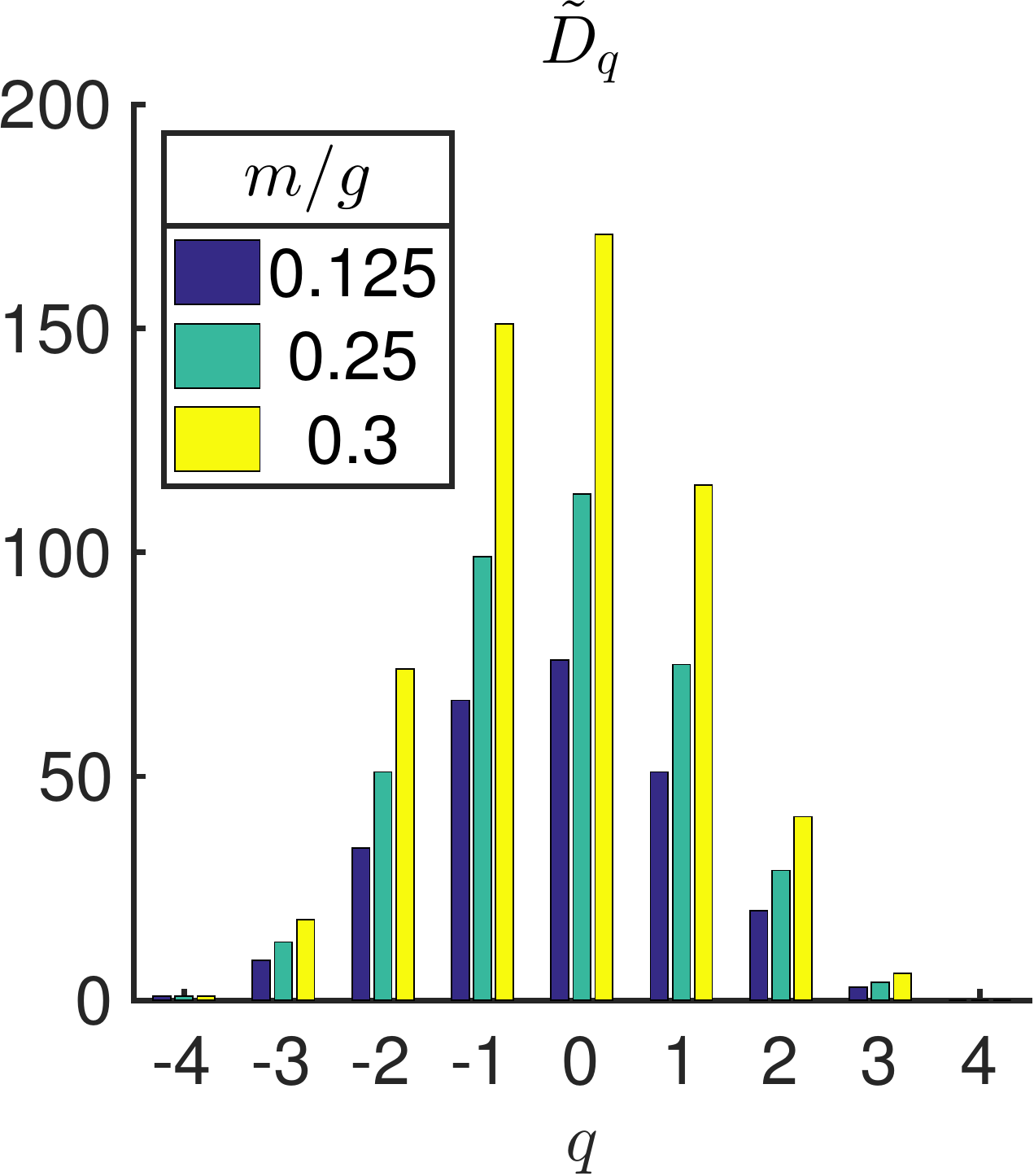}
\caption{\label{fig:DistrDqPhaseTransinm} $x =100$}
\end{subfigure}\hfill
\begin{subfigure}[b]{.24\textwidth}
\includegraphics[width=\textwidth]{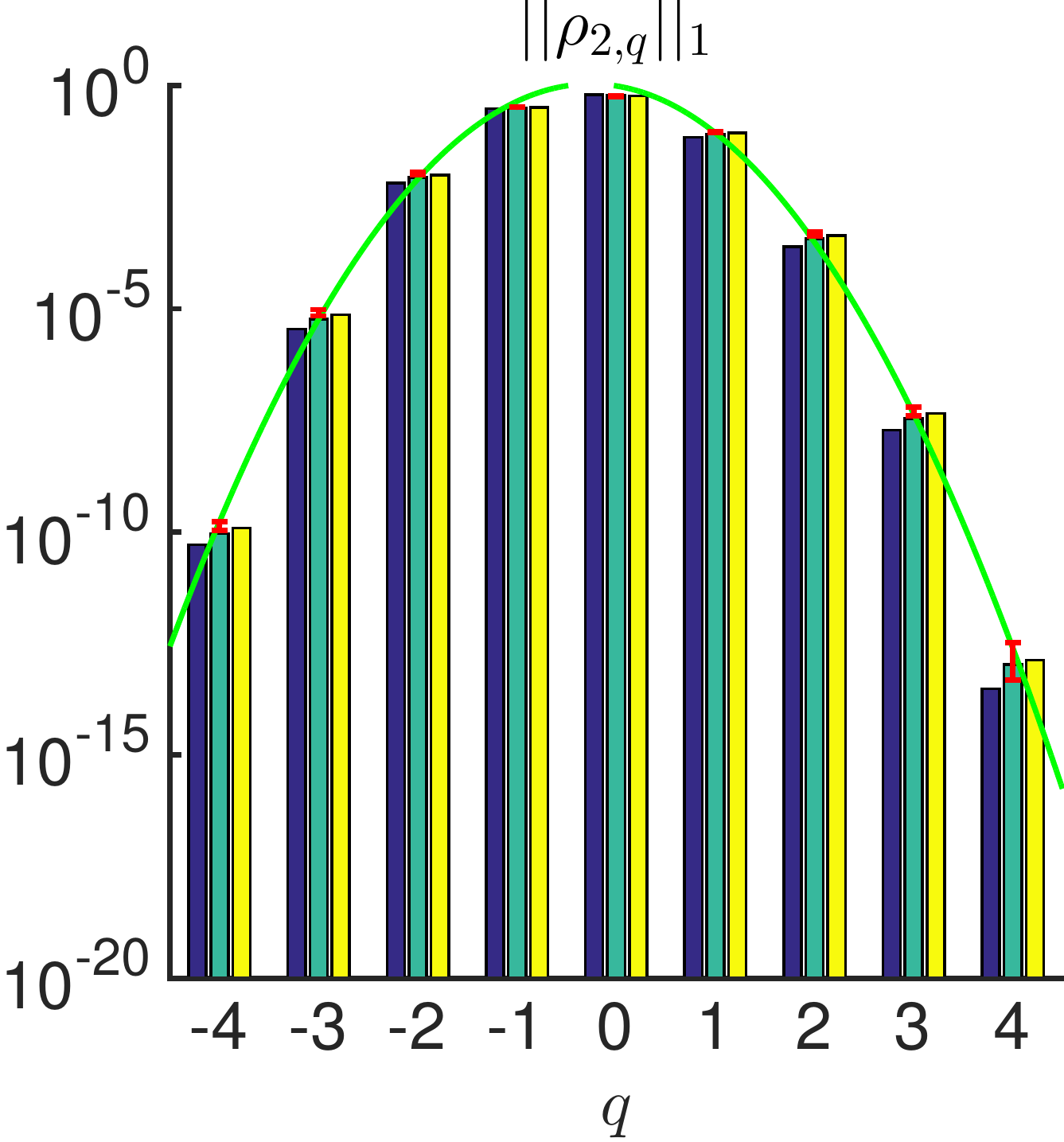}
\caption{\label{fig:DistrRhoqPhaseTransinm} $x = +\infty$.}
\end{subfigure}\vskip\baselineskip
\captionsetup{justification=raggedright}
\caption{\label{fig:DistrPhaseTransinm} $\alpha = 0.5$. Scaling of $\tilde{D}_q$ and $\vert\vert\rho_{2,q}\vert\vert_1$ when approaching the phase transition $m/g \rightarrow (m/g)_c \approx 0.33$. (a) $\tilde{D}_q$ increases in each of the eigenvalue sectors of $L(n)$ when getting close to $(m/g)_c$. Fortunately, it falls off very fast with $\vert q \vert$. (b) By performing a polynomial extrapolation of $\log_{10}(\vert\vert\rho_{2,q}\vert\vert_1)$ in the mass gap $\mathcal{E}_1$, we obtain an estimate for the value of $\vert\vert\rho_{2,q}\vert\vert_1$ at $m/g = (m/g)_c$ (red error bars). The green line represent a parabolic fit through these estimates, Eq.~(\ref{eq:rho2expdecPT}), and shows that $\vert\vert\rho_{2,q}\vert\vert_1$ falls off exponentially with $q^2$ at the phase transition.}
\end{figure}

Let us now investigate what happens when we approach the phase transition for $(m/g,\alpha) = ((m/g)_c \approx 0.33,1/2)$. At that point the system becomes gapless and the correlation length $\xi$ diverges, which leads again to the need of many variational parameters, see Eq.~(\ref{eq:scaleD}). Although MPS simulations are hard around the critical point, we were able to get very close to it. In Fig.~\ref{fig:DistrPhaseTransinm} we show $\tilde{D}_q$ (for $x = 100$) and $\vert \vert \rho_{2,q} \vert\vert_1$ (in the continuum limit) for $\alpha = 1/2$ and investigate their scaling towards $m/g \rightarrow (m/g)_c$.  

We observe that for $(m/g,\alpha) = (0.3,1/2)$ and $x = 100$, $\tilde{D}_q$ is large (e.g., $\tilde{D}_0 \approx 176$), but still easily manageable for a classical computer. In addition, when searching for the optimal MPS ground-state approximation close to the critical point, we also need a large amount of iterations Eq.~(\ref{eq:TDVPFlow}) to get the norm of the gradient below $\eta = 10^{-9}$. The TDVP takes a few weeks to obtain an optimal ground state. In contrast, for $m/g \lesssim 0.25$ and $\alpha \lesssim 0.48$ simulations take only a few hours until a day.  

In Fig.~\ref{fig:DistrRhoqPhaseTransinm}, we show the continuum estimates of $\vert\vert \rho_{2,q} \vert\vert_1$, obtained from our simulations for $x \in \{16,25,36,50,60,75,90,100\}$ as in Fig.~\ref{fig:log2qtoxinf}, for $m/g = 0.125, 0.25$ and $m/g = 0.3$. At the critical point, $(m/g) = (m/g)_c$, the system becomes gapless and it turns out that we can perform a linear extrapolation in the mass gap $\mathcal{E}_1$. In Fig.~\ref{fig:log2rhoqMassGap}, we show $\log_{10} \vert\vert \rho_{2,q=2} \vert\vert_1$ as a function of the mass gap $\mathcal{E}_1$ of the Schwinger model for $m/g = 0.125,0.25,0.3$ (we refer to subsection \ref{subsec:continuumLimit} for a discussion on how to obtain $\mathcal{E}_1$). One observes that they almost fit a straight line and, hence, we estimate the value of $\log_{10} \vert\vert \rho_{2,q=2} \vert\vert_1$ at the phase transition by the section of the linear fit with the $(\mathcal{E}_1 = 0)$-axis, see Fig.~\ref{fig:log2rhoqMassGap}. The estimates for $\vert\vert \rho_{2,q} \vert\vert_1$ for $(m/g) = (m/g)_c$ are now shown by the red error bars in Fig.~\ref{fig:DistrRhoqPhaseTransinm}. A parabolic fit though the points now gives (see green line Fig.~\ref{fig:DistrRhoqPhaseTransinm})
\be \label{eq:rho2expdecPT} \vert\vert \rho_{2,q} \vert \vert_1 \approx \exp\left(-1.60(6) q^2-0.81(2) q-0.0(2)\right),  \ee
which is very similar to Eq.~(\ref{eq:rho2expdec}). This shows that even at the phase transition we can safely truncate the Hilbert space of the gauge fields to a relatively small number of irreducible $U(1)$-representations. 

\begin{figure}
\begin{subfigure}[b]{.24\textwidth}
\includegraphics[width=\textwidth]{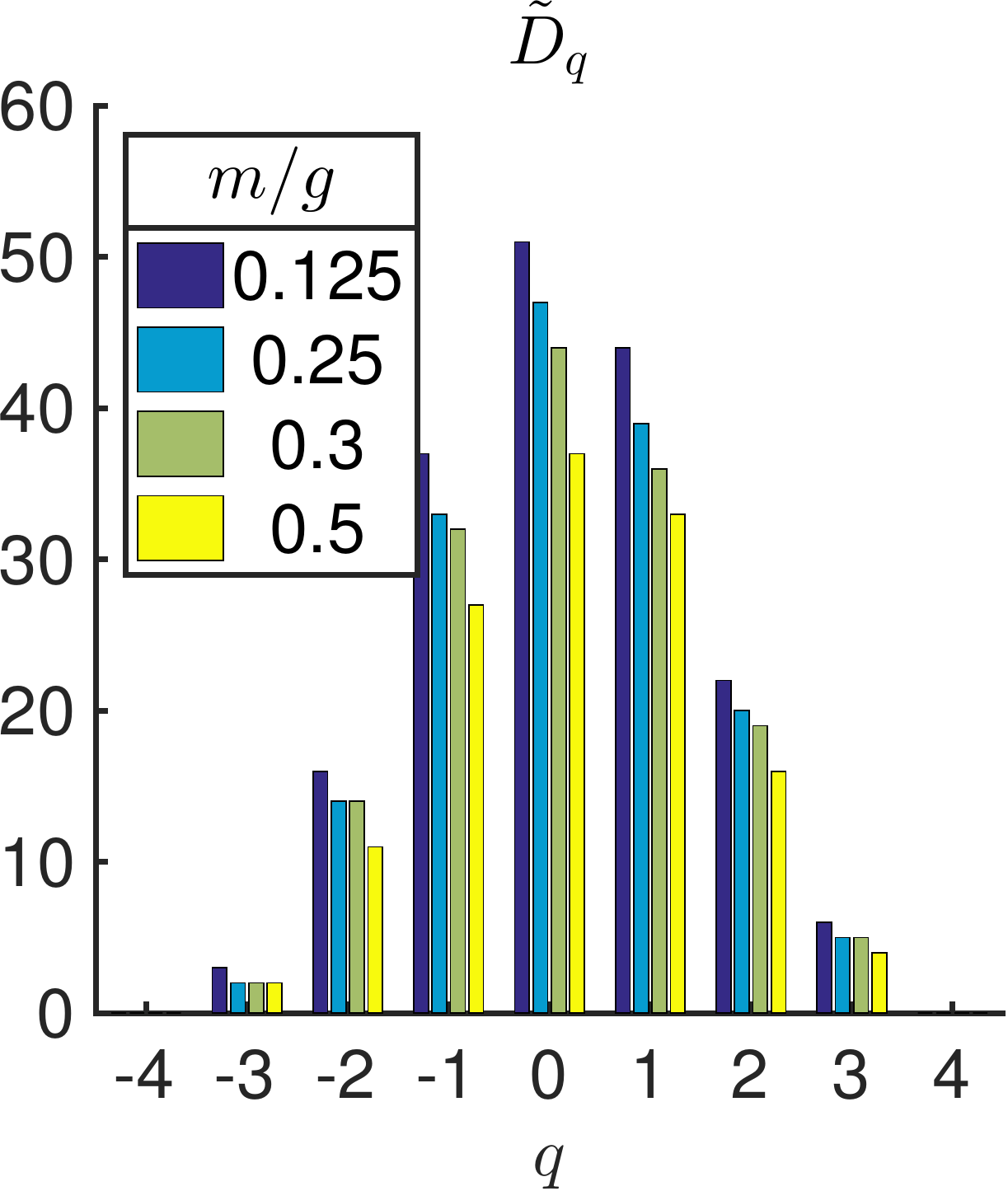}
\caption{\label{fig:DistrDqStrongToWeak}}
\end{subfigure}\hfill
\begin{subfigure}[b]{.24\textwidth}
\includegraphics[width=\textwidth]{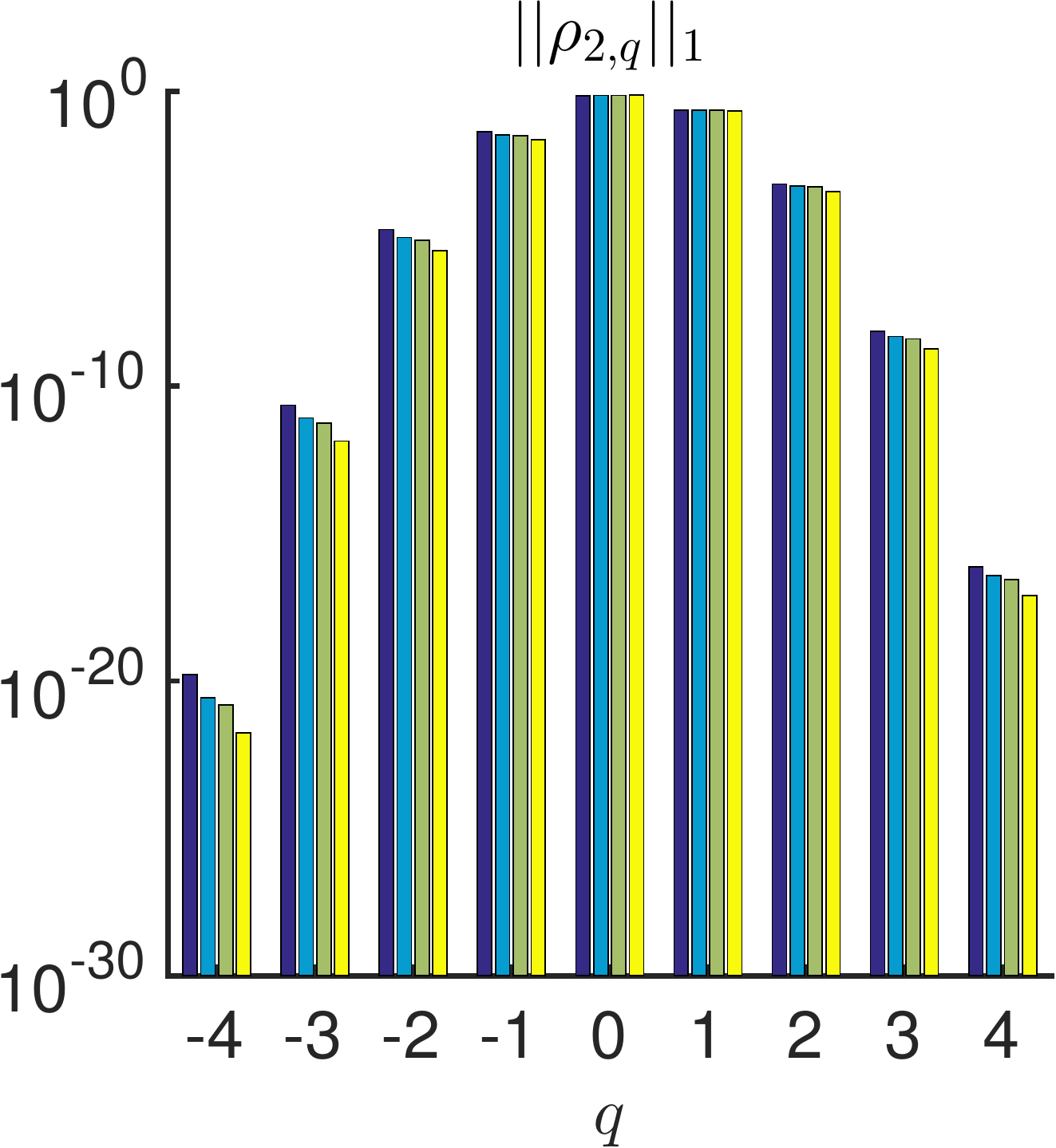}
\caption{\label{fig:DistrRhoqStrongToWeak}}
\end{subfigure}\vskip\baselineskip
\captionsetup{justification=raggedright}
\caption{\label{fig:DistrStrongToWeak} $x = 100, \alpha = 0.1$. Scaling of $\tilde{D}_q$ and $\vert\vert\rho_{2,q}\vert\vert_1$ when ranging from the strong-coupling regime ($m/g \ll 1$) to the weak-coupling regime ($m/g \gg 1$). (a) $\tilde{D}_q$ decreases in each of the eigenvalue sectors of $L(n)$ with increasing $m/g$. (b) $\vert\vert \rho_{2,q} \vert\vert_1$ falls of exponentially with $q^2$ and is almost independent of $m/g$.}
\end{figure}

\subsubsection{General dependence on $m/g$ and $\alpha$}
When $\alpha$ is sufficiently small the mass gap becomes larger when ranging from the strong-coupling ($m/g \ll 1$) limit to the weak-coupling limit ($m/g \gg 1$) (see subsection \ref{sec:SPspectrum}). For $\alpha = 0$, we found in \cite{Buyens2013} that we needed substantially smaller values of $D_q$ when $m/g$ increases. This is also what we observe in Fig.~\ref{fig:DistrDqStrongToWeak} for $\alpha = 0.1$: the number of Schmidt values above $10^{-16}$ decreases when $m/g$ increases. Note that this is the case for all the eigenvalue sectors $q$ of $L(n)$. This behavior is observed for all values of $\alpha$ smaller than $0.4$. Furthermore, in Fig.~\ref{fig:DistrRhoqStrongToWeak} we observe as before that $\vert\vert \rho_2(q)\vert\vert_1 \sim \exp(-q^2)$. This implies that the main contribution to the ground state expectation values comes from the small eigenvalue sectors of $L(n)$.  

\begin{figure}[t]
\begin{subfigure}[b]{.24\textwidth}
\includegraphics[width=\textwidth]{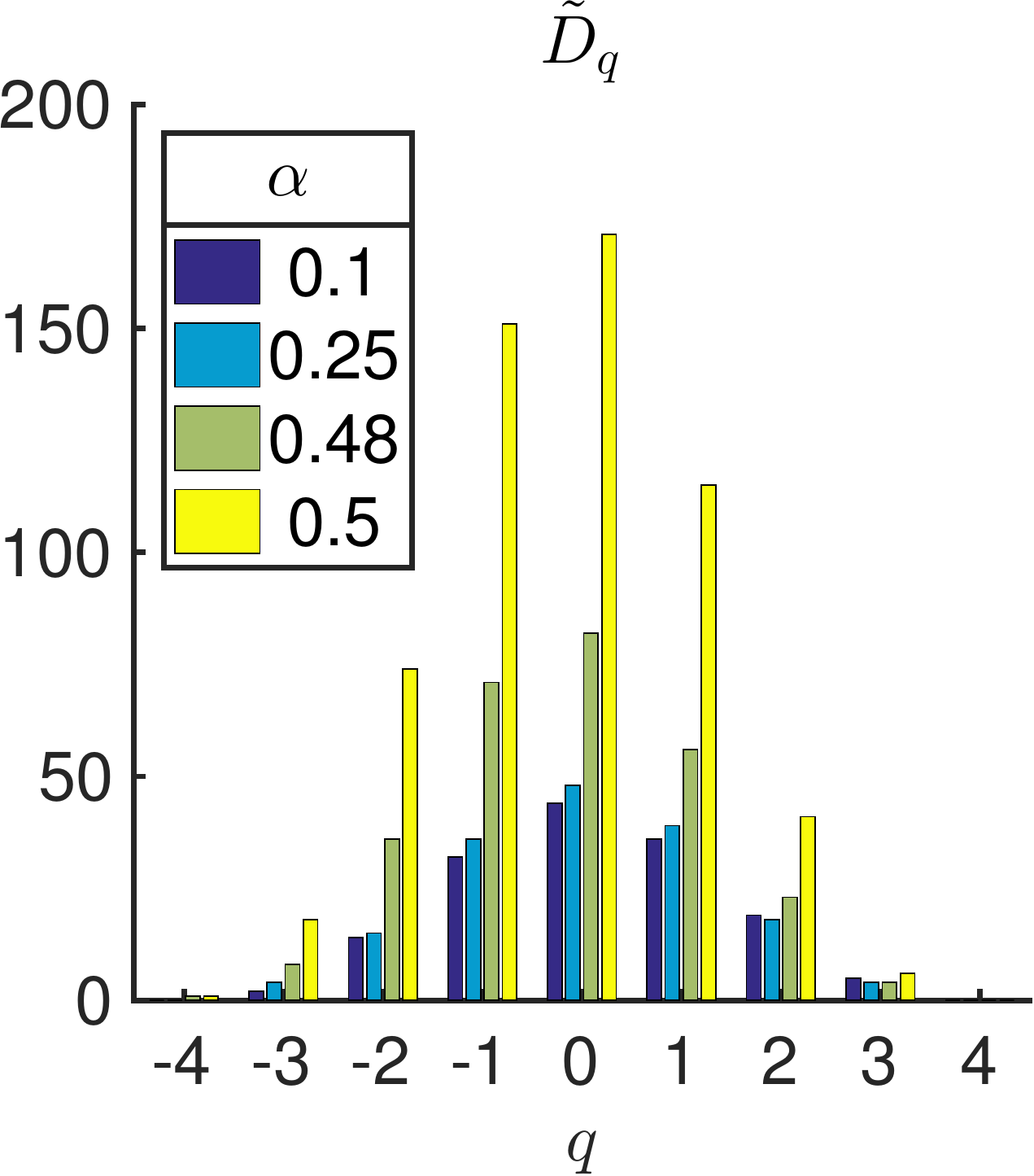}
\caption{\label{fig:DistrDqPhaseTransinalpha}}
\end{subfigure}\hfill
\begin{subfigure}[b]{.24\textwidth}
\includegraphics[width=\textwidth]{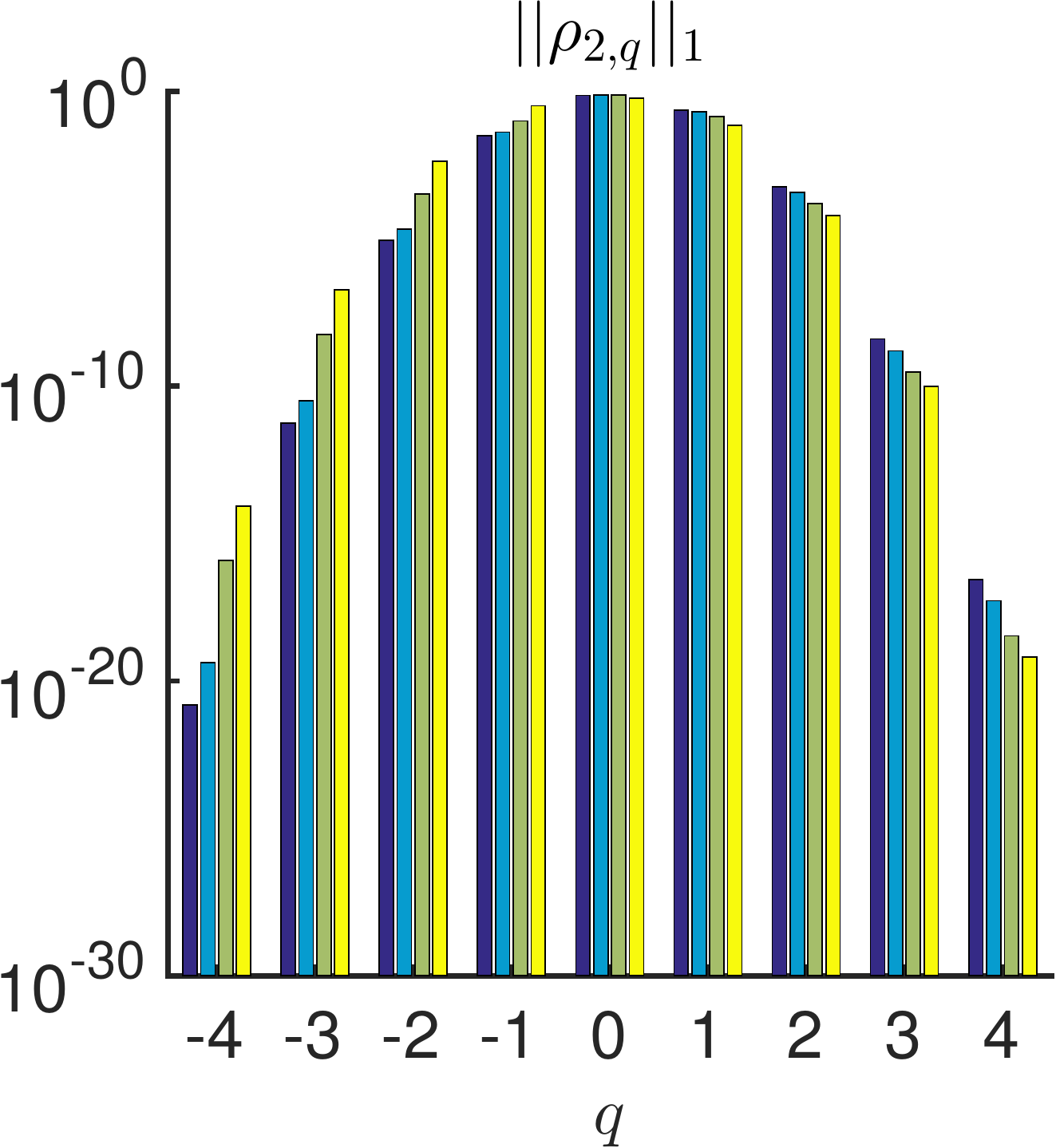}
\caption{\label{fig:DistrRhoqPhaseTransinalpha}}
\end{subfigure}\vskip\baselineskip
\captionsetup{justification=raggedright}
\caption{\label{fig:DistrPhaseTransinalpha} $m/g = 0.3, x = 100$. Scaling of $\tilde{D}_q$ and $\vert\vert\rho_{2,q}\vert\vert_1$ when varying $\alpha$. (a) $\tilde{D}_q$ increases in each of the eigenvalue sectors of $L(n)$ when $\alpha$ grows. Note also the difference between the values of $\tilde{D}_q$ for $\alpha = 0.48$ and $\alpha = 0.5$. (b) $\vert\vert \rho_{2,q} \vert\vert_1$ falls of exponentially with $q^2$. }
\end{figure}

For a fixed value of $m/g$, we find in general that we need more variational freedom when $\alpha$ increases, see Fig.~\ref{fig:DistrPhaseTransinalpha}. An explanation is that the mass gap decreases with increasing $\alpha$, see subsection \ref{sec:SPspectrum}. In particular, when $\alpha = 1/2$, $\tilde{D}_q$ becomes suddenly very large. Indeed there is a large difference between $\tilde{D}_q$ for $\alpha = 0.48$ and $\alpha = 0.5$ in Fig.~\ref{fig:DistrDqPhaseTransinalpha}. Here we also find that the contribution to the local expectation values mainly originates from the eigenvalue sectors of $L(n)$ with small $q$, see Fig.~\ref{fig:DistrRhoqPhaseTransinalpha}, confirming the general picture. 
\\
\\
In conclusion, the fast decay Eq.~(\ref{eq:rho2expdec}) and Eq.(\ref{eq:rho2expdecPT}) and the aforementioned results,
implies that for the study of the Schwinger model we only need to retain a few of the infinite number of $U(1)$-representations to obtain reliable results in the continuum limit. From a broader perspective, this holds optimism for the study of any lattice field theory in the Wilsonian formulation. As the Hamiltonian of a $SU(N)$ Yang-Mills theory has a quadratic electric field term \cite{Creutz1977},  generally referred to as the quadratic Casimir operator, we might expect that the contribution of each of the irreducible representations of $SU(N)$ to local expectation values also decays exponentially fast with its quadratic Casimir invariant. Hence, we expect that also for these theories we could safely truncate the Hilbert space of the gauge fields to a relatively small number of irreducible representations, not undermining the possibility of performing efficient tensor network simulations.  

\subsection{The continuum limit: $x \rightarrow + \infty$}\label{subsec:continuumLimit}
\begin{figure}[t]
\begin{subfigure}[b]{.24\textwidth}
\includegraphics[width=\textwidth]{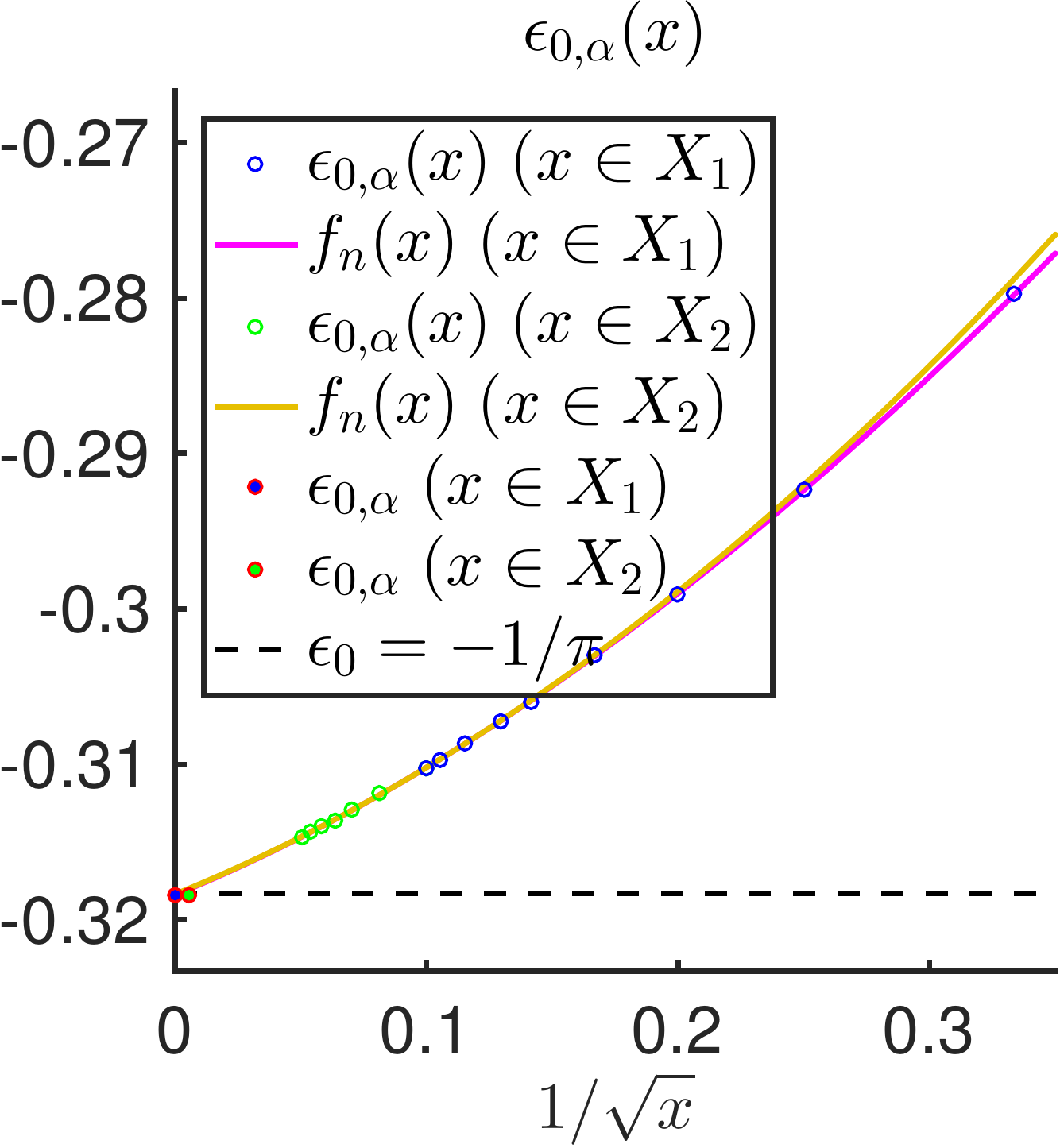}
\caption{\label{fig:extrapolationeps0}}
\end{subfigure}\hfill
\begin{subfigure}[b]{.24\textwidth}
\includegraphics[width=\textwidth]{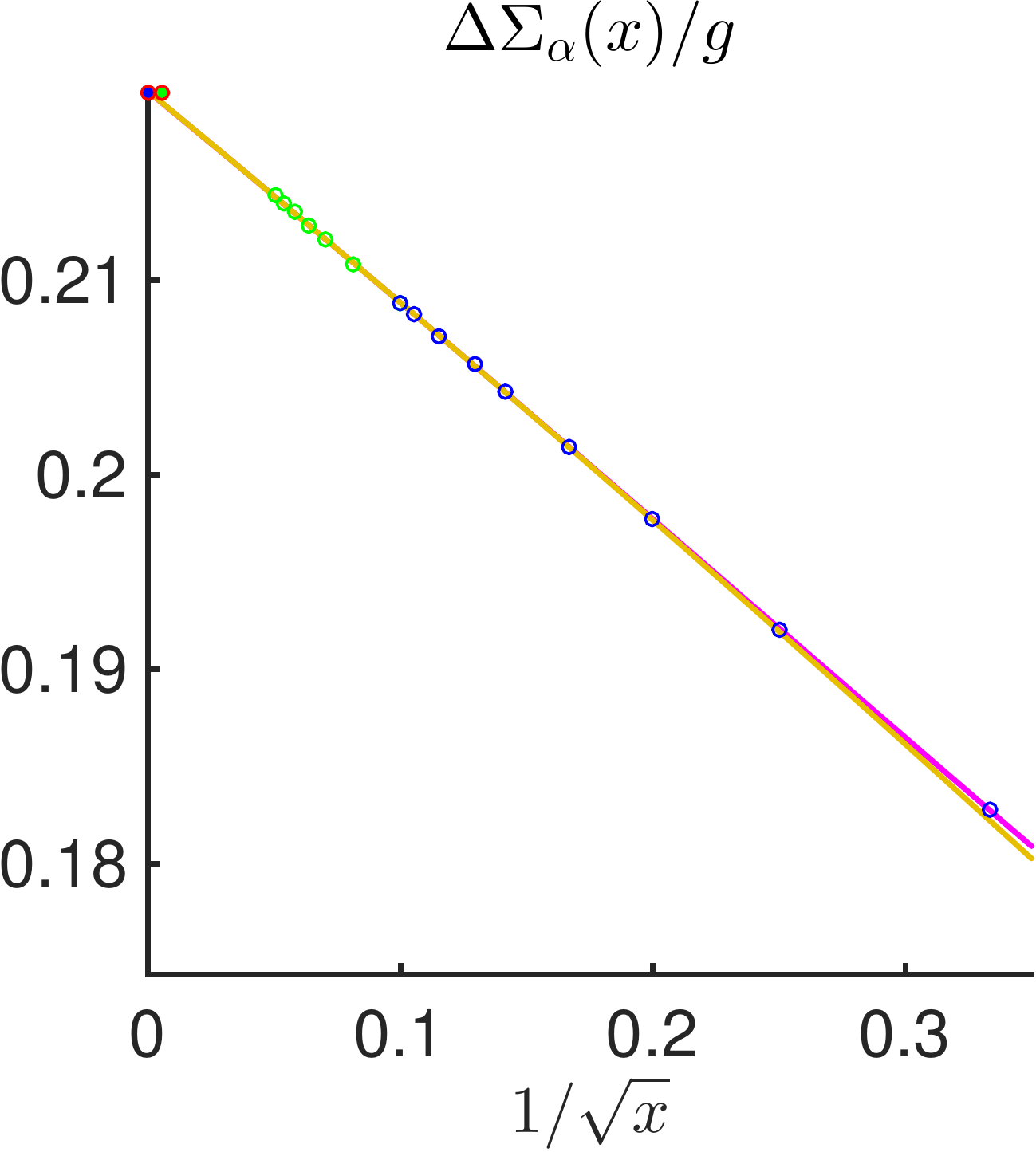}
\caption{\label{fig:extrapolationCC}}
\end{subfigure}\vskip\baselineskip
\begin{subfigure}[b]{.24\textwidth}
\includegraphics[width=\textwidth]{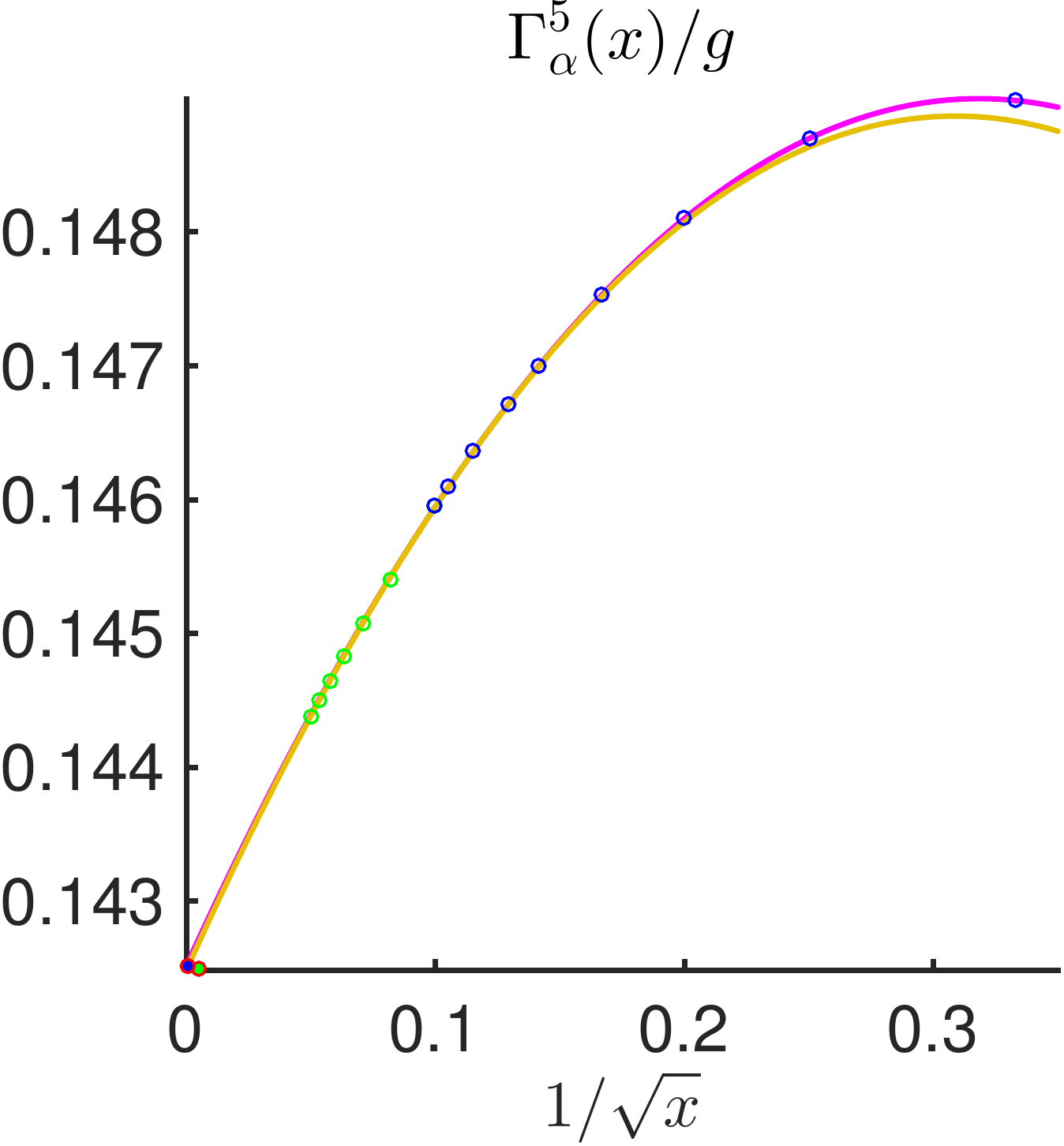}
\caption{\label{fig:extrapolationEntr}}
\end{subfigure}\hfill
\begin{subfigure}[b]{.24\textwidth}
\includegraphics[width=\textwidth]{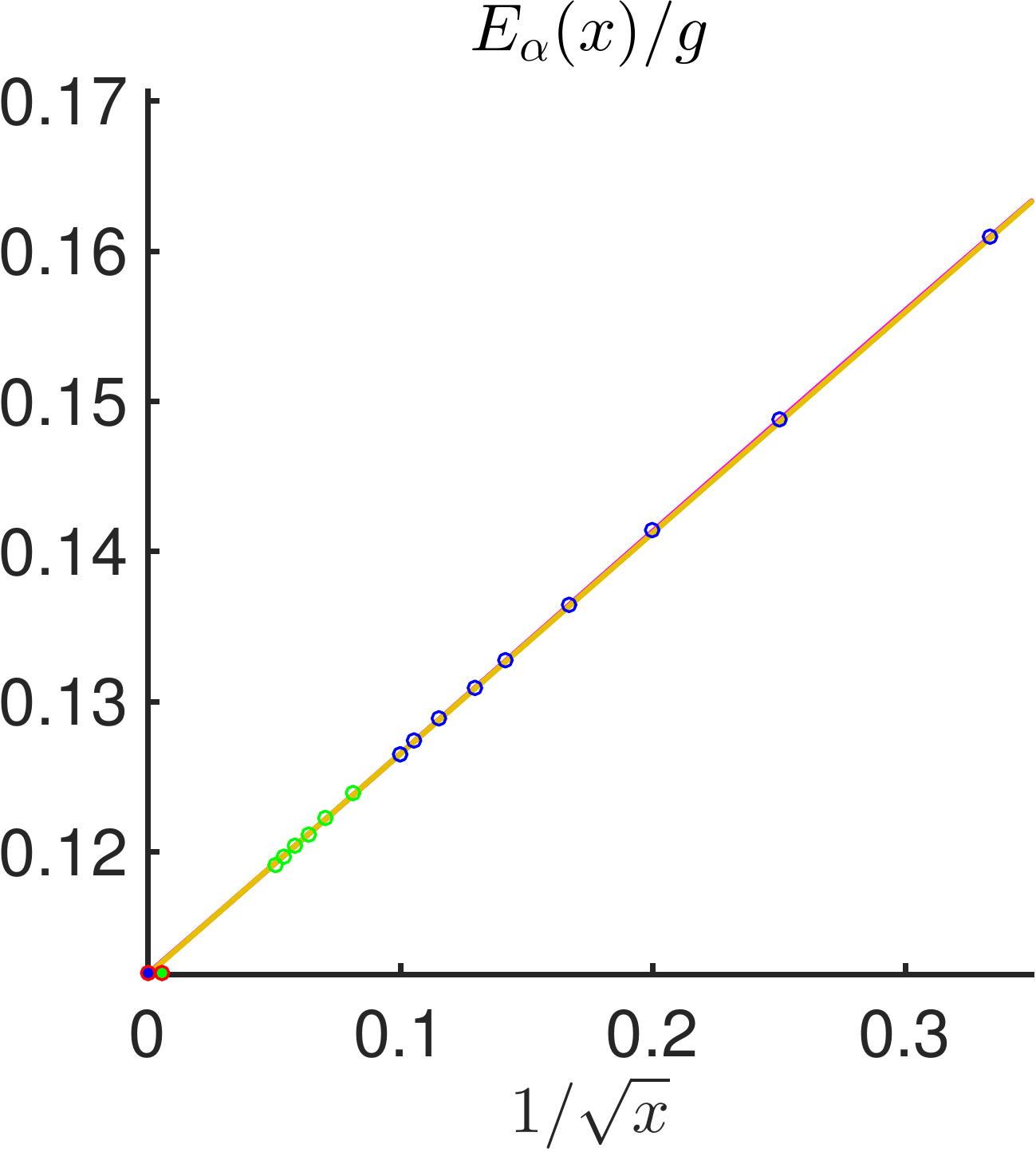}
\caption{\label{fig:extrapolationEL}}
\end{subfigure}\vskip\baselineskip
\captionsetup{justification=raggedright}
\caption{\label{fig:extrapolationGSexpVal} $m/g= 0.125, \alpha = 0.4$. Continuum extrapolation of ground-state expectation values. The blue circles are the data for $x = 9,16,25,50,60,75,90,100$ while the green circles represent the data for $x = 150,200,250,300,350,400$. The magenta line is the best polynomial fit in $1/\sqrt{x}$ through the data for $x \in X_1 = \{9,16,25,50,60,75,90,100\}$ while the yellow line is the best fit in $1/\sqrt{x}$ through the date for $x\in X_2 = \{90,100,150,200,250,300,350,400\}$. The intersection of these curves with the $(1/\sqrt{x} =0)$-axis gives the continuum estimate. In all cases, the continuum estimates obtained for $X_1$ and $X_2$ are in good agreement within an error $4 \times 10^{-4}$. 
(a) Energy density $\epsilon_0$ and comparison with the exact result $-1/\pi$ (dashed line). (b) Renormalized chiral condensate $\Delta\Sigma_\alpha$. Note that the figure confirms that $\Delta\Sigma_\alpha$ is a UV-finite quantity and, hence, we have properly renormalized it. (c) For the axial fermion current density $\Gamma_\alpha^5$, the cut-off effects at smaller values of $x$ are more severe and a higher order polynomial extrapolation is necessary. Note however that the continuum estimates for $x \in X_1$ and $x \in X_2$ agree. (d) The electric field $E_\alpha$. }
\end{figure}

In this subsection we discuss how to obtain the continuum limit of the ground-state expectation values and excitation energies which we have computed for 
\begin{subequations}\label{eq:defX1X2}
\be x \in X_1 =  \{9,16,25,36,50,60,75,90,100\}.\ee
In addition we quantify the uncertainty in our result which originates from the choice of fitting procedure. By performing a similar independent continuum extrapolation 
for 
\be x \in X_2 = \{90,100,150,200,250,300,350,400\},\ee
\end{subequations}
we show that our results are robust against the choice of fitting interval, and in particular, that the chosen $x-$range gives reliable continuum extrapolations. Finally we also check, where possible, our results against mass-perturbation theory \cite{Adam1997} and with the results of Byrnes \cite{Byrnes2002a,Byrnes2002b,Byrnes2003}.
\\
\\In Fig.~\ref{fig:extrapolationGSexpVal}, we show the energy density $\epsilon_{0,\alpha}(x) = \mathcal{E}_{0,\alpha}/2N\sqrt{x}$, the renormalized chiral condensate $\Delta\Sigma_\alpha(x)$, the axial fermion current density $\Gamma_\alpha^5 (x)$ and the electric field $E_\alpha(x)$ as a function of the lattice spacing $1/\sqrt{x}$ (in units $g = 1$) for $m/g = 0.125$ and $\alpha = 0.4$. 
As can be observed from the circles in Fig.~\ref{fig:extrapolationGSexpVal}, we have computed these quantities for the $x-$values in $X_1$ and $X_2$, see Eq.~(\ref{eq:defX1X2}). As has already been noticed in earlier studies \cite{Hamer1982,Byrnes2002a,Byrnes2002b,Byrnes2003,Buyens2013,Buyens2014,Buyens2015,Banuls2013a}, these quantities scale polynomially in $1/\sqrt{x}$ when approaching the continuum limit $x \rightarrow + \infty$. Therefore we propose to fit the data against the following polynomials in $1/\sqrt{x}$:
\begin{subequations}\label{eq:fitfunction}
\be\label{eq:fitfunctionaappa} f_1(x) = A_1 + B_1\frac{1}{\sqrt{x}}, \ee
\be\label{eq:fitfunctionbappb}f_2(x) = A_2 + B_2\frac{1}{\sqrt{x}} + C_2\frac{1}{x}, \ee
and
\be\label{eq:fitfunctioncappv}f_3(x) = A_3 + B_3\frac{1}{\sqrt{x}} + C_3\frac{1}{x} + D_3 \frac{1}{x^{3/2}}.\ee
\end{subequations}

By considering different sets of consecutive $x-$values and fitting them to $f_n$ ($n = 1,2,3$), we obtain several estimates for the continuum limit. Similar as in \cite{Buyens2016}, we take the median of the distribution of all these estimates weighed by $\exp(-\chi^2/N_{dof})$ to obtain a continuum estimate for each of the fitting functions $f_n$ and take the $15\%-85\%$ confidence interval to assign an error on this result for the choice of fitting interval. By comparing the different continuum estimates for each of the $f_n$ ($n = 1,2,3$) we obtain also an error for the choice of fitting function. We refer to appendix \ref{sec:ContinuumExtrapolationApp} for the technical details and to \cite{Buyens2016} for an even more extended discussion. 

\begin{figure}[t]
\begin{subfigure}[b]{.24\textwidth}
\includegraphics[width=\textwidth]{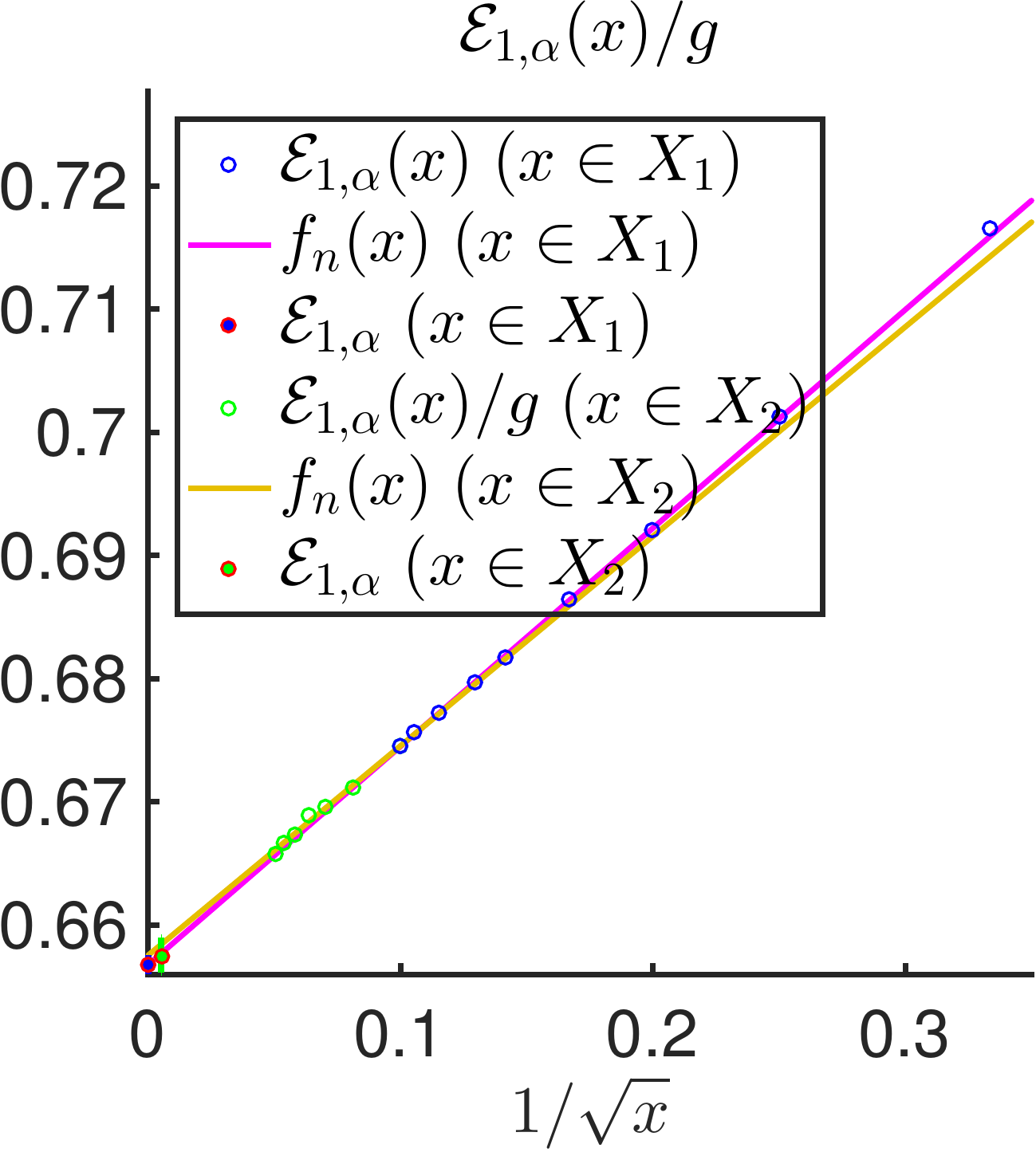}
\caption{\label{fig:extrapolationE1a} $\alpha = 0.25$.}
\end{subfigure}\hfill
\begin{subfigure}[b]{.24\textwidth}
\includegraphics[width=\textwidth]{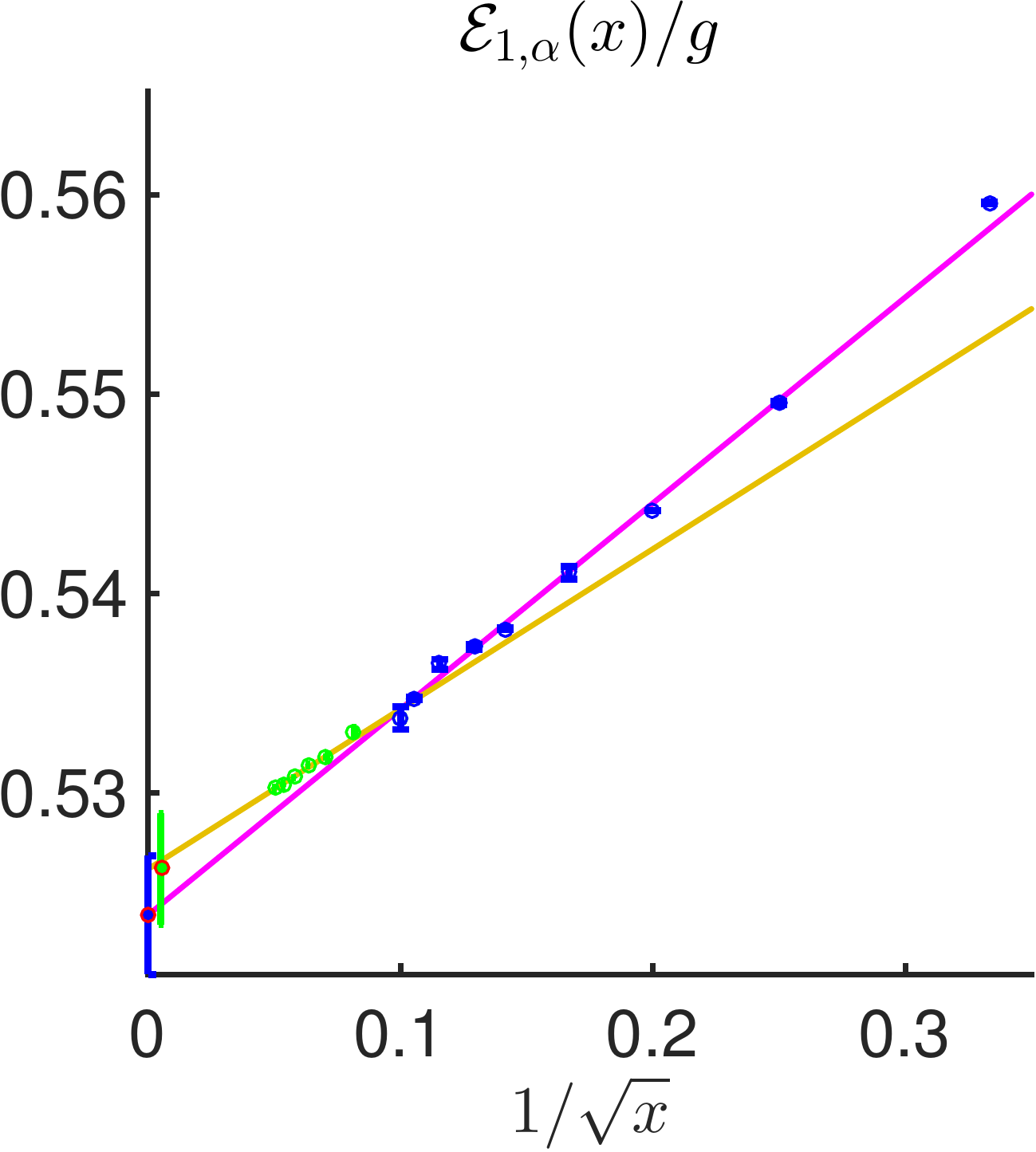}
\caption{\label{fig:extrapolationE1b} $\alpha = 0.35$.}
\end{subfigure}\vskip\baselineskip
\begin{subfigure}[b]{.24\textwidth}
\includegraphics[width=\textwidth]{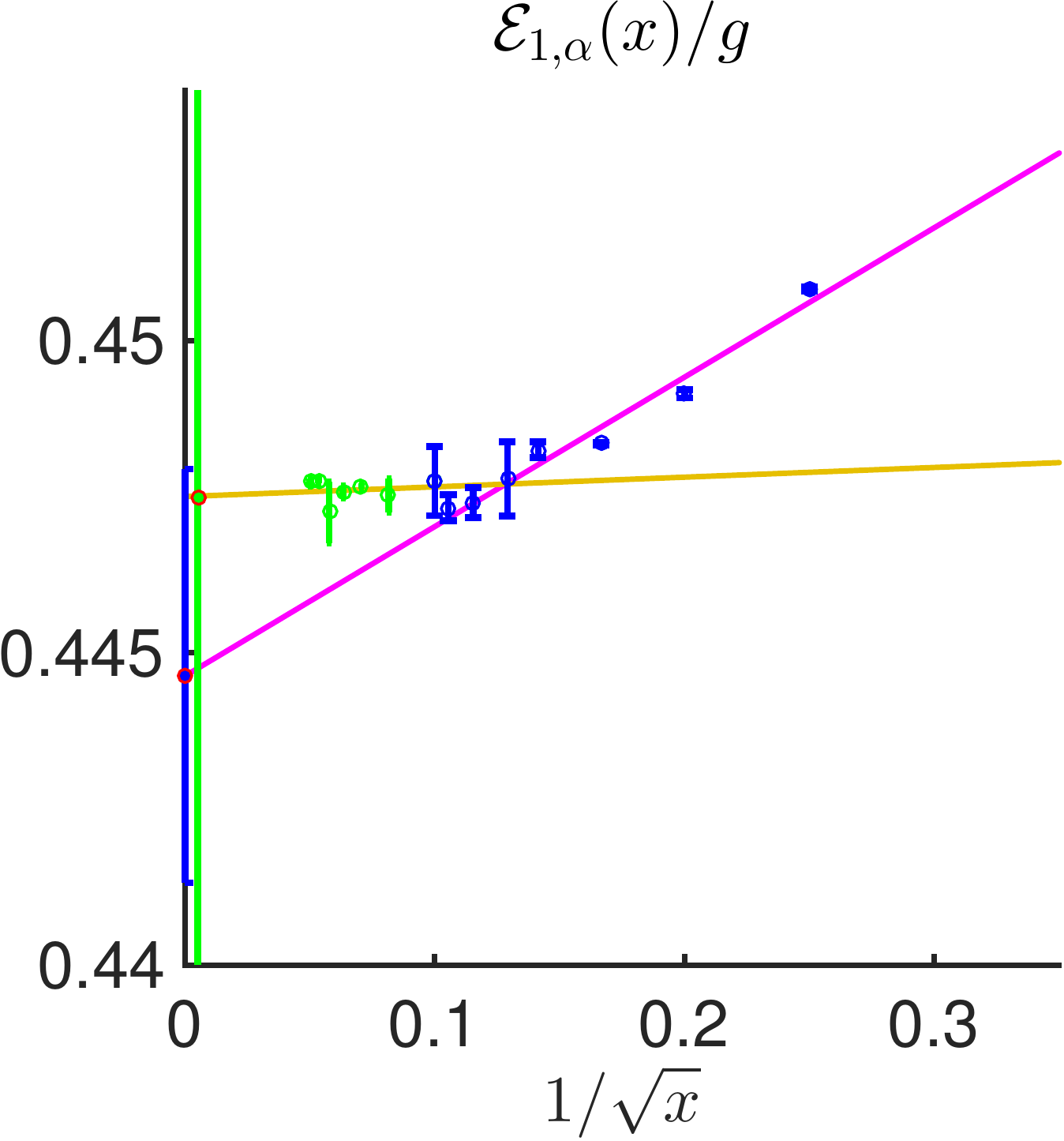}
\caption{\label{fig:extrapolationE1c} $\alpha = 0.4$.}
\end{subfigure}\hfill
\begin{subfigure}[b]{.24\textwidth}
\includegraphics[width=\textwidth]{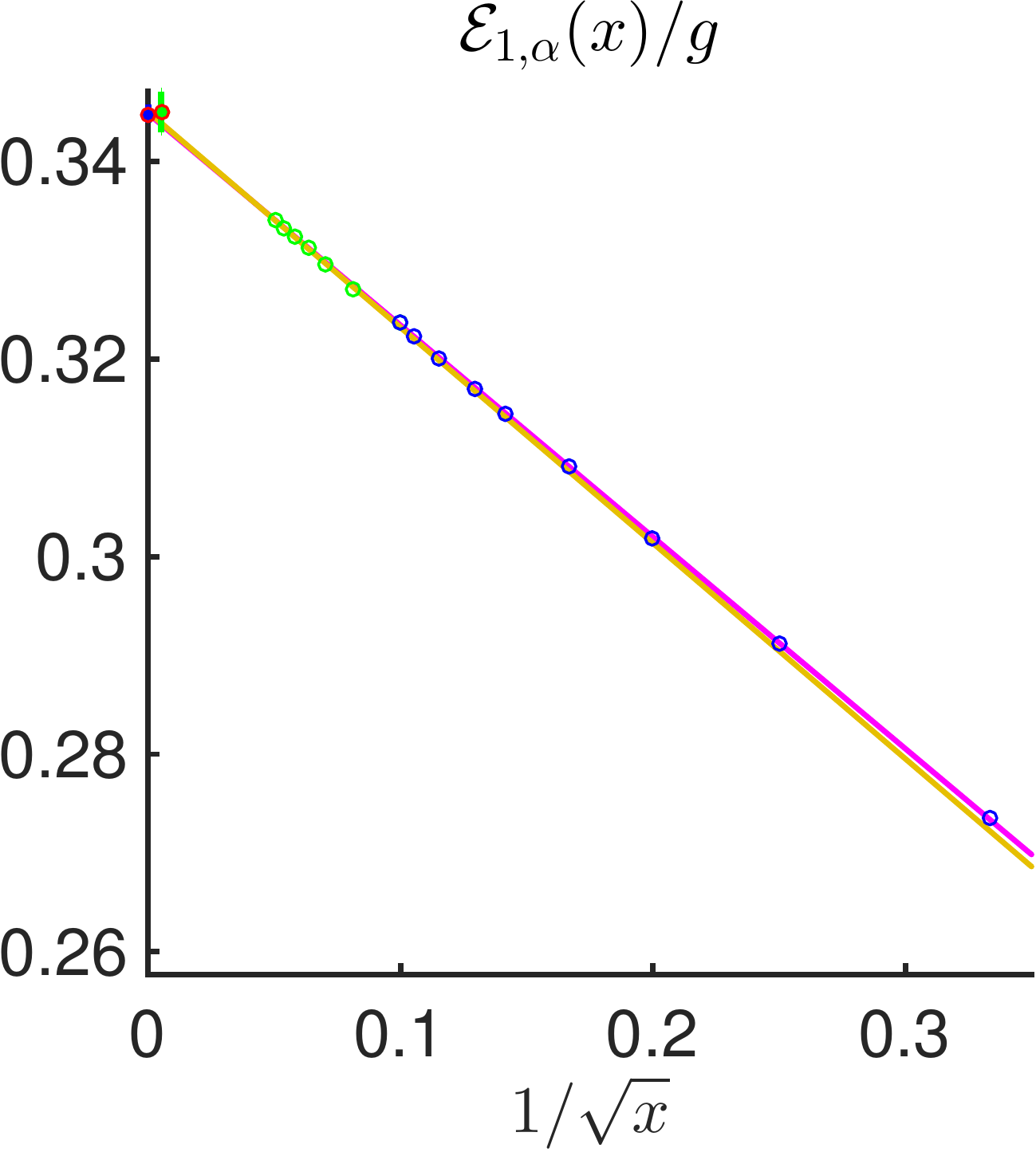}
\caption{\label{fig:extrapolationE1d} $\alpha = 0.5$.}
\end{subfigure}\vskip\baselineskip
\captionsetup{justification=raggedright}
\caption{\label{fig:extrapolationE1} $m/g= 0.125$. Similar as Fig.~\ref{fig:extrapolationGSexpVal} but now the continuum extrapolation of the mass gap $\mathcal{E}_1$ for different values of $\alpha$. One observes that the slope of $\mathcal{E}_{1,\alpha}(x)$ with respect to $1/\sqrt{x}$ changes when crossing $\alpha = 0.4$. Around $\alpha = 0.4$ a continuum extrapolation gives very large errors. Note however that the error bars from the continuum estimates obtained from $x \in X_1$ and $x \in X_2$ do overlap.  }
\end{figure}

In Fig.~\ref{fig:extrapolationGSexpVal} we show the fits that determine the continuum estimate (full line and filled circle at $1/\sqrt{x} = 0$). 
One observes that for all the quantities displayed in Fig.~\ref{fig:extrapolationGSexpVal}, the results are almost on top of each other. Note that as the error bars are very small, they are not drawn there. As another check, we show in Fig.~\ref{fig:extrapolationeps0} that the continuum estimate of the ground-state energy density $\epsilon_{0,\alpha}$ is very close to its real result $-1/\pi$ (dashed line) within an error of $1 \times 10^{-4}$ (for $x \rightarrow + \infty$: $H_\alpha/2\sqrt{x}$ becomes the Heisenberg $XY$ model). 

Fig. \ref{fig:extrapolationE1} shows the same as fig. \ref{fig:extrapolationGSexpVal}, but now for the energy $\mathcal{E}_{1,\alpha}$ of the first excited state and for different values of $\alpha$. One observes now that the slope of $\mathcal{E}_{1,\alpha}(x)$ with respect to $1/\sqrt{x}$ changes as $\alpha$ crosses 0.4. This makes a continuum extrapolation hard for $\alpha = 0.4$, see figs. \ref{fig:extrapolationE1c}, and, hence, introduces large errors for these values. Therefore we compute for $m/g = 0.125$ the excitation energies for $\alpha = 0.42$ instead of $\alpha = 0.4$. Note however that we do not face this problem for the ground-state expectation values, see fig. \ref{fig:extrapolationGSexpVal}, or when $\alpha$ is farther away from $\alpha = 0.4$, see figs. \ref{fig:extrapolationE1a} and \ref{fig:extrapolationE1d}. In particular, we find that also that the continuum estimates of the excitation energies, obtained independently from the sets $X_1$ and $X_2$, are in agreement with each other. 

As another check, we compare in Appedix \ref{subsec:compareEarlierStudies} our results with mass perturbation theory and with the results of \cite{Byrnes2002a,Byrnes2002b,Byrnes2003} for $\alpha = 0.5$. We find that our results agree in the appropriate regimes and, therefore, we can be confident that our procedure to obtain a continuum estimate from the simulations at non-zero lattice spacing $1/\sqrt{x}$ provides a reliable method. Therefore, we adopt this method to obtain continuum estimates of ground-state expectation values and excitation energies from our simulations with $x = 9,16,25,36,60,60,75,90,100$.

\section{Single particle spectrum}\label{sec:SPspectrum}
\begin{figure}[t]
\begin{subfigure}[b]{.24\textwidth}
\includegraphics[width=\textwidth]{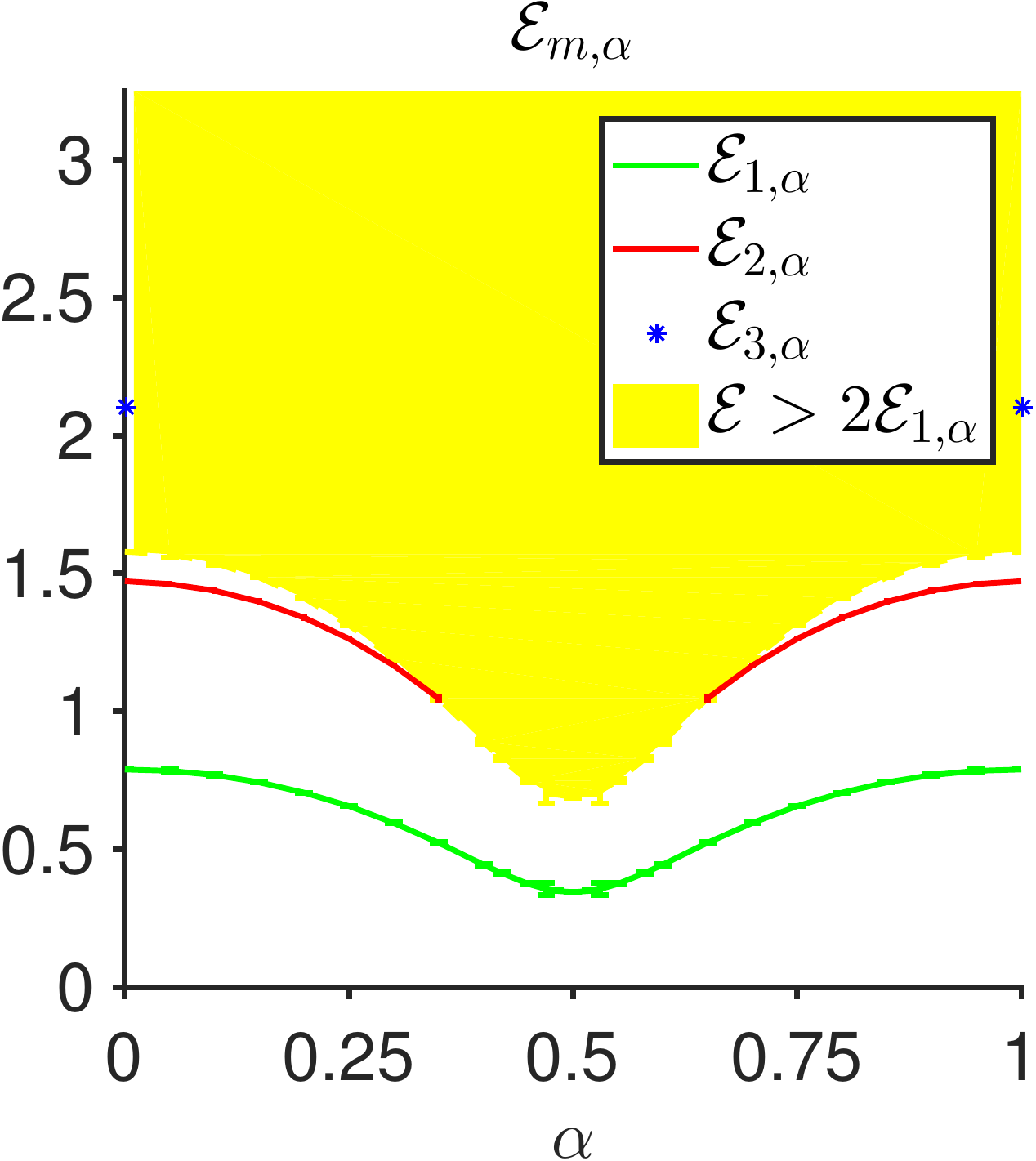}
\caption{\label{fig:ExcDiffBacka} $m/g = 0.125$.}
\end{subfigure}\hfill
\begin{subfigure}[b]{.24\textwidth}
\includegraphics[width=\textwidth]{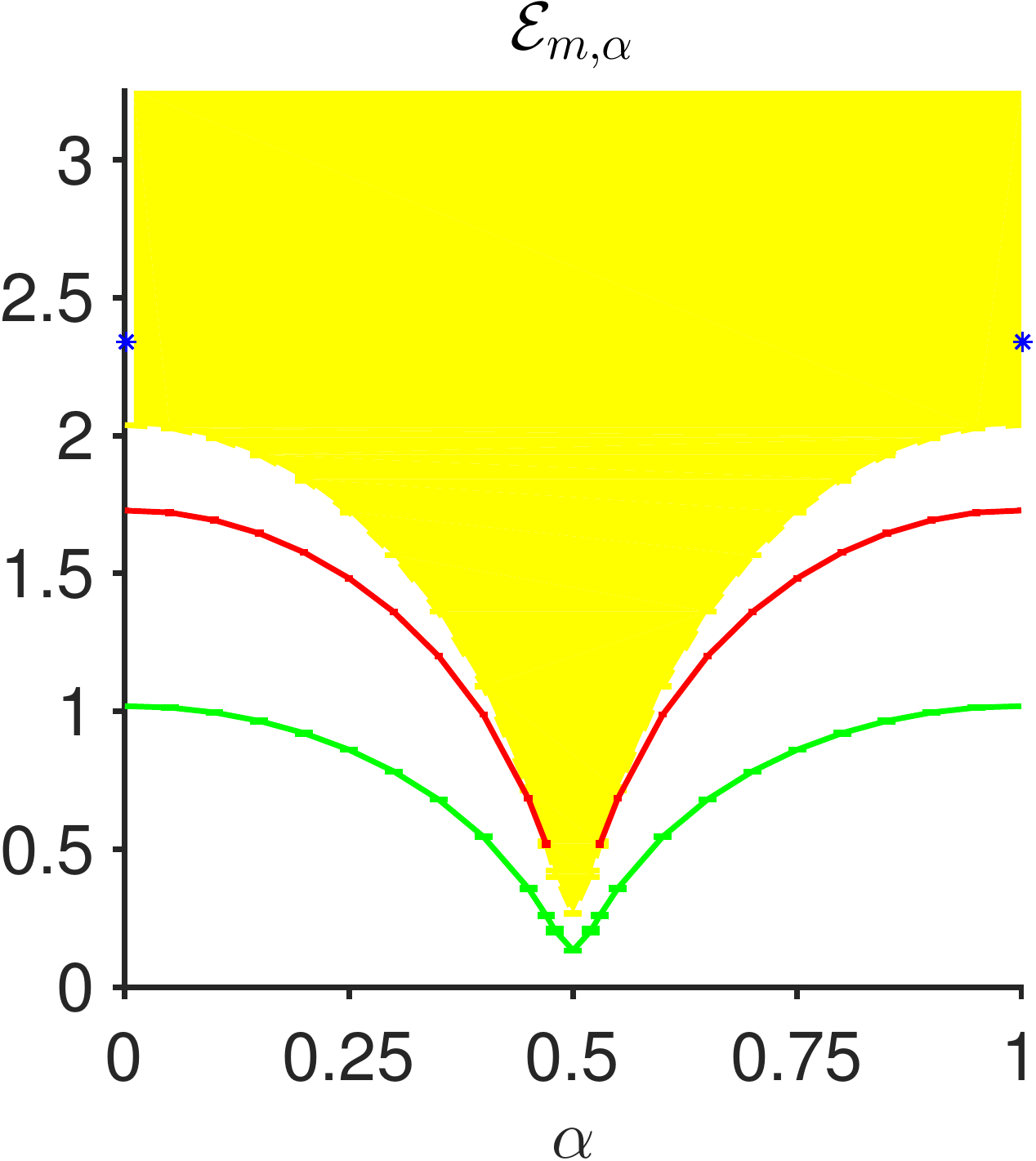}
\caption{\label{fig:ExcDiffBackb}$m/g = 0.25$.}
\end{subfigure}\vskip\baselineskip
\begin{subfigure}[b]{.24\textwidth}
\includegraphics[width=\textwidth]{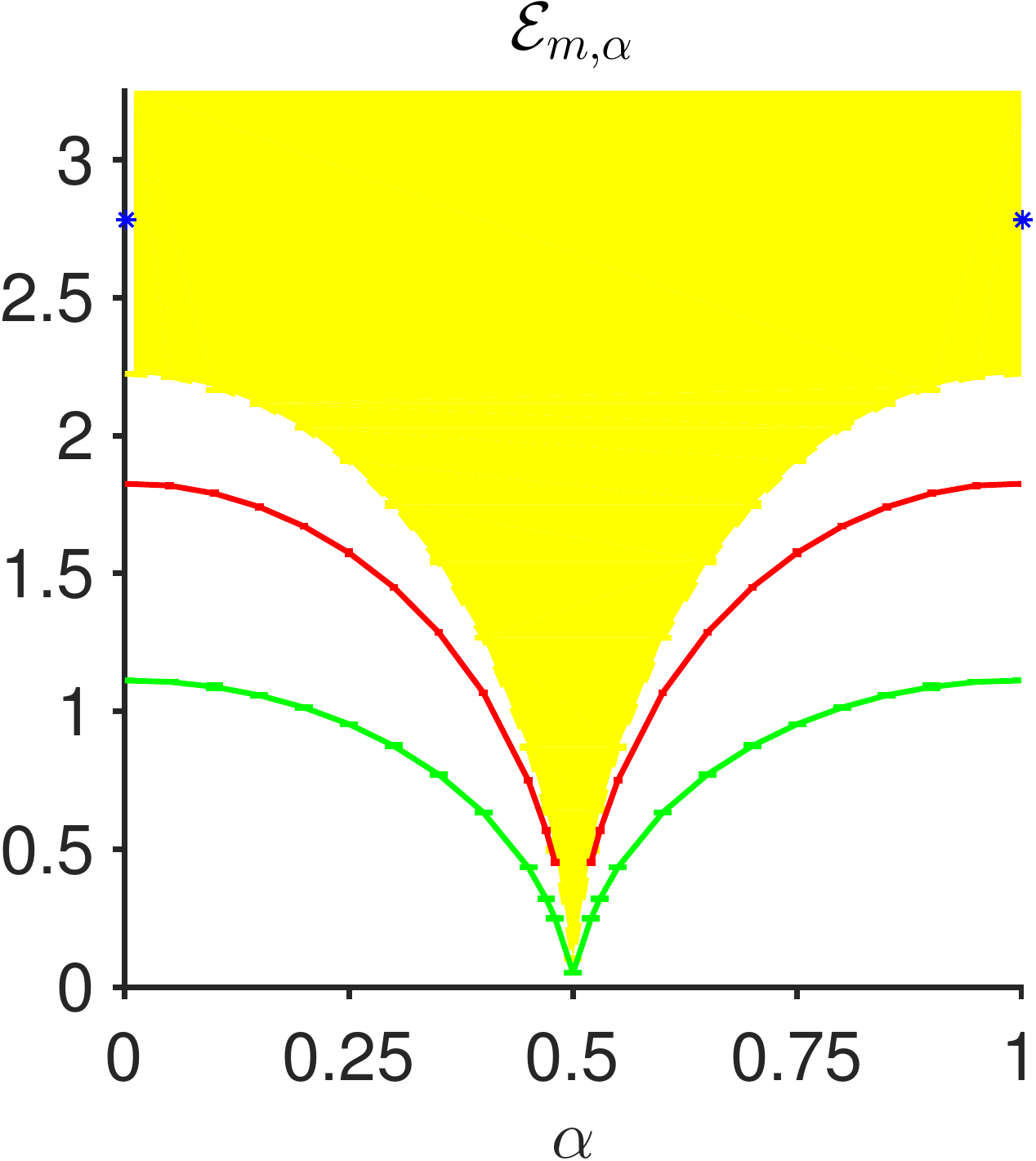}
\caption{\label{fig:ExcDiffBackc} $m/g = 0.3$.}
\end{subfigure}\hfill
\begin{subfigure}[b]{.24\textwidth}
\includegraphics[width=\textwidth]{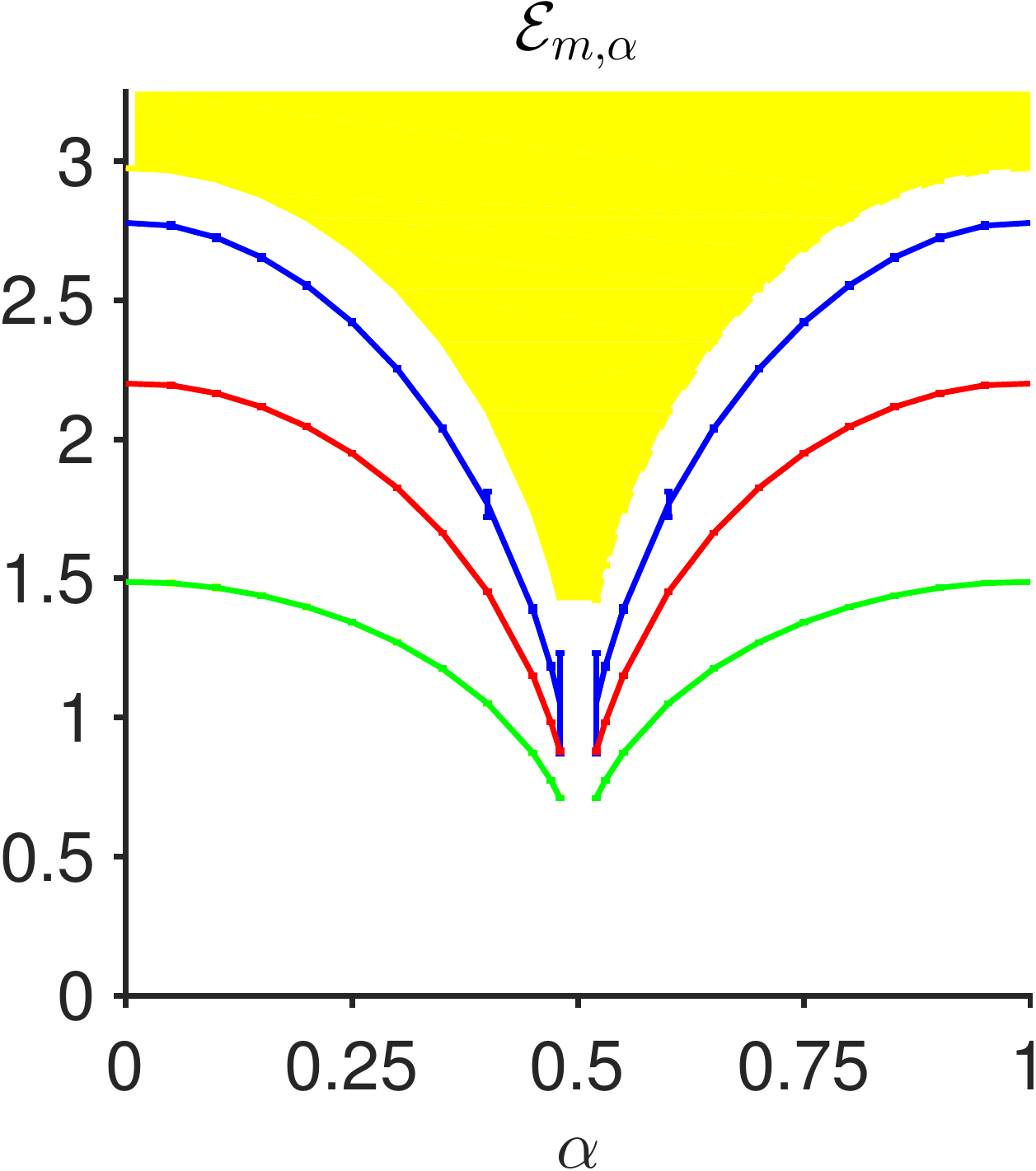}
\caption{\label{fig:ExcDiffBackd}$m/g = 0.5$.}
\end{subfigure}\vskip\baselineskip
\captionsetup{justification=raggedright}
\caption{\label{fig:ExcDiffBack} Energy of the single-particle excitations as a function of $\alpha$ for different values of $m/g$. The energy $\mathcal{E}_{1,\alpha}$ of the first single-particle excitation is shown with a green line, the energy $\mathcal{E}_{2,\alpha}$ of the second single particle excitation is shown with a red line and the energy $\mathcal{E}_{3,\alpha}$ of the third single-particle excitation is shown with a blue line. The yellow line shows the continuum spectrum consisting of multi-particle states with energy larger than $2\mathcal{E}_{1,\alpha}$. For $m/g = 0.125$, $0.25$, $0.3$ the excitation with energy $\mathcal{E}_{3,\alpha}$ corresponds only to a single-particle excitation for $\alpha = 0$ and $\alpha = 1$ (blue star).}
\end{figure}

Most of the ground-state properties have already been investigated in the context of confinement \cite{Buyens2015}. Therefore we present our results in Appendix \ref{subsec:ResultsGroundState}. Here we focus on the single-particle spectrum as a function of $\alpha$. 
\\
\\
As explained in subsection \ref{subsec:phaseDiagSchwinger}, for $\alpha = 0$, there are two single-particle excitations with $CT = -1$ and energies $\mathcal{E}_{1,\alpha}, \mathcal{E}_{3,\alpha}$ and one single-particle excitation with $CT = 1$ and energy $\mathcal{E}_{2,\alpha}$ with the hierarchy $\mathcal{E}_{1,\alpha} < \mathcal{E}_{2,\alpha} < \mathcal{E}_{3,\alpha}$, see \cite{Buyens2013}. For $m/g = 0.125,0.25,0.3$ we have that $\mathcal{E}_{1} < \mathcal{E}_{2} + \mathcal{E}_{3}$ and $\mathcal{E}_{3,\alpha} > 2\mathcal{E}_{1,\alpha}$ while for $m/g \gtrsim 0.5$ we have $\mathcal{E}_{3,\alpha} \leq 2\mathcal{E}_{1,\alpha}$. This means that for $m/g = 0.125,0.25,0.3$ the decay of $\mathcal{E}_{3,\alpha}$ into two elementary particles is only prevented by the $CT$ symmetry. When $0 < \alpha < 1/2$, the $CT$ symmetry is broken and this decay is no longer forbidden. This is indeed what we observe in the single-particle spectrum: for $\alpha > 0$, only the excitations with energy $\mathcal{E}_{1,\alpha}$ correspond to single-particle excitations, see fig. \ref{fig:ExcDiffBack}(a)-(c). Furthermore, we observe that the binding energy $\mathcal{E}_{bind}= 2\mathcal{E}_{1,\alpha} - \mathcal{E}_{2,\alpha}$ decreases as $\alpha$ tends towards $1/2$. 

For $m/g = 0.125$, see fig. \ref{fig:ExcDiffBack}, the second particle is stable until $\alpha \lesssim 0.35$. For $\alpha = 0.42$ our estimates are $\mathcal{E}_{1,\alpha} = 0.414(4)$ and $\mathcal{E}_{2,\alpha} = 0.852 (7)$, indicating that the second excited state is unstable against decay into two particles with energy $\mathcal{E}_{1,\alpha}$: $\mathcal{E}_{2,\alpha} > 2\mathcal{E}_{1,\alpha}$. When $\alpha \geq 0.35$ we have $\mathcal{E}_{2,\alpha}(x) > 2\mathcal{E}_{1,\alpha}(x)$ for all the $x-$values we used. We conclude that there are two single-particle excitations for $\alpha \lesssim 0.35$ and only one single-particle excitation for $\alpha \gtrsim 0.42$. This agrees qualitatively with mass perturbation theory, $m/g \ll 1$, where there are two single-particle excitations for $\alpha \leq 1/4$ and one single-particle excitation for $1/4 < \alpha \leq 1/2$ \cite{Coleman1976}.

For $m/g = 0.25$, see fig. \ref{fig:ExcDiffBackb}, our estimates for the energy $\mathcal{E}_{2,\alpha}$ were unstable against variation of the bond dimension $D$ for  $\alpha \geq 0.48$. The errors on $\mathcal{E}_{2,\alpha}$ were too large and prevent an extrapolation towards $x = \infty$. Nevertheless, in our simulations we have $\mathcal{E}_{2,\alpha}(x) < 2\mathcal{E}_{1,\alpha}(x)$ for all our $x-$values and the fact that $\mathcal{E}_{2,\alpha}(x)$ decreases as the bond dimension increases might suggest that this particle is still stable but with very small binding energy.  For $\alpha = 1/2$, the ground state is $CT$ invariant for $m/ g \leq (m/g)_c$ allowing us to classify the excitations according to their $CT$-number with the method similar as in \cite{Buyens2013}. We compute the excitation energies with and without classifying the states according to their $CT$ number for $(m/g,\alpha) = (0.25,1/2)$. In both cases, we found only one single-particle excitation. In the vector sector ($CT = -1$) all other states have energies that are larger than $3\mathcal{E}_{1,\alpha}$ and in the scalar sector ($CT = 1$) the energies were larger than $2\mathcal{E}_{1,\alpha}$. This corresponds to a theory with one single-particle excitation. Therefore we estimate the value of the electric background field where the second elementary particle disappears to be larger than $0.47$ but smaller than $0.5$ for $m/g = 0.25$. 

A similar picture arises for $m/g = 0.3$, see Fig.~\ref{fig:ExcDiffBackc}. Here we estimate that the second elementary particle disappears between $\alpha = 0.48$ and $\alpha = 0.5$. One also observes that the mass gap decreases as we approach the phase transition $(m/g, \alpha) \rightarrow ((m/g)_c, 1/2)$: for $(m/g,\alpha) = (0.3,1/2)$ our estimate for the mass gap is $\mathcal{E}_{1,\alpha} = 0.0527(5) $. 

We conclude that for $m/g \leq (m/g)_c$ and relative small values of $\alpha$ there are two single-particle excitations with energies $\mathcal{E}_1$ and $\mathcal{E}_2$. Above a certain value of $\alpha$, the excitation with energy $\mathcal{E}_2$ does not correspond to a single-particle excitation anymore and, hence, disappears in the continuum of the spectrum. This mechanism is best understood as the binding energy of the second excited state becoming too small to be stable against a decay into two elementary particles with smaller energy. Not surprisingly, we find that when approaching the phase transition that the mass gap becomes smaller. In particular, when $m/g$ is close to the critical mass, the mass gaps decreases more suddenly when approaching $\alpha = 1/2$ compared to the more smooth behavior for $m/g = 0.125$. 

This picture changes for $m/g \geq (m/g)_c$. For instance, for $m/g = 0.5$ we have for all values of $\alpha$ that $\mathcal{E}_{3,\alpha} < 2\mathcal{E}_{1,\alpha}$ and thus at least three single-particle excitations exist, see Fig.~\ref{fig:ExcDiffBackd}. When $\alpha \rightarrow 1/2$ we observe that the difference between the  energies $\mathcal{E}_{1,\alpha}$, $\mathcal{E}_{2,\alpha}$ and $\mathcal{E}_{3,\alpha}$ becomes smaller. This results in the fact that the ansatz Eq.~(\ref{eq:excAnsatz}) is less accurate in approximating the single-particle excitations. Indeed, for $\alpha \geq 0.45$ we observe large error bars for $\mathcal{E}_{3,\alpha}$. Anyway, we found that $\mathcal{E}_{1,\alpha}$, $\mathcal{E}_{2,\alpha}$ and $\mathcal{E}_{3,\alpha}$ were stable for all values of $x$. Furthermore, for $\alpha \geq 0.45$ we found even a fourth solution to the eigenvalue problem Eq.~(\ref{eq:excAnsatz}) that might correspond to a single-particle excitation. However, because its energy was very close to $2 \mathcal{E}_{1,\alpha}$ the errors on this energy using the ansatz Eq.~(\ref{eq:excAnsatz}) for fixed values of $x$ were too large to obtain a reliable continuum estimate. 

Our results thus show that the spectrum of $m/g = 0.5$ differs from the spectrum for $m/g \leq (m/g)_c$. For $\alpha = 1/2$, due to spontaneous symmetry breaking of the $CT$ symmetry, there are two vacua and kink excitations which connect these two vacua \cite{Byrnes2002a,Byrnes2002b,Byrnes2003}. Local excitations constructed on top of one of the two vacua are scattering states containing an equal number of kinks and antikinks. Away from $\alpha = 1/2$, the $CT$ symmetry is explicitly broken and only one of the two vacua survives as ground state, while individual kinks do no longer exist. The splitting in energy density between the two vacua acts as a linear attractive potential between kink-antikink pairs. As such, the elementary excitations on top of the ground state that we observe for $\alpha$ close to $1/2$, such as those with energies $\mathcal{E}_{1,\alpha},\mathcal{E}_{2,\alpha},\mathcal{E}_{3,\alpha}$, emerge as remnant of the symmetry breaking and can be thought of as kink-antikink bound states stabilized by the attractive interaction. As $\alpha \rightarrow 1/2$, the slope of the potential decreases and more and more bound states come closer together in the spectrum, below our limit of energy resolution, and finally make up the kink-antikink continuum for $\alpha = 1/2$.

\section{Conclusions}
In this paper we presented an overview of the low-energy properties of the Schwinger model in terms of the fermion mass $m/g$ and the electric background field $\alpha$, complementing earlier studies \cite{Coleman1976,Adam1997,Byrnes2002a,Byrnes2002b,Byrnes2003} for $\alpha = 0$ and $\alpha = 1/2$ with numerical MPS-simulations for $\alpha \in [0,1/2]$. We also investigated in great detail the influence of truncating the infinite dimensional Hilbert space of the gauge fields by quantifying the contribution of each of the irreducible $U(1)$-representations to ground state expectation values. The conclusion is that, even close to the continuum limit and a phase transition, this contribution falls of exponentially with the quadratic Casimir invariant of the representation. We expect the same conclusion to hold for any $SU(N)$ Yang-Mills gauge-theory, that is, that the infinite Hilbert space of the gauge fields poses no obstacle to study Yang-Mills theories in the Hamiltonian framework by means of tensor network methods.   

However, there are still formidable challenges for the TNS framework to overcome: possibly the biggest one is going to higher dimensions. The generalization of MPS to higher dimensions are the Projected Entangled-Pair States (PEPS) \cite{Verstraete2004b}. Although some interesting studies of gauge theories with PEPS have appeared \cite{Tagliacozzo2014,Haegeman2015,Zohar2015,Milsted2016,Zohar2016}, at present, the need for a large number of variational freedom when approaching the continuum limit, is still hindering a truly variational study of gauge field theories \cite{Lubasch2014}. Fortunately, in the last years the PEPS methods have significantly improved
\cite{Murg2007,Corboz2009,Jordan2008,Corboz2010,Kraus2010,Corboz2014,Vanderstraeten2015b,Phien2015,Corboz2016,Vanderstraeten2016b}. In particular, for some models the PEPS framework can already compete with state-of-the-art results of Monte-Carlo simulations \cite{Corboz2014}. This makes us confident that the TNS framework will provide a tool in the near future for the study of gauge field theories in the illusive regimes which are inaccessible with other methods.

\section*{Acknowledgements}
We acknowledge interesting discussions with M.C. Ba\~{n}uls, P. Silvi and L. Vanderstraeten. We are also grateful to the Mainz Institute for Theoretical Physics (MITP) for its hospitality and its partial support during completion of the work. This work is supported by an Odysseus grant from the FWO, a PhD-grant from the FWO (B.B), a post-doc grant from the FWO (J.H.), the FWF grants FoQuS and Vicom, the DFG via the SFB/TRR21, the ERC grants QUTE and ERQUAF, and the EU grant SIQS. Simone Montangero gratefully acknowledges the support of the DFG via a Heisenberg fellowship.

\appendix
\numberwithin{equation}{section}
\renewcommand\theequation{\Alph{section}.\arabic{equation}}
\section{Ground-state properties}
\subsection{The quantities and their lattice version}\label{sec:appQuantLatVer}
In this paper we consider the following quantities:
\begin{itemize}
\begin{subequations}\label{eq:quantDisc}
\item[-] The electric field $E_\alpha$:
\be E_\alpha  = \Braket{ E }_0  = \frac{g}{2}\Braket{L(1) + L(2) + 2\alpha}_0,\ee
\item[-]  The chiral condensate $\Sigma_\alpha$
\be \Sigma_\alpha   = \Braket{ \bar{\psi}\psi }_0 = g \frac{\sqrt{x}}{4}\Braket{-\sigma_z(1) + \sigma_z(2) + 2}_0,\ee
\item[-]  The axial fermion current density $\Gamma_\alpha^5$:
\bea \Gamma_\alpha^5   = & i \Braket{ \bar{\psi}\gamma^5 \psi }_0 \nonumber \\
= & g\displaystyle{\frac{\sqrt{x}}{4}\left(\Braket{\sigma^+(1)e^{i\theta(1)}\sigma^-(2) + h.c.}_0\right.} \nonumber \\
 	& \displaystyle{- \left.\Braket{\sigma^+(2)e^{i\theta(2)}\sigma^-(3) + h.c.}_0\right)},\eea
\end{subequations}
\end{itemize}
where $\braket{\ldots}_0$ denotes the expectation value with respect to the ground state of $H_\alpha$ with an electric background field $g\alpha$. As the chiral condensate is UV-divergent, we consider its renormalized version: if $\Sigma_\alpha$ is the chiral condensate of the ground state of $H_\alpha$ with electric background field $\alpha$ then we consider 
$$\Delta \Sigma_\alpha = \Sigma_\alpha - \Sigma_{\alpha = 0}, $$
with $\Sigma_{\alpha = 0}$ computed at the same value of $m/g$. 

Furthermore, the ground-state energy $\mathcal{E}_{0,\alpha} = \braket{H_\alpha}_0$ is IR-divergent and UV-divergent
\begin{subequations}\label{eq:endens}
\be \mathcal{E}_{0,\alpha} = g2N\sqrt{x}\epsilon_{0,\alpha} \ee
with $\epsilon_{0,\alpha}$ finite as $N \rightarrow + \infty$ and $x \rightarrow + \infty$. As $H_\alpha/2\sqrt{x}$ becomes the Heisenberg $XY$ model in the limit $x \rightarrow + \infty$ we have that
\be \epsilon_{0,\alpha} = -1/\pi \mbox{ for } x \rightarrow + \infty \ee
\end{subequations}
which is independent of $m/g$ and $\alpha$. Another possibility to renormalize the ground-state energy is to substract the zero-background contribution and consider the so-called string tension
$$\sigma_\alpha = g\sqrt{x}\frac{\mathcal{E}_{0,\alpha} - \mathcal{E}_{0,\alpha=0}}{2N} $$
which is also UV-finite. 

\subsection{Comparison with earlier studies}\label{subsec:compareEarlierStudies}
Adam \cite{Adam1997} showed in mass-perturbation theory $(m/g \ll 1)$ that
\begin{subequations}\label{eq:AdamMassPert}
\begin{widetext}
\be \mathcal{E}_{1,\alpha} =  \mu_0\sqrt{1+ 3.5621\frac{m}{\mu_0}\cos(2\pi \alpha) + \left(5.4807 - 2.0933\cos(4\pi\alpha)\right)\left(\frac{m}{\mu_0}\right)^2} + \mathcal{O}\left[\left(\frac{m}{g}\right)^3\right] \ee
\be E_\alpha = -2\pi\frac{m}{g}\tilde{\Sigma}\sin(2\pi \alpha) + \pi \left(\frac{m}{g}\right)^2\tilde{\Sigma}^2E_{+}\sin(4\pi\alpha) + \mathcal{O}\left[\left(\frac{m}{g}\right)^3\right] \ee
\be \Delta\Sigma_\alpha = -\tilde{\Sigma}(\cos(2\pi\alpha)-1) + \frac{m}{g}\frac{\tilde{\Sigma}^2}{2}E_+(\cos(4\pi\alpha) - 1) + \mathcal{O}\left[\left(\frac{m}{g}\right)^2\right]\ee
\be \Gamma_\alpha^5 = -\tilde{\Sigma}\sin(2\pi\alpha) - \frac{m}{g}\frac{\tilde{\Sigma}^2}{2}E_+\sin(4\pi\alpha)+ \mathcal{O}\left[\left(\frac{m}{g}\right)^2\right],\ee
\end{widetext}
\end{subequations}
with $\mu_0 = g/\sqrt{\pi}$, $\tilde{\Sigma} = -e^{\gamma}\mu_0/2\pi, \gamma \approx 0.5772$ (the Euler-Mascheroni constant) and $E_+ = -28.0038/g^2$.

\begin{figure}[t]
\begin{subfigure}[b]{.24\textwidth}
\includegraphics[width=\textwidth]{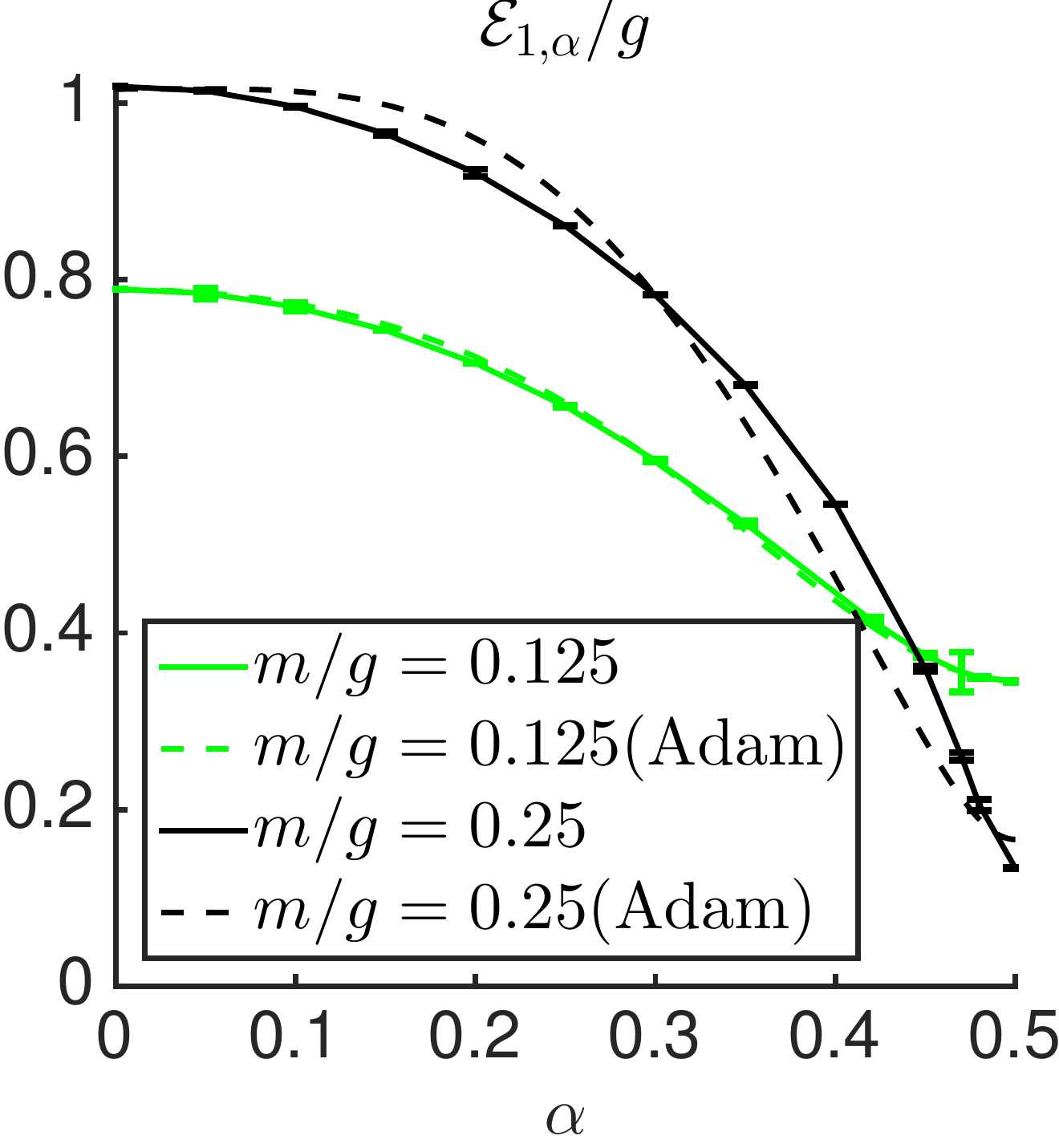}
\caption{\label{fig:checkExtrE0densm125e3}}
\end{subfigure}\hfill
\begin{subfigure}[b]{.24\textwidth}
\includegraphics[width=\textwidth]{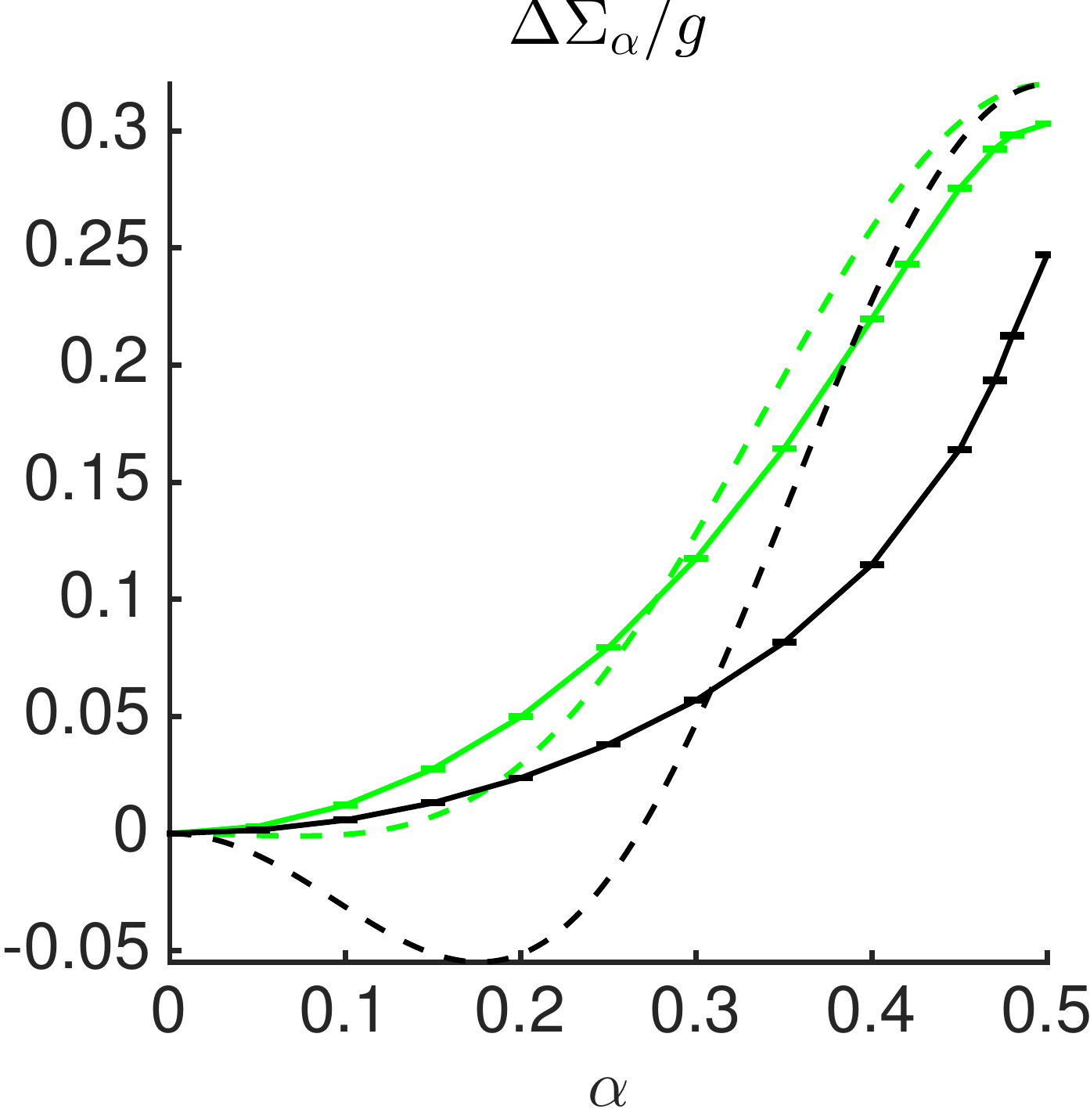}
\caption{\label{fig:checkExtrCCm125e3}}
\end{subfigure}\vskip\baselineskip
\begin{subfigure}[b]{.24\textwidth}
\includegraphics[width=\textwidth]{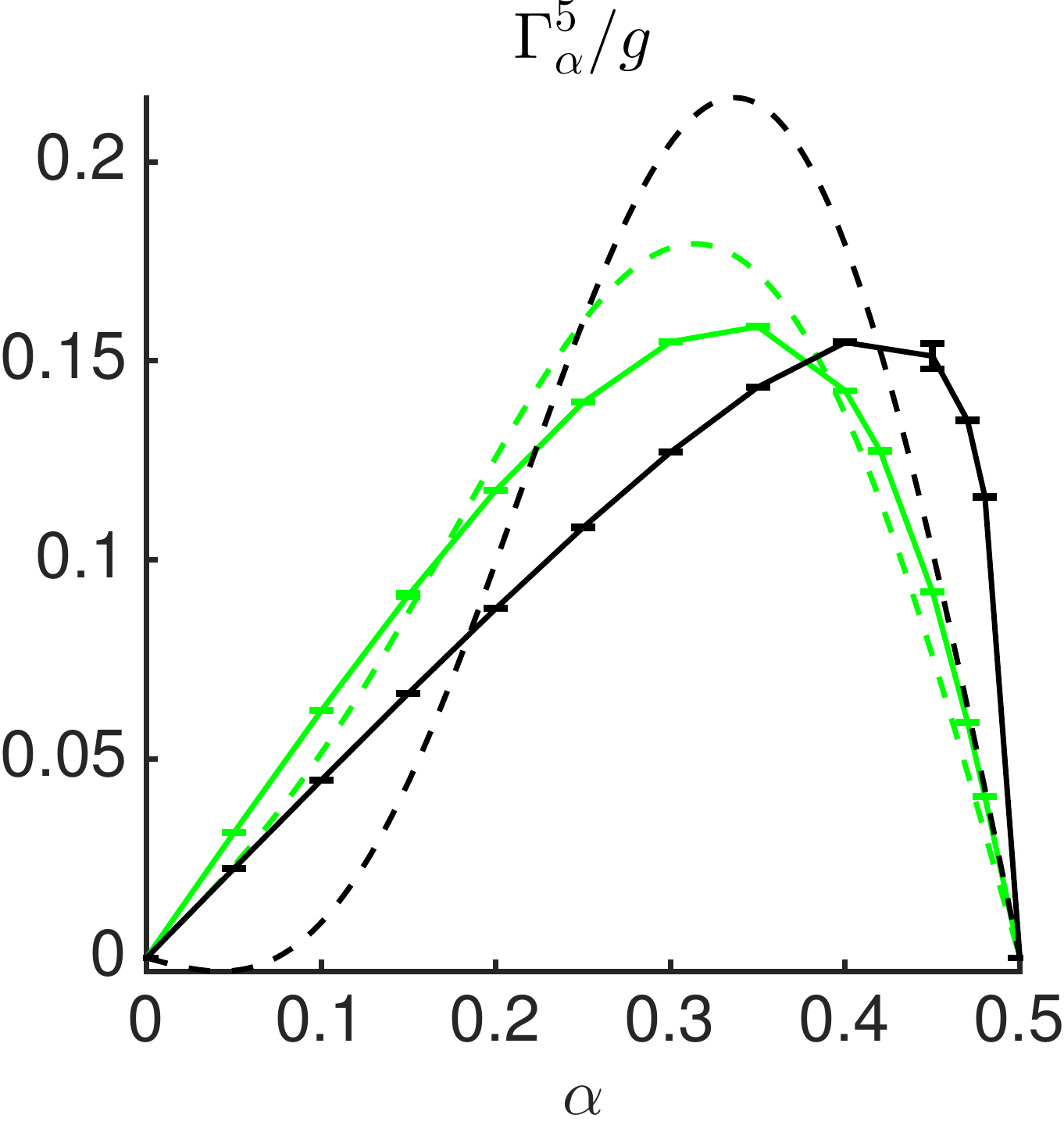}
\caption{\label{fig:checkExtrELm125e3}}
\end{subfigure}\hfill
\begin{subfigure}[b]{.24\textwidth}
\includegraphics[width=\textwidth]{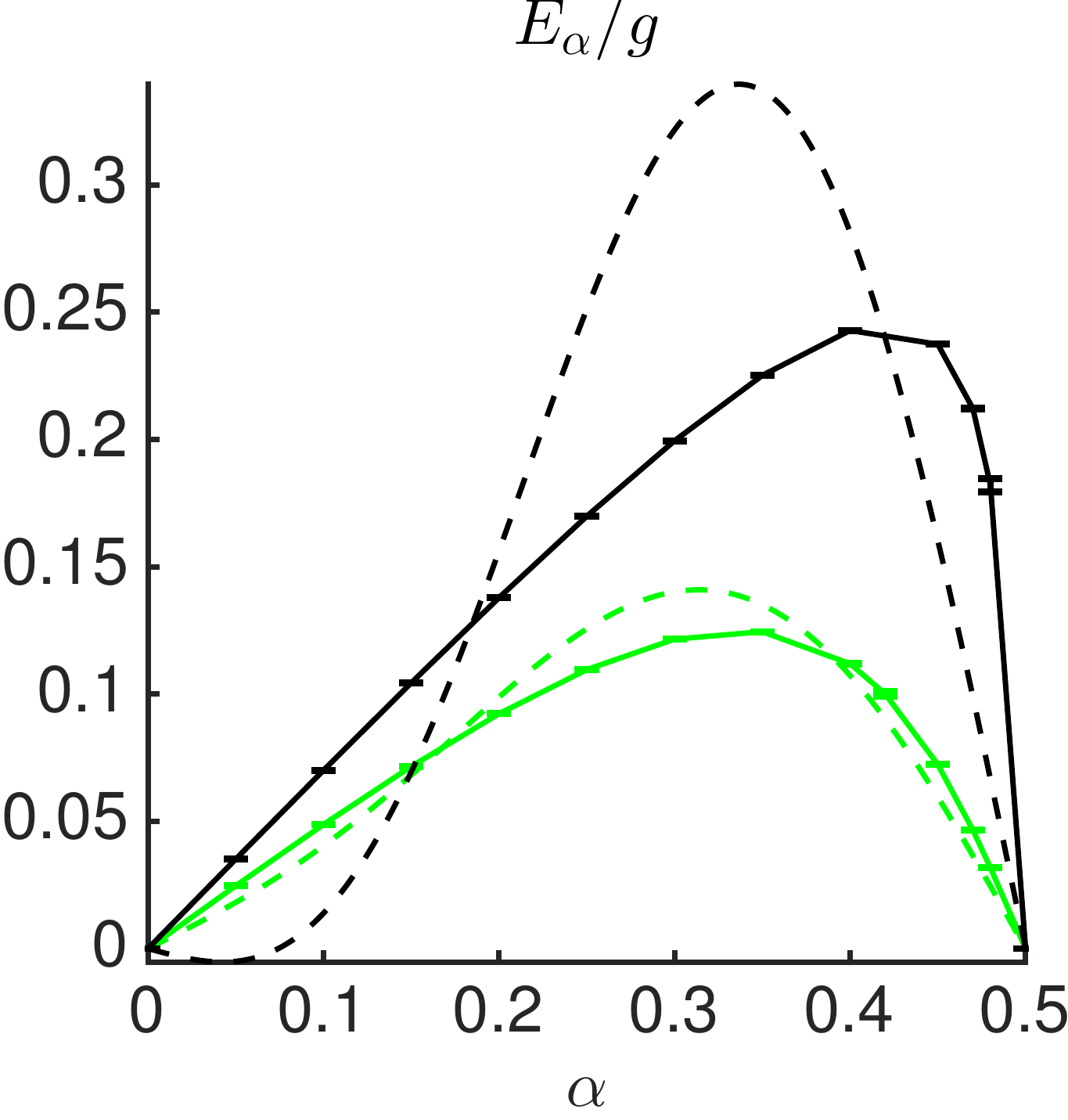}
\caption{\label{fig:checkExtrEntrm125e3}}
\end{subfigure}\vskip\baselineskip
\captionsetup{justification=raggedright}
\caption{\label{fig:checkExtrm125e3} Comparison of our results for $m/g = 0.125$ and $m/g = 0.25$ (full line) with the results in mass perturbation theory of Adam \cite{Adam1997}(dashed line) for different quantities. We observe convergence towards the perturbative results for $m/g \rightarrow 0$. In particular, for the mass gap $\mathcal{E}_1$ the result for $m/g = 0.125$ matches very well the predicted behavior in mass-perturbation theory.}
\end{figure}

In fig. \ref{fig:checkExtrm125e3} we compare our results (full line) with the perturbative results (dash line) for the quantities $\mathcal{E}_{1,\alpha}$, $\Delta\Sigma_\alpha$, $\Gamma_\alpha^5$ and $E_\alpha$ (dashed line) for $m/g = 0.125$ and $m/g = 0.25$. Although we are for $m/g = 0.125$ beyond the strong-coupling regime, we observe that our results converge towards the perturbative results as $m/g \rightarrow 0$. In particular, for the excitation energy $\mathcal{E}_{1,\alpha}$ the agreement is striking for $m/g = 0.125$.\\
\\
As another check, we compare in table \ref{table:compareByrnes} some quantities for $\alpha = 0.5$ with the results of Byrnes \cite{Byrnes2002a,Byrnes2002b,Byrnes2003}. When $m/g = 0.25,0.30$ the electric field $E_\alpha$ and the axial fermion current density $\Gamma_\alpha^5$ are zero due to the $CT$ symmetry. We recovered this in our numerical simulations for all our values of $1/\sqrt{x}$ up to $10^{-7}$. Therefore a continuum extrapolation is useless. For $m/g = 0.5$ and $\alpha = 0.5$, the elementary excitations are kinks which cannot be captured with the ansatz Eq.~(\ref{eq:excAnsatz}). The lowest solutions to the generalized eigenvalue equation Eq.~(\ref{eq:genEigExc}) correspond to excitations with at least twice the energy of the kinks and, hence, are also not faithfully represented by the ansatz Eq.~(\ref{eq:excAnsatz}). Therefore we do not have a reliable estimate for the mass gap for $\alpha = 0.5$ and $m/g = 0.5$.

\begin{table}[t]
\begin{tabular}{| c| l||   c | c | c |}
\hline 
$m/g$&  & $\mathcal{E}_1$ & $E_\alpha$ & $\Gamma_\alpha$\\
\hline
$0.25$ & Buyens &   0.1338(7) & -  & - \\
 	    & Byrnes \cite{Byrnes2003} & 0.134(2)  &  -  & -\\
	    \hline
$0.3$ &Buyens&  0.0527(5)  & -  & -\\
 	 & Byrnes \cite{Byrnes2003}  & 0.05(2)&   - &   -\\
	 \hline
$0.5$ &Buyens & -  & 0.4206(2)   & 0.136(2) \\
 	 & Byrnes \cite{Byrnes2003} & 0.246(3)&   0.421(1) &   0.135(2)\\
\hline
\end{tabular}
\captionsetup{justification=raggedright}
\caption{\label{table:compareByrnes} $ \alpha = 0.5$. Comparison with the results of Byrnes \cite{Byrnes2003} for $m/g = 0.25, 0.3, 0.5$ and $\alpha = 0.5$. For $m/g = 0.125$ and $m/g = 0.25$ the ground is $CT$ invariant and, hence, $E_\alpha = \Gamma_\alpha^5 = 0$. In our numerics we recovered this up to $10^{-7}$ and, hence, a continuum extrapolation makes no sense. For $m/g = 0.5$ and $\alpha = 0.5$ the elementary excitations are kinks which cannot be approximated by the ansatz Eq.~(\ref{eq:excAnsatz}). Therefore we do not have an estimate for that.  }
\end{table}

\begin{figure}[t]
\includegraphics[scale = 0.25]{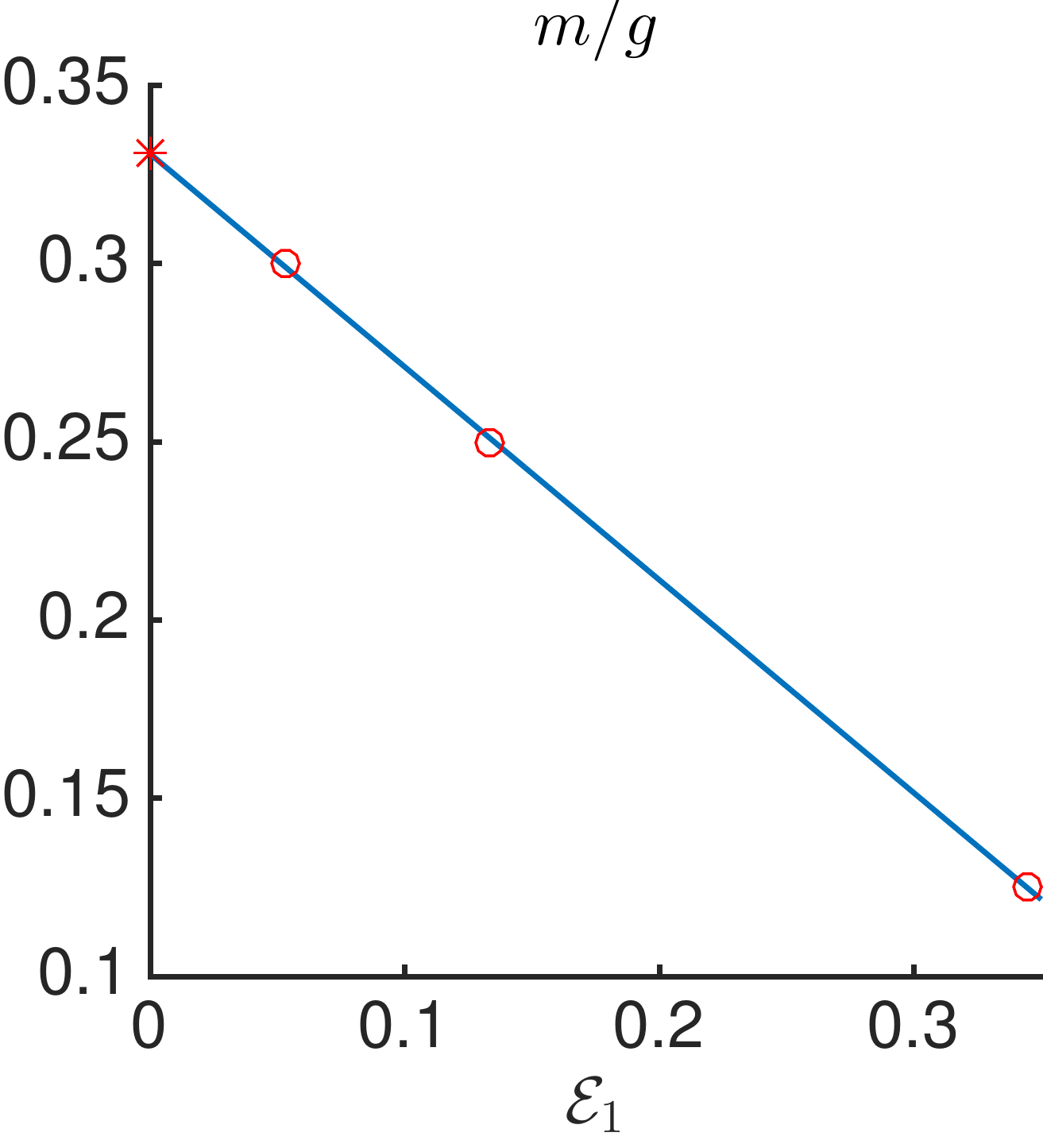}
\captionsetup{justification=raggedright}
\caption{\label{fig:determineCriticallMass} We show here the mass gap $\mathcal{E}_1$ as a function of $m/g$ for $m/g = 0.125,0.25,0.3$ (red circles). A linear fit (blue line) enables us to extrapolate the curve to $\mathcal{E}_1 = 0$ which gives us the estimate $(m/g)_c \approx 0.3308\ldots $ (red star) for the critical mass. }
\end{figure}

We were also able to obtain a rough estimate for the critical mass $(m/g)_c$. Therefore we fitted $m/g$ against $\mathcal{E}_1$ for $m/g = 0.125,0.25,0.3$. As can be observed from fig. \ref{fig:determineCriticallMass}, $\mathcal{E}_1$ behaves almost linear in $m/g$ \cite{Byrnes2002a,Byrnes2002b,Byrnes2003}. Hence, the critical mass $(m/g)_c$ is obtained by the intersection of the linear fit with the ($\mathcal{E}_1 = 0$)-axis. Indeed, the mass gap vanishes at the phase transition. A linear fit gives yields $(m/g)_c = 0.3308 \ldots$ which is in agreement with the result of Byrnes, $(m/g)_c \approx 0.3335(2)$, up to $3\times 10^{-3}$. By performing additional simulations for $m/g \in [0,0.3]$ we could improve this results, but this falls beyond the scope of this paper. 

\subsection{Results}\label{subsec:ResultsGroundState}
\begin{figure}[t]
\begin{subfigure}[b]{.24\textwidth}
\includegraphics[width=\textwidth]{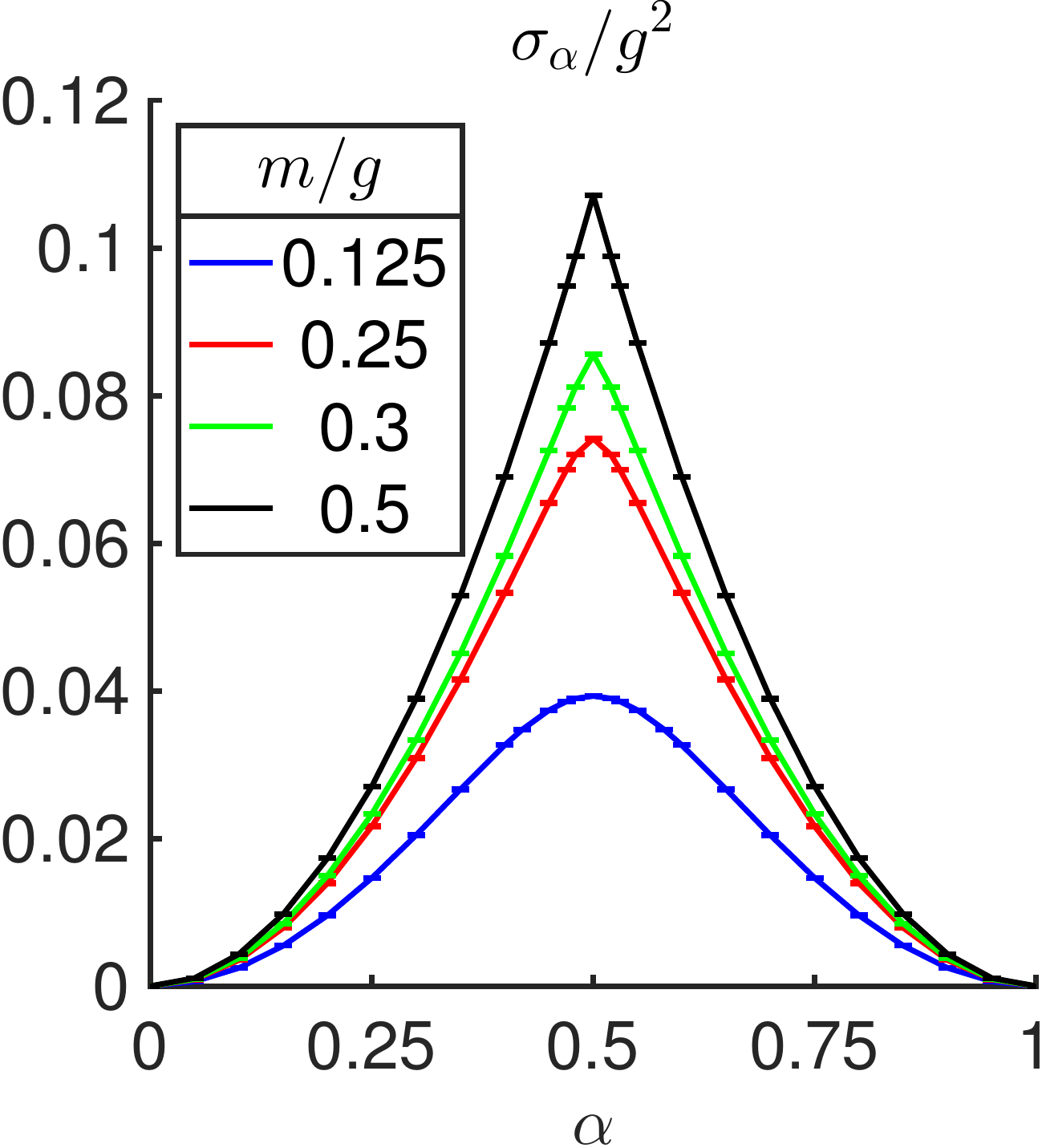}
\caption{\label{fig:GSexpVala}}
\end{subfigure}\hfill
\begin{subfigure}[b]{.24\textwidth}
\includegraphics[width=\textwidth]{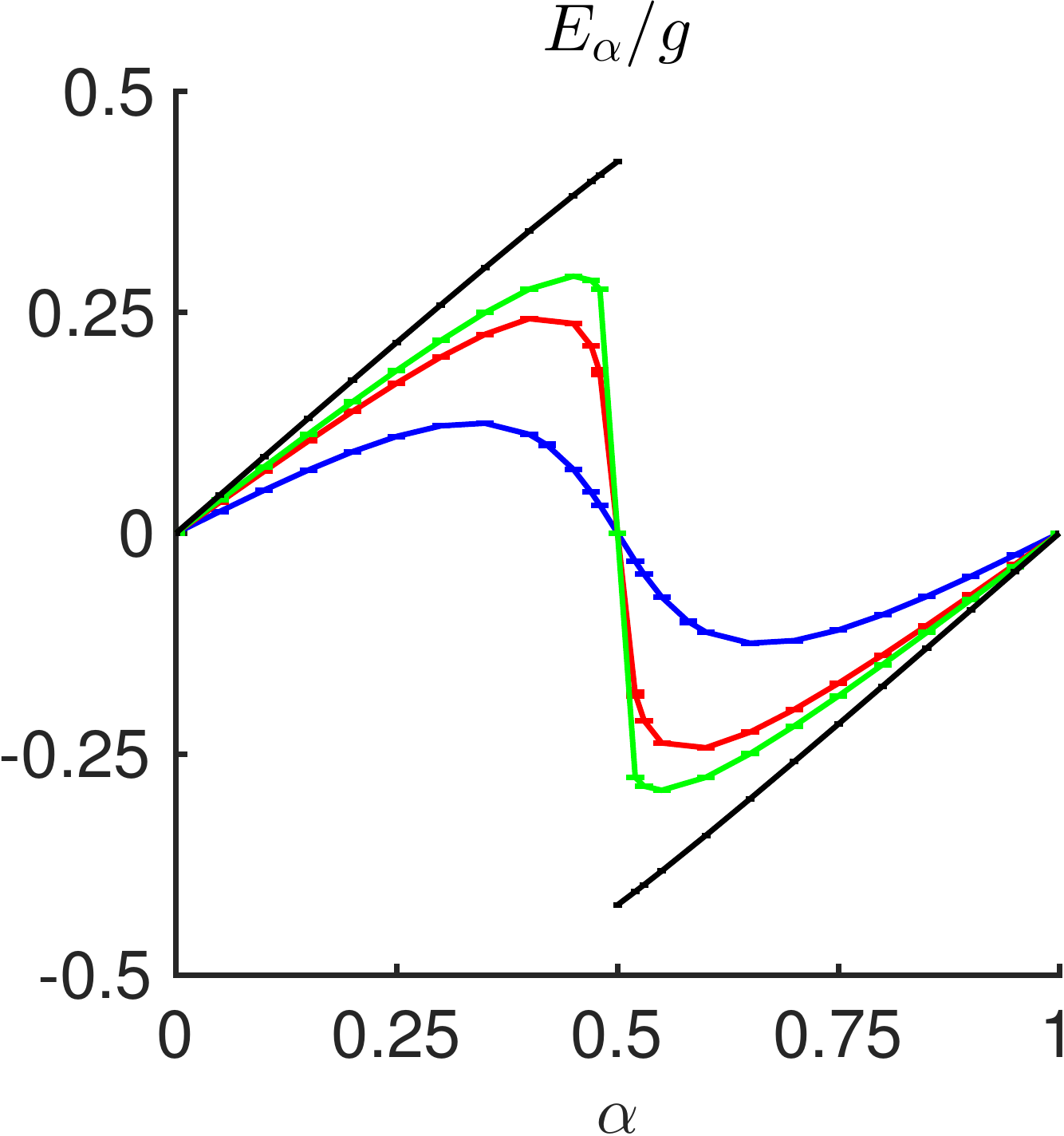}
\caption{\label{fig:GSexpValb}}
\end{subfigure}\vskip\baselineskip
\begin{subfigure}[b]{.24\textwidth}
\includegraphics[width=\textwidth]{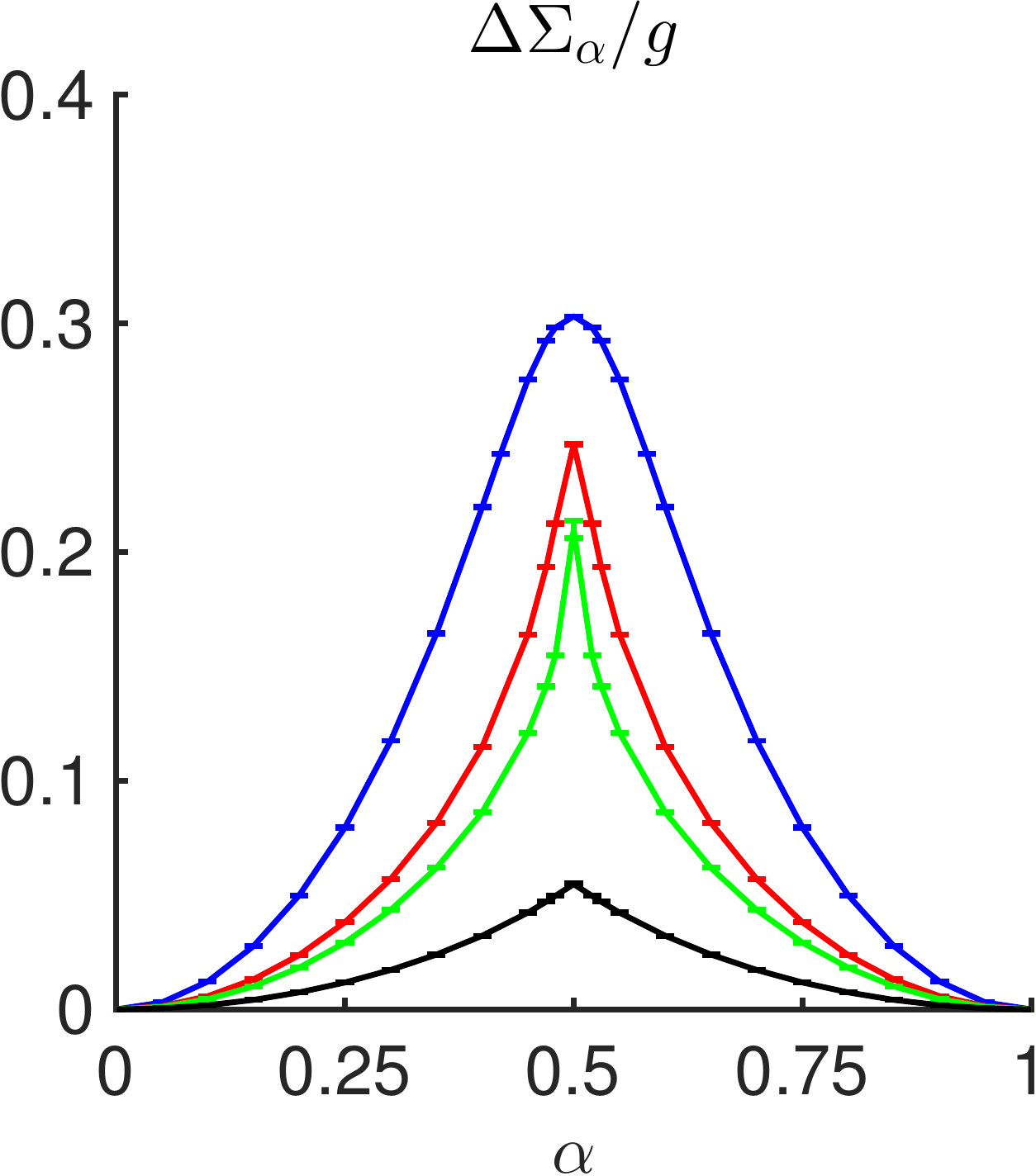}
\caption{\label{fig:GSexpValc}}
\end{subfigure}\hfill
\begin{subfigure}[b]{.24\textwidth}
\includegraphics[width=\textwidth]{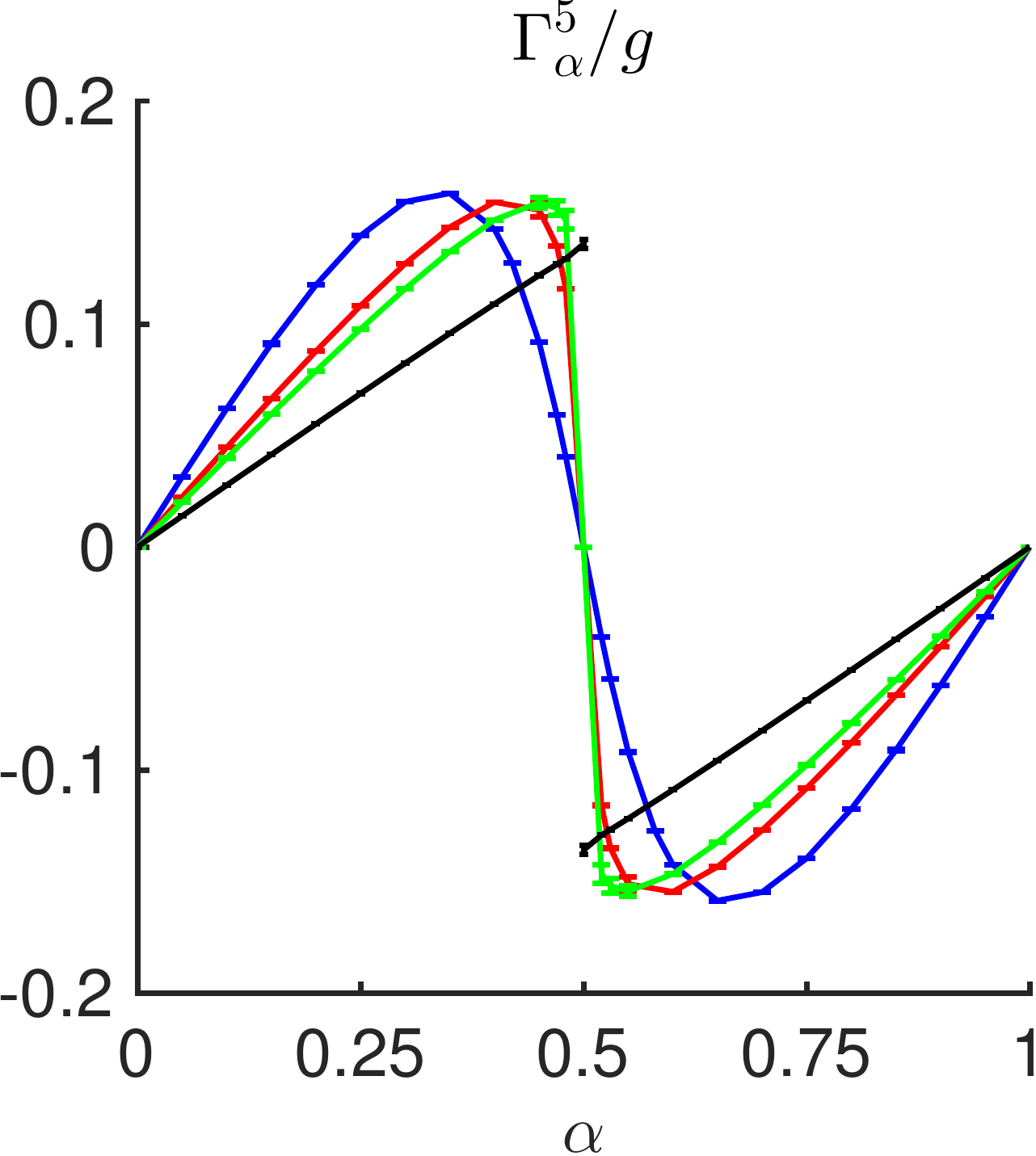}
\caption{\label{fig:GSexpVald}}
\end{subfigure}\vskip\baselineskip
\captionsetup{justification=raggedright}
\caption{\label{fig:GSexpVal}Results for the ground state for $\alpha \in [0,1]$ for $m/g = 0.125, 0.25, 0.3, 0.5$. (a) String tension $\sigma_\alpha$. (b) Electric field $E_\alpha$. (c) Renormalized chiral condensate $\Delta\Sigma_\alpha$. (d) Axial fermion current density $\Gamma_\alpha^5$.   }
\end{figure}

In \cite{Buyens2015} we found that the string tension $\sigma_\alpha$, see fig. \ref{fig:GSexpVala}, interpolates smoothly between the behavior in the strong-coupling limit for small values of $m/g$ and the weak-coupling limit for large values of $m/g$. In particular, for $m/g = 0.5$ we find that the string tension is non-differentiable for $\alpha = 1/2$ which is a consequence of the spontaneous breaking of the $CT$ symmetry. Indeed, an order parameter for this spontaneous symmetry breaking is the electric field $E_\alpha$, see fig. \ref{fig:GSexpValb}, which is related to the string tension by
$$E_\alpha = \frac{\partial \sigma_\alpha}{\partial \alpha}. $$
Hence, the discontinuity of $E_\alpha$ at $\alpha = 1/2$ implies that $\sigma_\alpha$ is non-differentiable at $\alpha = 1/2$. This holds for all values of $m/g \geq (m/g)_c$. Similarly, we find that the renormalized chiral condensate $\Delta \Sigma_\alpha$, see fig. \ref{fig:GSexpValc}, which is related to the string tension by
$$ \Delta \Sigma_\alpha =  \frac{\partial \sigma_\alpha}{\partial m}, $$
is non-differentiable at $\alpha = 1/2$ for $(m/g) \geq (m/g)_c$. 

Note that for $m/g = 0.125$ and $m/g = 0.25$ it is hard to see with the naked eye whether $\sigma_\alpha$ and $\Delta\Sigma_\alpha$ is differentiable at $\alpha = 1/2$. However, the differentiability follows from the fact that the electric field is zero and continuous at $\alpha = 1/2$. 

Finally, in fig. \ref{fig:GSexpVald} we show the axial fermion current density. This quantity also switches sign under a $CT$ transformation and, hence, serves as an order parameter as well. In fact, one observes that $\Gamma_\alpha^5$ correlates with $E_\alpha$. However note that $E_\alpha$ increases with $m/g$ while $\Gamma_\alpha^5$ decreases with $m/g$. Similarly, $\Delta \Sigma_\alpha$ correlates with $\sigma_\alpha$, but again, $\sigma_\alpha$ increases with $m/g$ while $\Delta\Sigma_\alpha$ decreases with $m/g$. 

\section{Reduced density matrix of a MPS}\label{app:RedDensMPSSchwingerModel}
Here, we briefly discuss how to compute the reduced density matrices of a MPS of the form Eq.~(\ref{eq:gaugeMPSFinalForm}):
\begin{subequations}\label{eq:gaugeMPSFinalFormApp}
\be\ket{\Psi[a]} = \sum_{\bm{\kappa}}v_L^\dagger \left(\prod_{n = 1}^N A_{\kappa_{2n-1},\kappa_{2n}}\right)v_R \ket{\bm{\kappa}},\ee
$N \rightarrow + \infty$, $\kappa_n = (s_n,p_n)$, $s_n \in \{-1,1\}, p_n \in \mathbb{Z}[p_{min},p_{max}]$, with
\bea\label{eq:gaugeMPSFinalFormAppb} [A_{s_1,p_1,s_2,p_2}]_{(q,\alpha_q);(r,\beta_r)} &= \delta_{p_1,q + (s_1 -1)/2} \delta_{p_2,q + (s_1+s_2)/2} \nonumber
\\ & \delta_{p_2,r} [a_{q,s_1,s_2}]_{\alpha_q,\beta_r}\eea
\end{subequations}
where $a_{q,s_1,s_2} \in \mathbb{C}^{D_q \times D_{q + (s_1 + s_2)/2}}$. 

We assume that the state is proper normalized, i.e. the largest eigenvalue of the transfer matrix
$$\mathbb{E} = \sum_{\kappa_1,\kappa_2} A_{\kappa_1,\kappa_2}\otimes \overline{A_{\kappa_1,\kappa_2}}, $$  
equals one and the matrices $\Lambda_L$ and $\Lambda_R$ corresponding to the left and right leading eigenvector are positive definite. Moreover, Eq.~(\ref{eq:gaugeMPSFinalFormAppb}) implies that
$$[\Lambda_L]_{(q,\alpha_q);(r,\beta_r)} = \delta_{q,r} [\lambda_{L,q}]_{\alpha_q,\beta_r}, $$
$$[\Lambda_R]_{(q,\alpha_q);(r,\beta_r)} = \delta_{q,r} [\lambda_{R,q}]_{\alpha_q,\beta_r}  $$
for $\lambda_{R,q},\lambda_{L,q} \in \mathbb{C}^{D_q \times D_q}$ positive definite matrices. 
Consider an operator $O$ of the form
$$O = \sum_{n = 1}^{N} T^{2n-2}oT^{-2n+2}, $$
where $o$ acts on the effective sites $1$ and $2$ (i.e. sites $1,2,3,4$ and links $1,2,3,4$) and where $T$ is the translation operator (over one site). If $O$ is gauge-invariant, i.e. for all $n$: 
$$[O,G(n)] = 0, G(n) = L(n) - L(n-1) - \frac{\sigma_z(n) + (-1)^n}{2},$$
then
\begin{subequations}\label{eq:redDensMatr2app}
\bea \frac{1}{2N}\Braket{\Psi[\bar{a}] \vert O \vert \Psi[a]}   =&  \mbox{tr}\left(\rho_{2}[a]\cdot o\right)  \nonumber
\\  = & \displaystyle{\sum_{q = p_{min}}^{p_{max}} \mbox{tr}\left(\rho_{2,q}[a]\cdot o_q \right) }\nonumber \eea
where $\rho_{2,q}[a]$ and $o_q \in \mathbb{C}^{2^{\otimes 4} \times 2^{\otimes 4}}$ have components
\begin{widetext}
\bea \Braket{s_1,s_2,s_3,s_4 \vert \rho_{2,q}[a] \vert t_1,t_2,t_3,t_4}  =&  
\mbox{tr}\left(\lambda_{L,q}a_{q,t_1,t_2}a_{q+(t_1+t_2)/2,t_3,t_4}\lambda_{R,q + (t_1+t_2+t_3+t_4)/2} [a_{q+(s_1+s_2)/2,s_3,s_3}]^\dagger [a_{q,s_1,s_2}]^\dagger \right)  \nonumber \\
&\delta_{t_1+t_2+t_3+t_4,s_1 + s_2 + s_3 + s_4}\eea
\end{widetext}
($s_k, t_k \in \{-1,1\}, k = 1,2,3,4$) and 
\begin{widetext}
\be \Braket{s_1,s_2,s_3,s_4 \vert o_q \vert t_1,t_2,t_3,t_4} = \Braket{s_1,p_1, s_2, p_2, s_3, p_3, s_4, p_4 \vert o \vert t_1,r_1,t_2, r_2, t_3, r_3, t_4, r_4} \delta_{s_1+s_2+s_3+s_4, t_1 + t_2 + t_3 + t_4}.  \ee 
\end{widetext}
with 
\be p_1 = q + \frac{s_1 -1}{2}, r_1 = q +  \frac{t_1 -1}{2}\ee
\be p_2 = q +  \frac{s_1 + s_2}{2}, r_2 = q +  \frac{t_1 + t_2}{2}, \ee
\be p_3 = q +  \frac{s_1 + s_2 + s_3 -1}{2}, r_3 = q +  \frac{t_1 + t_2 + t_3 -1}{2}, \ee
\be p_4 = q +  \frac{s_1 + s_2 + s_3 + s_4}{2} = q +  \frac{t_1 + t_2 + t_3 + t_4}{2} = r_4, \ee
\end{subequations}
$p_k,r_k \in \mathbb{Z}[p_{min},p_{max}]$; $s_k,t_k \in�\{-1,1\}$. 

We find that the contribution of each of the eigenvalue sectors $q$ of $L(n)$ to 
this expectation value equals
\begin{subequations}\label{eq:boundTrHolder}
\be \mbox{tr}[\rho_{2,q}[a]\cdot o_q] \ee
for which the magnitude is bounded by (H\"older's inequality) 
\be \vert \mbox{tr}[\rho_{2,q}[a]\cdot o_q] \vert \leq \vert\vert \rho_{2,q}[a] \vert \vert_1 \cdot \vert\vert o_q \vert\vert_{\infty}.\ee
\end{subequations}
Note that $\vert\vert o_q \vert\vert_{\infty}$ equals the largest singular value (i.e. the largest eigenvalue of $O(q)$ in magnitude). For instance, to compute the expectation value of the electric field,
$$E = \frac{g}{2}\Braket{\Psi[\bar{a}]\vert L(1) + L(2) + 2\alpha \vert \Psi[a]},$$ 
we have
$$\vert\vert o_q \vert\vert_{\infty} \leq g (\vert q \vert + 1 + \vert \alpha \vert) .$$
For the expectation value of the electric field squared $E^2$,
$$E^2 = \frac{g^2}{2}\Braket{\Psi[\bar{a}]\vert (L(1) + \alpha)^2 + (L(2) + \alpha)^2 \vert \Psi[a]},$$ 
we find similarly 
$$\vert\vert o_q \vert\vert_{\infty} \leq g^2 \left(\vert q \vert + 1 + \vert \alpha \vert \right)^2.$$
For the Hamiltonian $H_\alpha$, Eq.~(\ref{eq:equationH}), we have
$$\vert\vert o_q \vert\vert_{\infty} \leq \frac{g}{2\sqrt{x}}\left(\vert q \vert + 1 + \vert \alpha \vert \right)^2 + m + g\frac{\sqrt{x}}{2}. $$

We conclude that for the quantities we are interested in (electric field, energy,$\ldots$) that $\vert\vert o_q \vert\vert_{\infty}$ scales at most polynomially $q$. Provided that $\vert \vert \rho_{2,q} [a] \vert \vert_1$ decreases fast (e.g. exponentially) with $q$, it follows from Eq.~(\ref{eq:boundTrHolder}) that we can indeed conclude that the contribution of the eigenvalue sectors $q$ of $L(n)$ for large $\vert q \vert$ is negligible.  

\section{Continuum extrapolation of the quantities}\label{sec:ContinuumExtrapolationApp}
In this appendix we explain how we performed the continuum extrapolation of all the quantities discussed in section \ref{subsec:continuumLimit}. We employ the method used in \cite{Buyens2016} which is based on the methods discussed in \cite{Banuls2013a,Banuls2016}. 

Consider a quantity $\mathcal{O}(x)$ for which we compute its values for 
$$x = x_1,\ldots, x_M.$$ 
The goal is to obtain a continuum value $\mathcal{O} = \lim_{x \rightarrow +\infty}\mathcal{O}(x)$ and to estimate a reliable error on this extrapolation. For the quantities we considered here we observed that they behave polynomially (see for instance figs. \ref{fig:extrapolationGSexpVal} and \ref{fig:extrapolationE1} in main text) as a function of $1/\sqrt{x}$, therefore we fit our data against the following polynomials in $1/\sqrt{x}$: 
\begin{subequations}\label{eq:fitfunctionapp}
\be\label{eq:fitfunctionaappa} f_1(x) = A_1 + B_1\frac{1}{\sqrt{x}} \ee
\be\label{eq:fitfunctionbappb}f_2(x) = A_2 + B_2\frac{1}{\sqrt{x}} + C_2\frac{1}{x} \ee
and
\be\label{eq:fitfunctioncappv}f_3(x) = A_3 + B_3\frac{1}{\sqrt{x}} + C_3\frac{1}{x} + D_3 \frac{1}{x^{3/2}}.\ee
\end{subequations}

Let us discuss in more detail how we obtain a continuum estimate for each of the fitting ans\"atze $f_n$ (subsection \ref{subsec:confitanz}) and a final continuum estimate (subsection \ref{subsec:finCon}).

\subsection{Obtaining a continuum estimate for the fitting ansatz $f_n$}\label{subsec:confitanz}
\noindent For every type of fitting ansatz, i.e. a particular $f_n$ $(n = 1,2,3)$ Eq.~(\ref{eq:fitfunctionapp}), we determine an estimate $\mathcal{O}^{(n)}$ for the continuum value and an error $\Delta^{(n)}\mathcal{O}$ which originates from the choice of fitting interval. Given our dataset $\{(x_j,\mathcal{O}(x_j)): j = 1,\ldots, M\}$ of $M$ points. We perform all possible fits of $f_n$ against at least $n+5$ consecutive data points where the coefficients ${A_n,B_n,C_n,D_n}$ ($C_n = 0$ if $n < 2$, $D_n = 0$ if $n < 3$) are estimated using an iterative generalized least-squares algorithm.

By taking at least $n+5$ consecutive data points we reduce the problem of overfitting: the fitted function $f_n$ fits the considered points extremely well, but fails to fit the overall data. Furthermore we also discard the fits that give statistically insignificant coefficients (p-value $\geq$ 0.05). In practice, this means that we discard the fits $f_n$ where the error on one of its coefficients $(A_n,B_n,C_n,\ldots)$ is larger than approximately half of its value. 

For every fit $\theta$ of $f_n$ against a subset of at least $n+5$ consecutive $x-$values, say $\{x_j\}_{j \in \mbox{fit}\theta}$, which produces statistically significant coefficients we obtain values
$$A_n^{(\theta)},B_n^{(\theta)},C_n^{(\theta)},D_n^{(\theta)},$$
with $C_n^{(\theta)} = 0$ for $n < 2$ and  $D_n^{(\theta)} = 0$ for  $n < 3$,
and a corresponding fitting function $g_\theta(x)$.
$$g_\theta(x)  = A_n^{(\theta)} +  B_n^{(\theta)}\frac{1}{\sqrt{x}} + C_n^{(\theta)}\frac{1}{x} + D_n^{(\theta)} \frac{1}{x^{3/2}}$$

All the values $A_n^{(\theta)}$ are an estimate for the continuum value of $\mathcal{O}$ for the fitting ansatz $f_n$. Let us denote with $\{A_n^{(\theta)}\}_{\theta = 1\ldots R_n}$ all the $A_n$'s obtained from a fit $\theta$ against $f_n$ which produces significant coefficients with
$$A_n^{(1)} \leq A_n^{(2)} \leq \ldots \leq A_n^{(R_n)} .$$
For each fit $\theta$ we also compute its $\chi^2$ value:
\be \label{eq:chisq} \chi_\theta^2 = \sum_{j \in \mbox{fit}\theta} \left(\frac{g_\theta(x_j) - \mathcal{O}(x_j)}{\Delta\mathcal{O}(x_j)}\right)^2 \ee
where $\Delta\mathcal{O}(x_j)$ is a measure for the error in $\mathcal{O}(x_j)$ originating from taking a finite value for the virtual dimensions $D_q$. For the ground-state expectation values we take $\Delta \mathcal{E}_0$, see Eq.~(\ref{eq:defHsqE0}), while for the excitation energies $\mathcal{E}_{1,\alpha}$ and $\mathcal{E}_{2,\alpha}$ we take $\Delta \mathcal{E}_m$, see Eq.~(\ref{eq:defHsqExcm}). When our dataset is large enough the quantity $\chi_\theta^2/N_{dof}^{\theta}$, with $N_{dof}^{\theta}$ the number of degrees of freedom of the fit (here the number of data points used in the fit minus $n+2$), gives an indication whether $g_\theta$ fits the dataset well ($\chi_\theta^2/N_{dof}^{\theta} \ll 1$) or not ($\chi_\theta^2/N_{dof}^{\theta} \gg 1$). 

If we have at least 10 fits $\theta$ with $\chi_\theta^2/N_{dof}^{\theta} \leq 1$ we can obtain a reliable continuum estimate by taking the median of $\{A_n^{(1)},  \ldots, A_n^{(R_n)}\}$ weighted by $\exp(-\chi_\theta^2/N_{dof}^{\theta})$, see also \cite{Banuls2013a,Buyens2016}. More specifically we build the cumulative distribution $X_\theta$,  
$$ X_\theta = \frac{\sum_{\kappa = 1}^\theta  \exp(-\chi_\kappa^2/N_{dof}^{\kappa})}{\sum_{\kappa = 1}^{R_n} \exp(-\chi_\kappa^2/N_{dof}^{\kappa})},$$
and take as our continuum estimate $\mathcal{O}^{(n)}$ for the fitting ansatz $f_n$:  
$\mathcal{O}^{(n)} = A_n^{(\theta_0)}$ where $\theta_0$ corresponds to the value for which $X_{\theta_0}$ is the closest to $1/2$, i.e.
$$\theta_0 = \mbox{arg}\min_\theta \vert X_{\theta} - 1/2 \vert. $$
The systematic error $\Delta^{(n)}\mathcal{O}$ from the choice of $x$-interval comes from the $\% (68,3)$-confidence interval, it is computed as
$$ \Delta^{(n)}\mathcal{O} = \frac{1}{2}\left(A_n^{(\theta_2)} - A_n^{(\theta_1)}\right)$$
with 
$$\theta_1 = \mbox{arg}\min_\theta \vert X_{\theta} - 0.85 \vert,  \theta_2 = \mbox{arg}\min_\theta \vert X_{\theta} - 0.15 \vert. $$

If we have less than 10 fits $\theta$ with $\chi_\theta^2/N_{dof}^{\theta} \leq 1$, only a few fits dominate the histogram of the $\chi^2$-distribution. Therefore we adopt the more conservative approach from \cite{Banuls2016}. We only consider the fits with statistically significant coefficients and with $\chi_\theta^2/N_{dof}^{\theta} \leq 1$; the corresponding continuum estimates are
$$A_n^{(1)} \leq A_n^{(2)} \leq \ldots \leq A_n^{(R'_n)}, \mbox{ with }R'_n \leq R_n.$$
Of these estimates we take the $A_n^{(\theta_0)}$ which corresponds to the $\theta$ for which the mean squared of the variances $\Delta \mathcal{O}$ is minimal, i.e.
$$\theta_0 = \mbox{arg}\min_\theta \frac{1}{\vert \mbox{fit} \theta \vert }\left(\sqrt{\sum_{j \in \mbox{fit} \theta}\left(\Delta\mathcal{O}(x_j)\right)^2}\right). $$
As the systematic error originating from the choice of fitting range we take the difference in magnitude of this estimate with the most outlying $A_n^{(\theta)}$ (for the same type of fitting ansatz):
$$ \Delta^{(n)}\mathcal{O} = \max_{1 \leq \theta \leq R'_n} \vert A_n^{(\theta_0)} - A_n^{(\theta)} \vert.$$

\subsection{Final continuum estimate and uncertainty}\label{subsec:finCon}
\noindent Using the method discussed in subsection \ref{subsec:confitanz} we now have three estimates for $\mathcal{O}$ ($\mathcal{O}^{(1)},\mathcal{O}^{(2)}$ and $\mathcal{O}^{(3)}$) corresponding to the fitting functions $f_1$, $f_2$ and $f_3$. As our final estimate we take the estimate from the fitting function $f_{n_0}$ which had the most statistically significant fits with $\chi_\theta^2/N_{dof}^{\theta} \leq 1$. The error originating from the choice of fitting function is then computed as the maximum of the difference with the continuum estimates from the other fitting functions. As our final result we report $\mathcal{O} = \mathcal{O}^{(n_0)}$ and the error $\Delta \mathcal{O}$ is the maximum of
\begin{itemize}
\item[i.] $\max\left(\max_{j}\delta\mathcal{O}(x_j)\right),$ where 
$$\delta\mathcal{O}(x_j) = \max_{n = 1,2,3}\vert \vert \mathcal{O}(x_j)[a_n] -  \mathcal{O}(x_j)[a_n] \vert \vert.$$ 
$\mathcal{O}(x_j)[a_n]$ is the expectation value of $\mathcal{O}(x_j)[a_n]$ with respect to the MPS ground-state approximation $\ket{\Psi[a_n]}$ (see Eq.~(\ref{eq:gaugeMPSFinalForm})) obtained with the parameters $\epsilon$ and $p_{max}$ as shown in Eq.~(\ref{eq:epsandpmax}). In particular, for the excitations energies $\mathcal{E}_m$ we find $\delta\mathcal{O}(x_j) = \delta \mathcal{E}_m(x_j)$ as defined in Eq.~(\ref{eq:deltaEmsmall}),
\item[ii.] the error originating from the choice of $x$-range: $\Delta^{(n_0)}\mathcal{O},$
\item[iii.] the error originating from the choice of fitting ansatz: $\max_{n = 1,2,3}\vert \mathcal{O} - \mathcal{O}^{(n)}\vert.$
\end{itemize}

\bibliography{paperSMEB.bib}

\end{document}